\newcommand{\arXiv}{}

\ifdefined\arXiv
   \documentclass[acmsmall,nonacm=true]{acmart}
\else
   \documentclass[acmsmall]{acmart}
\fi

\usepackage{graphicx} % Required for inserting images
\usepackage{forest}
\forestset{qtree/.style={for tree={align=center}}}
\usepackage{subcaption}
\usepackage{booktabs}
\usepackage{makecell}
\usepackage{draftwatermark}
\usepackage{multirow}
\usepackage{enumitem}
\usepackage{array}
\usepackage{rotating}
\SetWatermarkText{Preprint version}
\SetWatermarkScale{0.6}
\SetWatermarkColor[gray]{0.8}

\begin{document}

\title{Design Methodologies for Deep Learning Accelerators on Heterogeneous Architectures}

\author{Serena Curzel}
\email{serena.curzel@polimi.it}
\author{Fabrizio Ferrandi}
\email{fabrizio.ferrandi@polimi.it}
\author{Leandro Fiorin}
\email{leandro.fiorin@polimi.it}
\author{Daniele Ielmini}
\email{daniele.ielmini@polimi.it}
\author{Cristina Silvano}
\email{cristina.silvano@polimi.it}
\affiliation{%
  \institution{Politecnico di Milano}
  \country{Italy}
}
            
\author{Francesco Conti}
\email{f.conti@unibo.it}
\author{Luca Bompani}
\email{luca.bompani5@unibo.it}
\author{Luca Benini} 
\email{luca.benini@unibo.it}
\affiliation{%
\institution{Università di Bologna}
            \country{Italy}
}

\author{Enrico Calore}
\email{enrico.calore@fe.infn.it}
\additionalaffiliation{%
    \institution{Università degli Studi di Ferrara}
    \country{Italy}
}
\affiliation{%
    \institution{INFN}
    \country{Italy}
}
\author{Sebastiano Fabio Schifano}
\email{schsst@unife.it}
\additionalaffiliation{%
    \institution{INFN}
    \country{Italy}
}
\affiliation{%
    \institution{Università degli Studi di Ferrara}
    \country{Italy}
}
\author{Cristian Zambelli} 
\email{cristian.zambelli@unife.it}
\affiliation{%
\institution{Università degli Studi di Ferrara}
            \country{Italy}
            }

\author{Maurizio Palesi}
\email{maurizio.palesi@unict.it}
\author{Giuseppe Ascia}
\email{giuseppe.ascia@unict.it}
\author{Enrico Russo} 
\email{enrico.russo@phd.unict.it}
\affiliation{%
\institution{Università degli Studi di Catania}
            \country{Italy}}

\author{Valeria Cardellini}
\email{cardellini@ing.uniroma2.it}
\author{Salvatore Filippone}
\email{salvatore.filippone@uniroma2.it}
\author{Francesco Lo Presti} 
\email{lopresti@info.uniroma2.it}
\affiliation{%
\institution{Università degli Studi di Roma Tor Vergata},
\country{Italy}}

\author{Stefania Perri}
\email{stefania.perri@unical.it}
\affiliation{%
\institution{Università degli Studi della Calabria}
  \country{Italy}
}

\renewcommand{\shortauthors}{Serena Curzel et al.}

\begin{abstract}
Given their increasing size and complexity, the need for efficient execution of deep neural networks has become increasingly pressing in the design of heterogeneous High-Performance Computing (HPC) and edge platforms, leading to a wide variety of proposals for specialized deep learning architectures and hardware accelerators.
The design of such architectures and accelerators requires a multidisciplinary approach combining expertise from several areas, from machine learning to computer architecture, low-level hardware design, and approximate computing.
Several methodologies and tools have been proposed to improve the process of designing accelerators for deep learning, aimed at maximizing parallelism and minimizing data movement to achieve high performance and energy efficiency.
This paper critically reviews influential tools and design methodologies for Deep Learning accelerators, offering a wide perspective in this rapidly evolving field.
This work complements surveys on architectures and accelerators by covering hardware-software co-design, automated synthesis, domain-specific compilers, design space exploration, modeling, and simulation, providing insights into technical challenges and open research directions.
\end{abstract}

\maketitle

\begin{figure}[t]
    \centering
\includegraphics[width=0.7\linewidth]{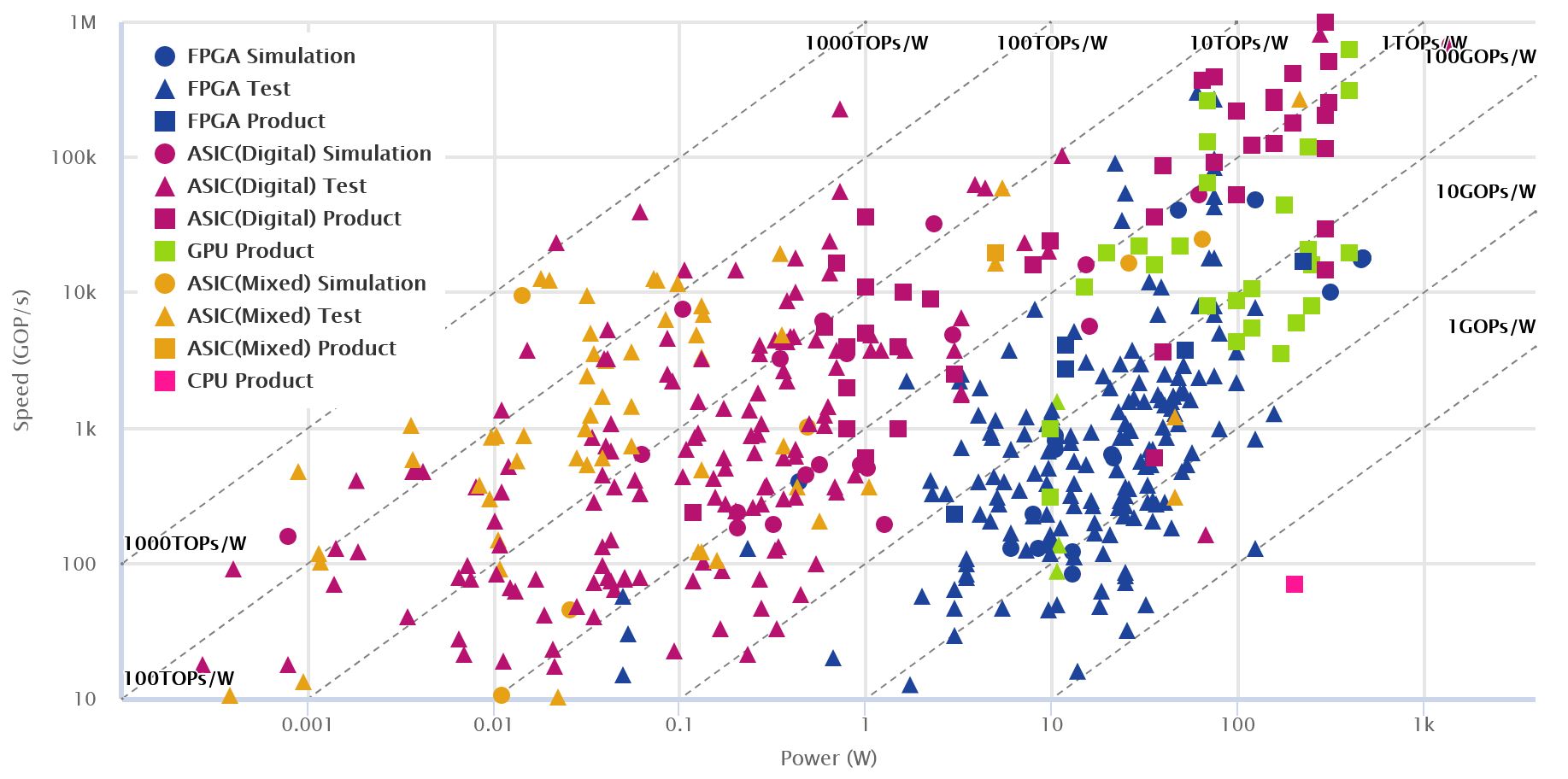}
    \caption{Speed and power consumption of state-of-the-art DNN accelerators~\cite{silvano2023survey}.}
    \label{fig:survey-graph}
\end{figure}

\section{Introduction}

Over the past few years, deep learning (DL) has contributed to remarkable progress in applications such as computer vision, natural language processing, speech and audio processing, recommendation systems, autonomous systems, environmental sciences, and many more.
% Deep Learning refers to a subset of machine learning that uses deep neural networks (DNNs) with multiple layers of artificial neurons to mimic human brain behavior and learn how to execute a task from large amounts of data.
The size and complexity of modern deep neural networks (DNNs) require ever-increasing amounts of computational and memory resources, leading to an unsustainable level of power consumption if executed on traditional general-purpose High-Performance Computing (HPC) systems.
Therefore, when implementing a DL application, developers must possess not only knowledge of abstract algorithms and training methods but also the ability to map the DNN model to an efficient, possibly heterogeneous, hardware architecture.
The rise of transformer-based Large Language Models (LLMs) has further exacerbated these challenges by increasing model size, memory requirements, and computational cost~\cite{transformerSurvey2023}.

Designing hardware accelerators for DL requires expertise in machine learning, computer architecture, and low-level hardware design; such a multidisciplinary approach has fostered research into new design automation tools, compilers, and methodologies that help developers explore a vast design space to reach their performance and power consumption goals (Figure \ref{fig:survey-graph}).
Previous surveys have reviewed DL accelerators from an architectural perspective, classifying platforms such as GPUs, TPUs, ASICs, FPGAs, RISC-V-based accelerators, in-memory computing architectures, neuromorphic processors, and emerging technologies, with emphasis on architectural organization, performance, energy efficiency, and deployment targets~\cite{chen2020engineering, akkad2023embedded, silvano2023survey}.
Compiler surveys have focused on the structure of DL compiler stacks, intermediate representations, frontend/backend optimizations, and code generation for heterogeneous hardware~\cite{9222299,zhang2023compiler}.
This paper provides a different point of view: a broad, system-level, and cross-layer perspective on methodologies for DL accelerator design, integrating traditionally fragmented design phases (partitioning, modeling, simulation, compilers, HLS, and approximate computing).
Rather than organizing the literature review primarily by accelerator architecture or target device, we organize it following the methodological steps that support the design and deployment of DL accelerators on heterogeneous systems.

%This survey thus highlights various approaches that support the compilation and mapping of a DNN model to existing hardware accelerators, including GPUs, Tensor Processor Units, custom designs implemented on Field Programmable Gate Arrays (FPGAs) or Application-Specific Integrated Circuits (ASICs), Neural Processing Units, In-Memory-Computing architectures, and co-processors based on the open hardware RISC-V architecture.
%Moreover, it describes methodologies supporting the implementation of new accelerators, such as High-Level Synthesis (HLS), application partitioning and mapping, design space exploration, and architectural simulation.
%In a scenario where the design space of possible accelerator architectures is vast, and each of them has different characteristics in terms of performance, power consumption (Figure \ref{fig:survey-graph}), but also monetary cost and programmability, choosing the right design tools is fundamental to achieving optimal results.

Although we do not claim to provide a comprehensive survey of all DL accelerator architectures, we aim to present a wide perspective on the design methodologies and tools that support accelerator design, evaluation, and deployment.
Most of the described methodologies and tools are explicitly designed for convolutional neural networks (CNNs) and DNNs; however, as they target the acceleration of common linear algebra operators and computational patterns, they may also be applied to the transformers at the core of LLMs and to graph neural networks (GNNs).
We hope that such an overview will be useful for a wide range of readers, whether researchers, computer architects, hardware or software developers, and that the list of references can serve as a starting point for further in-depth study.

The paper is organized into sections, each focused on a different topic and ordered according to their level of abstraction, from the highest (modeling, partitioning, and compilation) to the lowest (implementation of the hardware); Figure \ref{fig:structure} summarizes the main topics addressed in each section.
Finally, Sections \ref{sec:open} and \ref{sec:end} conclude the paper and present some remarks on emerging research trends.

\forestset{
  dir tree/.style={
    for tree={
      parent anchor=south west,
      child anchor=west,
      anchor=mid west,
      inner ysep=-3.5pt,
      grow'=0,
      align=left,
      edge path={
        \noexpand\path [draw, \forestoption{edge}] (!u.parent anchor) ++(1em,0) |- (.child anchor)\forestoption{edge label};
      },
     font=\sffamily,
      if n children=0{}{
        delay={
          prepend={[,phantom, calign with current]}
        }
      },
      fit=rectangle,
      before computing xy={
        l=2em
      }
    },
  }
}

\begin{figure}[t]
\resizebox{\textwidth}{!}{%
\centering
\small
\subfloat{
\label{fig:contribution_left}
\begin{forest}
  dir tree
  [
    [\textbf{\pmb{\S}~\ref{sec:hardware_software_co_design} Application Partitioning and Mapping}
    %[Hardware-aware neural architecture search]
    [Distributed training] %[Data / model / pipeline / hybrid parallelism]
    [Inference and deployment orchestration]
    ]
    [\textbf{\pmb{\S}~\ref{sec:modeling_simulation_exploration} Modeling / Simulation / Profiling / Exploration}
      [High-level frameworks]
      [Cycle-accurate simulators]
      [FPGA performance modeling]
      [In-Memory Computing modeling and simulation]
    ]
  ]
\end{forest}}
\subfloat{
\label{fig:contribution_right}
\begin{forest}
  dir tree [
    [\textbf{\pmb{\S}~\ref{sec:compilers} Deep Learning Compilers}
      [Compilers for HPC and edge systems]
      [Compilers for IMC architectures]
    ]
    [\textbf{\pmb{\S}~\ref{sec:HLS_Design_based_methodologies} HLS-based Design Methodologies}
      [HLS tools]
      [HLS-based design flows for DL]
    ]
    [\textbf{\pmb{\S}~\ref{sect:ApproxComp} Approximate Computing}
    [Algorithmic techniques]
    [Architecture- and gate-level design]
    ]
  ]
\end{forest}}
}
\caption{Organization of the paper.}
\label{fig:structure}
\end{figure}
\section{Application Partitioning and Mapping}
\label{sec:hardware_software_co_design}

%Hardware-aware design of DNNs has recently received increased attention to tackle system heterogeneity issues.
%Hardware-aware neural architecture search (HW-NAS)~\cite{chitty-venkata2023} takes hardware characteristics like latency, power, or area into account in the early stages of the design of a neural network, and it has also become a central aspect in automating the process of designing efficient architectures for DNN applications. 
%While traditional HW-NAS optimizes the neural network for a fixed accelerator, modern approaches perform hardware-software co-design to optimize both the network architecture and accelerator microarchitecture~\cite{jiang2020hardware}. %and improve energy efficiency and performance. 
%Different methodologies have been exploited to search for the optimal model architecture, ranging from reinforcement learning (RL) to evolutionary algorithms; for example, RL has been used to search for an accelerator configuration on FPGAs and optimize it for latency and area~\cite{abdelfattah2020best,jiang2020hardware}. 

%However, even if the DNN model architecture has been carefully optimized, 
Training large DL models and serving them for inference on a distributed HPC infrastructure composed of multiple, possibly heterogeneous compute nodes is a complex task, requiring designers to optimize the complete system stack: from DNN algorithms to model optimization and compression~\cite{han2016deep,jacob2018quantization}, to deployment, down to the design of the underlying hardware architecture itself. 
This full-stack perspective shifts the focus from isolated accelerators to the coordinated orchestration of heterogeneous systems, bridging algorithmic abstractions with the constraints of modern computing platforms. 
This section reviews approaches and frameworks for distributing, partitioning, and mapping DL training and inference applications on multiple processing nodes in the underlying computing infrastructure. 
We define \textit{application partitioning} as the process of dividing a model's computational graph and/or its associated data workload into smaller components, while \textit{mapping} refers to assigning, scheduling, and placing those components, and managing the resulting data movement, onto the underlying hardware topology.
While many %distributed 
training and inference systems have been proposed, this section focuses on a representative set of frameworks that illustrate different approaches to DL application partitioning and mapping.

\subsection{Distributed Training Strategies}

In the context of a distributed infrastructure with an ever-increasing number of available GPU-accelerated nodes, parallelization is the first solution to accelerate the execution of DNN applications.
%, as these models present many opportunities for concurrent execution.
In techniques based on \textit{data parallelism}, a set of workers (e.g., GPUs) load an identical copy of the DL model; the input data is partitioned into non-overlapping subsets, with each worker processing its portion independently~\cite{krizhevsky2012}.  
%The data it needs to process is split into non-overlapping portions, and each of them is fed into the model replicas of a worker~\cite{krizhevsky2012}. 
%
The main advantage of data parallelism is that it is model-agnostic, i.e., it can be applied to any DL model architecture without requiring knowledge of the model’s internal structure. 
It also scales well for models with high computational intensity but relatively few parameters, such as CNNs. 
However, during training, workers must synchronize to update model parameters; this typically involves aggregating gradients
%, often via an all-reduce operation 
to maintain consistency across replicas. 
%This parameter synchronization introduces a significant communication bottleneck~\cite{JQLA18}. To mitigate these overheads and avoid single-point-of-failure issues, synchronization can be implemented in a decentralized manner~\cite{Mayer2020}, though this often incurs higher communication costs due to the peer-to-peer exchanges. 
%To optimize these exchanges in distributed environments, collective communication operations such as \texttt{all-reduce} are widely employed. 
%Frameworks like Horovod~\cite{sergeev18horovod} and PyTorch Distributed Data Parallel (DDP)~\cite{li20pytorchdistributed} leverage high-performance all-reduce algorithms to combine local parameter updates %(e.g., gradients) 
%from all nodes using a reduction operation %(such as summation) 
%and then distribute the aggregated result back to every node. 
%This ensures that all workers maintain a synchronized model state without relying on a central parameter server. 
Despite these optimization frameworks, the scalability of data parallelism is inherently constrained because the entire model must fit 
within the memory of a single accelerator. 
%, leaving performance vulnerable to hardware limits, network jitter, and stragglers.

\textit{Model parallelism} techniques partition the DL model, and each GPU worker executes %a different part of it.  
only a portion of the overall computational graph. 
Effective model parallelism requires two distinct, yet tightly coupled, decision processes. First, partitioning involves the decomposition of the computational graph into sub-graphs, where the primary objective is to balance the computational load while minimizing the communication-to-computation ratio, often formulated as a min-cut problem on the DNN 
%directed acyclic graph (DAG)
graph~\cite{MM17}. Second, mapping addresses the placement of these sub-graphs onto a heterogeneous architecture, taking into account  specific device capabilities. 
%(e.g., peak FLOPS, memory bandwidth, and interconnect latency) 
In heterogeneous design, this mapping operates at two granularities: 
\textit{macro-mapping}, which handles the spatial placement of partitions across heterogeneous infrastructure nodes based on device capabilities, 
and \textit{micro-mapping}, which orchestrates the fine-grained assignment of operators to specific execution engines and hardware units within a single accelerator.
This need for multi-granularity orchestration is exemplified by recent platforms, such as AMD Versal adaptive SoCs, which integrate CPUs, FPGAs, AI engines, and high-bandwidth memory on a single chip. These heterogeneous architectures require a shift from uniform spatial partitioning to asymmetric mapping strategies
%, as shown by automatic partitioning frameworks like AP-DRL
~\cite{li26apdrl}. 
%At the system level, workloads must be partitioned according to the capabilities and organization of the available compute resources, while at the architectural level, data movement and operator execution must be carefully orchestrated across heterogeneous processing elements to maximize performance and avoid memory bottlenecks.
%Furthermore, modern mapping frameworks must go beyond macro-level device assignment to support fine-grained operator-to-engine mapping. 
%This includes selecting the most suitable execution units within an accelerator, such as matrix or vector engines, for each operator to maximize hardware utilization. 
While macro-level orchestration focuses on balancing workloads across devices, low-level optimization and code generation 
for these specialized units are typically handled by DL compilers, as discussed in Section~\ref{sec:compilers}.

%A major challenge of model parallelism is how to split the model into partitions that are assigned to the parallel workers~\cite{MM17}, and 
DNN workload partitioning across different devices was initially a manual process driven by human experts. 
An automated approach for solving this spatial mapping and placement problem is to use reinforcement learning~\cite{MPL+17,MGP+18}: starting from some initial partitioning, permutations are performed, and performance is measured to learn a placement policy that can then be adjusted for new workloads via transfer learning \cite{ABVG+19, ZRA+19} or used to bootstrap a genetic algorithm~\cite{PGN+20}. 
However, these methods are often computationally expensive, as they need to evaluate large numbers of placements and measure the runtime of several training/inference steps. 
Alternatively, the problem can be treated as an offline optimization problem of finding good partitions and schedules. 
This includes classic results in scheduling on multiple machines and/or devices~\cite{LLKS93, KL70, PY90, SW99, ST93}, 
%~\cite{LLKS93, Gra66, KL70, PY90, SW99, ST93}, 
as well as DNN-specific scheduling works~\cite{JQLA18, NHP+19}. 
These algorithms use profiled compute time of each DNN layer or operator, data-transfer requirements between nodes in a graph, and details on the target system infrastructure, such as machine and network properties.
Research has demonstrated that with well-defined cost models, the objective function obtained through such methodologies closely matches real performance \cite{NHP+19,JZA19}. 

\textit{Pipeline parallelism} combines model parallelism with data parallelism.
In pipeline parallelism, the model is divided among available workers, assigning a group of consecutive DNN layers or operators to each of them, and then overlapping the computation and communication of different inputs in a pipelined fashion. 
This approach significantly reduces the volume of inter-worker communication compared to data-parallel synchronization. 
Well-known frameworks supporting pipeline parallelism include GPipe~\cite{YY19} and PipeDream~\cite{NHP+19}. 
While pipelining is a simple and widely adopted idea, DNN training poses an important challenge not present in the inference phase: in fact, DNN training is bi-directional, as the forward pass is followed by a backward pass through the same layers in reverse order, which uses state and intermediate results from the forward pass.
This may result in low hardware efficiency unless careful optimizations are introduced~\cite{NHP+19}.
Proposals related to pipeline training can be classified according to the temporal aspect, by differentiating between \textit{synchronous} vs. \textit{asynchronous} training.
The former requires executing gradient synchronizations between adjacent training iterations to ensure convergence~\cite{YY19}, but it suffers from a significant memory consumption, which can be partially mitigated by re-computation. 
Asynchronous training, instead, inserts micro-batches into the pipeline concurrently to achieve maximum throughput~\cite{NHP+19}.

For DNNs, several frameworks attempt to find \textit{hybrid} solutions that combine the strengths of various parallelism techniques to mitigate their respective drawbacks. 
For example, layer-wise parallelism strategies~\cite{JQLA18} apply different parallelization techniques to each DNN layer rather than the same parallelization strategy to all layers. 
The optimal technique for each DNN layer is determined through a dynamic programming-based graph search. 
Similarly, DAPPLE~\cite{fan2021dapple} is a training framework for large-scale DNNs that combines data and pipeline parallelism and employs pipeline execution schedules such as 1F1B (one-forward, one-backward) to improve device utilization while reducing memory consumption.  
Alternatively, FlexFlow~\cite{JZA19} automates parallelization by exploring a generalized multi-dimensional search space. 
Using stochastic search and performance simulation, it identifies hybrid parallelization strategies that outperform traditional layer-wise heuristics.

%\subsection{Large-Scale Parallelism Strategies}

Recent advances in transformer-based architectures have driven the adoption of more granular, hybrid parallelization strategies. 
Megatron-LM~\cite{narayanan21megatron} has established \textit{tensor parallelism} as a standard for efficient large-scale training: 
%on GPU clusters. 
unlike pipeline parallelism, which focuses on layer-wise distribution, %inter-layer partitioning 
tensor parallelism operates at a fine-grained intra-layer granularity by partitioning weight tensors and matrix multiplications across multiple devices. 
%This strategy addresses the memory capacity limitations of single GPU devices by distributing large weight matrices 
%across accelerators within a node. 
%However, this approach increases the demand for frequent collective communication, such as \textit{all-reduce} operations, making the throughput of the underlying high-speed hardware interconnects (e.g., NVLink) a scalability bottleneck.
Tensor parallelism has since been extended to large-scale multi-GPU, multi-server clusters.
%through \textit{3D parallelism}, which combines data, pipeline, and tensor parallelism to train models with trillions of parameters. 
Such systems often employ sophisticated pipeline schedules, such as interleaved 1F1B scheduling~\cite{narayanan21megatron}, to reduce pipeline bubbles and improve device utilization, while relying on topology-aware placement to optimize communication over interconnects such as NVLink and InfiniBand. 
Overall, these developments mark a shift from heuristic-based mapping toward framework-driven orchestration that jointly considers software parallelization strategies and the underlying HPC topology. 
For a comprehensive taxonomy of distributed LLM training, we refer the reader to the survey by Zhao et al.~\cite{zeng25survey}.

\subsection{Inference Optimization and Mapping Strategies}

Although the fundamental principles of partitioning and mapping remain similar, optimization objectives differ between DL training and inference: while the former aims to maximize throughput and convergence efficiency, the latter focuses on minimizing latency and supporting high request concurrency and efficient resource utilization under strict service-level constraints. We summarize the key strategies, strengths, limitations, and engineering trade-offs of inference mapping and partitioning strategies in Table~\ref{tab:inference_comparison}.

\begin{table}[t]
\centering
\caption{Taxonomy and trade-offs of inference partitioning and mapping strategies.}
\label{tab:inference_comparison}
\footnotesize
\resizebox{1\columnwidth}{!}{%
\begin{tabular}{lccccc}
\toprule
\textbf{Strategy} & 
\textbf{Methods and frameworks} & 
\textbf{Mapping level} & 
\textbf{Strengths} & 
\textbf{Limitations} & 
\textbf{Trade-offs} \\
\midrule

\textbf{Intra-chip}
& Graph planning and & Operator/core & Fine-grained accelerator utilization & Compile-time mapping complexity & Efficiency vs. \\
& spatial mapping & & Deterministic dataflow execution & On-chip routing constraints & compilation overhead\\
& AIE4ML~\cite{danopoulos26aie4ml} & & & Memory hierarchy limitations & \\
\midrule

\textbf{Intra-node}
& Tensor parallelism and & Tensor/operator & High accelerator utilization & Communication bottlenecks & Kernel efficiency vs.  \\
& kernel optimization & & Reduced memory movement  & & communication overhead \\
& TensorRT-LLM~\cite{nvidia_tensorrt_llm} & & Communication-computation overlap & & \\
\midrule

\textbf{Inter-device}
& Layer/subgraph partitioning & Graph/layer & Scalability beyond single node & Network latency and & Latency vs. \\
& Neurosurgeon~\cite{kang17neurosurgeon}, & & Adaptive heterogeneous mapping & partition imbalance  & communication cost \\
& PipeEdge~\cite{hu22pipeedge}, ScalPipe~\cite{wang26scalpipe}, & & Geo-distributed resources  & Runtime profiling overhead & \\
& Helix~\cite{mei25helix} & & & & \\ 
\midrule

%\textbf{Memory-aware}
%& KV-cache management and & Memory/data & Improved memory efficiency & Memory management  & Capacity vs. access latency \\
%& IO-aware optimization & & Reduced data movement & overhead & \\
%& PagedAttention~\cite{kwon23paged} & & Improved memory locality & & \\
%FlashAttention~\cite{flashattention}
%\midrule
\textbf{Memory-aware}
& KV-cache management & Memory/data & Improved memory efficiency & Memory management & Capacity vs. access latency \\
& Hierarchical memory placement  & & and locality & overhead & \\
%& memory placement and offloading & & and locality & overhead & \\
& PagedAttention~\cite{kwon23paged}, & & Reduced data movement & & \\
& FlexGen~\cite{sheng23flexgen} & & & & \\
\midrule
%\textbf{Algorithm-aware}
%& Workload decomposition and & Model/phase & Workload asymmetry  & Synchronization and & Latency reduction vs. \\
%& phase-aware execution & & exploitation & scheduling complexity & coordination overhead \\
%& SpecInfer~\cite{miao24specinfer}, & & Improved resource utilization & & \\
%& DistServe~\cite{zhong24distserve}, Splitwise~\cite{patel23splitwise} & & & & \\
%\midrule

\textbf{Runtime} & Adaptive placement and routing & System/cluster & Dynamic resource allocation & Scheduling overhead & Flexibility vs. \\
\textbf{orchestration} & INFaaS~\cite{romero21infaas}, & & Workload adaptation & Management complexity & control overhead\\
& GreenServ~\cite{ziller26greenserv} & & Multi-objective optimization & & \\
\bottomrule
\end{tabular}
}
\end{table}

At the \textit{intra-node} level, optimized inference engines such as NVIDIA FasterTransformer~\cite{nvidia_fastertransformer} and its successor TensorRT-LLM~\cite{nvidia_tensorrt_llm} improve execution efficiency through graph- and kernel-level optimizations combined with tensor-parallel execution. 
%These frameworks leverage hardware-software co-design techniques, including operator fusion, which combines adjacent operations in the computation graph to reduce execution overhead and memory movement. 
%Instead of considering the transformer model as a sequence of  monolithic layers, they apply tensor-slicing strategies to partition individual layers across multiple devices and exploit specialized accelerator execution units. Furthermore, communication primitives such as \texttt{all-reduce} and \texttt{all-gather} can be overlapped with computation to reduce synchronization overhead and satisfy inference latency requirements.

When models exceed the memory capacity of a single accelerator, distributed inference introduces \textit{inter-device} partitioning and mapping challenges. 
Unlike training, where workloads are often statically defined, inference systems must accommodate variable request patterns, dynamic batch sizes, and heterogeneous resource availability. Consequently, distributed inference frameworks partition computational graphs into subgraphs or pipeline stages that are mapped across heterogeneous devices based on their compute capability, memory capacity, and communication characteristics, while satisfying latency constraints~\cite{shuvo23}.
Early collaborative inference frameworks explored this problem by partitioning DNN execution between resource-constrained edge devices and cloud servers.
For example, Neurosurgeon~\cite{kang17neurosurgeon} pioneered automatic DNN partitioning by determining layer placement between mobile devices and cloud resources to optimize latency and energy consumption. 
%More recent edge-cloud inference frameworks extend this approach by considering heterogeneous accelerators, geographically distributed resources, and dynamic system conditions. 
%As surveyed by Wang et al.~\cite{wang25survey}, on-device AI systems combine hardware-aware partitioning with lightweight optimization techniques to distribute workloads across mobile CPUs, GPUs, NPUs, and cloud accelerators. Hao et al.~\cite{hao26dnn} review mathematical models and distributed toolchains that enable adaptive edge-cloud partitioning under changing runtime conditions.
Instead of uniformly distributing model components across devices, heterogeneous inference systems increasingly adopt device-aware partitioning strategies that account for differences in compute capability, memory bandwidth, and network connectivity. 
These approaches assign workload partitions according to the performance characteristics of individual devices, reducing pipeline imbalance and avoiding inefficient resource utilization. 
For example, collaborative inference frameworks such as PipeEdge~\cite{hu22pipeedge} optimize layer placement across heterogeneous edge clusters by balancing computation latency with network variability. 
Similarly, ScalPipe~\cite{wang26scalpipe} dynamically partitions and maps model subgraphs across heterogeneous devices using lightweight runtime heuristics to adapt to changing resource conditions. 
Helix~\cite{mei25helix} formulates the partitioning and placement of transformer-based models across heterogeneous GPU clusters as a max-flow optimization problem, jointly optimizing computation, memory, and network resources.

As LLM inference becomes more widespread, partitioning and mapping become increasingly constrained by memory capacity and communication bandwidth. 
%The autoregressive generation process introduces sequential dependencies between tokens, while the key-value (KV) cache, which stores previously computed attention states to avoid redundant computation, introduces substantial memory requirements. 
Efficient distributed inference then requires not only computational partitioning, but also coordinated management of model states and memory resources across heterogeneous devices, i.e., \textit{memory-aware} mapping strategies. 
%can overcome accelerator memory limitations. 
Localized memory optimization techniques, such as PagedAttention~\cite{kwon23paged}, improve KV-cache management by reducing fragmentation and enabling more efficient memory allocation. 
Other approaches, such as FlexGen~\cite{sheng23flexgen}, formulate a linear programming optimization problem to determine the placement of model weights, activations, and KV-cache data across GPU memory, host memory, and secondary storage, maximizing inference throughput under memory and bandwidth constraints.
%Large-scale distributed inference extends these ideas across heterogeneous memory hierarchies. 
%Therefore, emerging systems  explore multi-tier memory management approaches that migrate model states and KV-cache data across accelerator memory, host memory, and remote storage to increase the effective capacity of the inference system while balancing communication overhead.
%Memory-aware mapping is closely coupled with \textit{communication-aware} partitioning. 

%MoE models introduce an additional dimension of communication-aware partitioning. 
%Although sparse expert activation reduces computation, dynamic expert routing creates load imbalance and communication overhead across devices. 
%Consequently, expert placement, routing, and caching strategies partition experts across devices while balancing memory capacity, computation demand, and inter-device communication~\cite{rajbhandari22expert}.

At the accelerator level, partitioning and mapping can also occur within a single heterogeneous chip. 
Unlike distributed edge-cloud systems, which rely on runtime orchestration, adaptive SoCs %system-on-chip architectures 
require compile-time spatial mapping of computation and communication. 
For example, AIE4ML~\cite{danopoulos26aie4ml} enables fine-grained mapping of neural network workloads onto AMD Versal AI Engine architectures through graph planning and deterministic data routing over a 2D processing fabric. 
Such approaches must jointly optimize computation placement, memory movement, and on-chip communication to avoid routing congestion and maximize accelerator utilization.

%In addition to hardware-level optimizations, algorithmic innovations introduce new opportunities for distributed inference partitioning. 
%Speculative decoding accelerates autoregressive generation by combining a lightweight draft model with a larger target model that verifies generated tokens. 
%In distributed settings, this creates opportunities to separate token generation and verification across heterogeneous resources. 
%Frameworks such as SpecInfer~\cite{miao24specinfer} exploit this asymmetry by distributing draft and verification workloads and applying topology-aware mapping to reduce communication overhead while improving decoding efficiency.

The increasing diversity of inference workloads has also motivated \textit{phase-aware} resource partitioning:
%(also known as \textit{disaggregated inference}).  
since LLM inference consists of distinct prefill and decode phases with different computational characteristics, considering them as separate workloads enables more efficient resource allocation. 
%The prefill phase is typically compute-intensive, whereas token-by-token decoding is often constrained by memory bandwidth and KV-cache access. 
Systems such as DistServe~\cite{zhong24distserve} and Splitwise~\cite{patel23splitwise} exploit this asymmetry through inference disaggregation, separating compute-intensive and memory-intensive components and allocating them to resources optimized for their respective requirements.
More broadly, heterogeneous inference systems are increasingly constrained by memory capacity and bandwidth rather than %raw 
compute capability. 
Therefore, mapping strategies must account for heterogeneous memory hierarchies, including high-bandwidth memory (HBM), host memory, and emerging memory technologies such as CXL-attached memory, to minimize costly data movement. 
This extends the IO-aware optimization principles introduced by FlashAttention~\cite{flashattention}, which optimizes data movement between on-chip SRAM and HBM, toward distributed settings where memory movement across devices and nodes becomes a dominant bottleneck. 
%As analyzed by Yuan et al.~\cite{yuan24llminference}, inference bottlenecks vary across execution phases: prompt processing is generally compute-bound, whereas autoregressive decoding is primarily memory-bandwidth-bound. Therefore, effective workload mapping must adapt to phase-specific bottlenecks and employ performance models, such as Roofline-based analysis, to guide partitioning decisions before deployment.

Finally, the growing complexity of heterogeneous inference deployments has motivated the development of automated orchestration frameworks that abstract hardware complexity through runtime policies. 
These systems jointly optimize model placement, resource allocation, and request routing across heterogeneous infrastructures. 
Frameworks such as INFaaS~\cite{romero21infaas} and GreenServ~\cite{ziller26greenserv} address this challenge by dynamically selecting deployment configurations based on workload requirements and resource availability, extending application mapping from static partitioning to adaptive system-level orchestration. 
%Specifically, INFaaS leverages integer programming to select suitable model variants and deployment configurations for heterogeneous hardware environments, while GreenServ employs contextual multi-armed bandits to learn adaptive model routing policies that balance accuracy and energy efficiency at the request level in real time.
\section{Modeling, Simulation, Profiling and Exploration}
\label{sec:modeling_simulation_exploration}

To effectively design a hardware accelerator for DL, it is essential to have access to powerful modeling tools that can provide detailed insights into the power consumption, performance, and area (PPA) requirements of the accelerator.
These tools enable designers to experiment with various design choices and configurations for different technologies at different levels of abstraction and to optimize their designs for specific PPA metrics. 
In this section, we will explore some of the most popular and effective tools available for modeling hardware accelerators for DL and discuss their key features and capabilities. 
Figure~\ref{fig:modeling_simulation_exploration} provides a taxonomy of tools, frameworks, and methodologies discussed in the following subsections; Table~\ref{tab:simulators}, instead, compares them according to their integration with standard DL frameworks, modeling strategy, target technology, whether they focus on the accelerator itself or the entire SoC, functional and non-functional attributes they are capable of estimating, and estimation error.

\begin{figure}[t]
    \centering
    {
        \tiny
\resizebox{\textwidth}{!}{%
    \begin{forest}
rounded/.style={ellipse,draw},squared/.style={rectangle,draw},
    qtree,
        [{\textbf{Modeling, Simulation, Profiling, and Exploration}}
            [{\textit{High-level modeling}}
                [{
                MLPAT~\cite{tang_dossa18}, MAESTRO~\cite{kwon_micro20},\\ 
                        Timeloop~\cite{parashar_ispass19}, ZigZag~\cite{mei_tc21} \\
                        LAMBDA~\cite{russo_percom21}, 
                        DNN-Chip Predictor~\cite{zhao2021dnnchip}\\
                        Interstellar~\cite{yang_asplos20}, 
                        SMAUG~\cite{xi_taco20},\\
                        DeFiNES~\cite{mei_hpca23}, Sparseloop~\cite{wu_micro22}, \\
                        Gemini~\cite{cai_hpca24},
                        QADAM~\cite{inci2022qadam}, QAPPA~\cite{inci2022qappa}\\
                        Gemmini~\cite{genc_dac21}, Juraci~et~al.~\cite{juracy_tcs22}
                     }]
                ],
                [{\textit{Cycle-accurate simulation}}
                    [{
                        SCALE-SIM~\cite{samajdar_ispass20}\\
                        STONNE~\cite{munozmartinez2020stonne}\\
                        SimuNN~\cite{cao_jestcs20}\\
                        AccTLMSim~\cite{kim2020transactionlevel}
                    }]
                ],
                [{\textit{ FPGA modeling and simulation}}
                    [{
                        Shuhai~\cite{shuhai}, 
                        HPCChallenge~\cite{hpcc-fpga}\\
                        HPCG Benchmark~\cite{hpcg-fpga}\\
                        Da~Silva~et~al.~\cite{roofline-fpga-hls}, 
                        Siracusa~et~al.~\cite{roofline-fpga-cad,roofline-fpga-cad2}\\
                        Muralidharan~et~al.~\cite{roofline-multibench-fpga}\\
                        ERT~\cite{ert-opencl-fpga,ert-opencl-fpga2}                    
                    }]
                ],
                [{\textit{Tools for In-Memory Computing}}
                    [{
                        DNN+NeuroSim~\cite{peng2019iedm}\\
                        SySCIM~\cite{shadmehri2022date}\\
                        MemTorch~\cite{Lammie2022}\\
                        MNSIM~\cite{xia2016date}\\
                        Reiser~et~al.~\cite{reiser2023newcas}
                    }]
                ]
            ]
        \end{forest}
        }}
    
    \caption{Tools and methodologies discussed in Section~\ref{sec:modeling_simulation_exploration}.}
    \label{fig:modeling_simulation_exploration}
\end{figure}

\begin{table}[ht]
    \caption{Comparison of modeling and simulation tools analyzed in Section \ref{sec:modeling_simulation_exploration}.}
    \label{tab:simulators}
    \resizebox{1\columnwidth}{!}{%
    \begin{tabular}{lcccccc}
    \toprule
    & \makecell{Integration with\\NN frameworks} & \makecell{Model\\type} & \makecell{Full\\SoC} & \makecell{Evaluation \\metrics} & \makecell{Target} & \makecell{Estimation\\error}\\
    \midrule
    MLPAT~\cite{tang_dossa18} & No & Analytical & No & PPA & ASIC & \makecell{$<5$\% area\\$<$10\% power}\\
    MAESTRO~\cite{kwon_micro20} & No & Empirical & No & Performance & ASIC & 5\% \\
    Timeloop~\cite{parashar_ispass19} & No & \makecell{Analytical/\\Empirical} & No & PPA & ASIC & 5\% \\
    ZigZag~\cite{mei_tc21} & No & Analytical & No & PPA & ASIC & $<$5\% \\
    FactorFlow \cite{factorflow} & No & Analytical & No & \makecell{Energy, Latency, Utilization} & ASIC & $<$5\%\\
    LAMBDA~\cite{russo_percom21} & No & \makecell{Analytical/\\Empirical} & No & PPA & ASIC & 5\% \\
    DNN-Chip Predictor~\cite{zhao2021dnnchip} & No & Analytical & No &  \makecell{Performance\\Energy} & FPGA/ASIC & $<$18\% \\
    Gemmini~\cite{genc_dac21} & No & Simulation & \makecell{Yes +\\OS support} & Performance & FPGA/ASIC & -- \\
    Interstellar~\cite{yang_asplos20} & No & Analytical & No & PPA & ASIC & 2\% \\
    SCALE-SIM~\cite{samajdar_ispass20} & No & Empirical & Yes & Performance, Area & ASIC & -- \\
    STONNE~\cite{munozmartinez2020stonne} & Caffe & \makecell{Cycle level\\simulation} & Yes & Performance & ASIC & $<$3\% \\
    SimuNN~\cite{cao_jestcs20} & TensorFlow & \makecell{Cycle level\\simulation} & Yes & PPA & FPGA/ASIC & -- \\
    AccTLMSim~\cite{kim2020transactionlevel} & No & \makecell{Cycle level\\simulation} & Yes & Performance & ASIC & 3\% \\
    Juracy \emph{et al.}~\cite{juracy_tcs22} & TensorFlow & \makecell{Cycle level\\simulation} & No & PPA & ASIC & $<$7\% \\
    DNN-NeuroSim \cite{peng2019iedm} & TensorFlow, PyTorch &  \makecell{Instruction accurate\\simulation} & Yes & PPA & ASIC & - \\
    SySCIM \cite{shadmehri2022date} & No & \makecell{Circuit level\\simulation} & No & Accuracy & ASIC & $<$4\% accuracy \\
    Memtorch \cite{Lammie2022} & PyTorch & \makecell{Analytical/\\Empirical} & Yes & PPA & ASIC & - \\
    MNSIM \cite{xia2016date} & No & \makecell{Cycle level\\simulation} & Yes & PPA & ASIC & - \\
    SMAUG~\cite{xi_taco20} & TensorFlow, PyTorch & Simulation & Yes & PPA & ASIC & 10\% \\
    DeFiNES~\cite{mei_hpca23} & No & Analytical & No & Energy, Latency & ASIC & \makecell{3\% latency\\6\% energy}\\
    Sparseloop~\cite{wu_micro22} & No & Analytical & No & Performance, Energy & ASIC & $<$8\% \\
    Gemini~\cite{cai_hpca24} & No & Analytical & No & \makecell{Performance, Energy\\Monetary cost} & Chiplet & -\\
    \bottomrule
    \end{tabular}
    }
\end{table}

\subsection{High-Level Frameworks for Design and Exploration}
\label{ssec:exploration}

%State-of-the-art frameworks for modeling, simulation, and design-space exploration (DSE) of DL accelerators enable researchers to identify optimal architectures for specific DL workloads and accelerate the design process.
This subsection focuses on high-level design-space exploration (DSE) frameworks, i.e., tools that model accelerators as architectural templates with configurable parameters rather than low-level hardware implementations, allowing fast DSE before committing to a specific target technology and moving to slower, more precise simulators.

Analytical DSE frameworks combine configurable architecture models with analytical cost models to estimate performance, area, and energy.
Representative examples include MLPAT~\cite{tang_dossa18}, which enables comprehensive modeling of machine learning accelerators incorporating various architectural components such as systolic arrays, memory hierarchies, and dataflow, and MAESTRO~\cite{kwon_micro20}, which provides a domain-specific language for describing neural network processing engines, aiding in the analysis of power/performance trade-offs.
However, these tools need to be updated to support emerging neural network paradigms, such as spiking neural networks and transformer-based architectures.

A key issue to explore the vast design space of DL accelerators is balancing accuracy and computational efficiency.
Frameworks such as Timeloop~\cite{parashar_ispass19}, LAMBDA~\cite{russo_percom21}, ZigZag~\cite{mei_tc21}, and FactorFlow~\cite{factorflow} progressively improve the accuracy and flexibility of workload-to-hardware mapping by incorporating communication models, uneven loop mappings, and faster mapping algorithms.
%Timeloop~\cite{parashar_ispass19} is a comprehensive framework consisting of a fast and accurate performance, area, and energy model, along with a mapper that searches for optimal workload-to-architecture mappings; configurable templates provide a flexible approach to architectural exploration, enabling the evaluation of multiple configurations in a short time frame.
%LAMBDA~\cite{russo_percom21} extends the Timeloop framework to support additional configurable parameters modeling communication networks and DNN compression, allowing designers to explore a wider range of architectural and microarchitectural choices.
%ZigZag~\cite{mei_tc21} introduces the ability to evaluate uneven loop mappings, where operands at shared memory levels are not necessarily bound to the same memory level for each loop index.
%This modular framework integrates cost models, architecture generators, and mapping strategies.
%FactorFlow \cite{factorflow} improves the mapping efficiency and reaches accurate results in less time than ZigZag.
Despite recent advances, current tools struggle to capture the full complexity of heterogeneous systems, and precise modeling of multi-accelerator and distributed inference scenarios remains an open research challenge.

To further accelerate the design process, predictive modeling frameworks enable rapid estimation of accelerator performance before hardware implementation.
For example, DNN-Chip Predictor~\cite{zhao2021dnnchip} uses analytical performance formulations to predict energy consumption, throughput, and latency.
%Its support for different algorithm-to-hardware mappings makes it a versatile tool for early-stage optimization. 
However, the increasing heterogeneity of DL workloads raises concerns about the generalization of predictive models, indicating the need for adaptive techniques that can dynamically adjust to new architectures and workloads.
Recent contributions also explore AI-assisted DSE: an LLM-augmented multi-modal fusion approach for SoC design space exploration~\cite{luo_iccad25} illustrates the growing adoption of foundation models to assist architectural exploration and optimization, potentially reducing the cost of navigating large design spaces. Likewise, LaZagna~\cite{youssef_iccad25} provides an open-source framework for flexible 3D FPGA architectural exploration, addressing the growing relevance of heterogeneous and three-dimensional integration technologies.
Together, these efforts suggest a transition from traditional analytical DSE methodologies toward AI-assisted exploration and architectural co-optimization frameworks. Effectiveness, scalability, and generalization of these approaches across diverse accelerator architectures and workloads remain to be explored.

Efficient workload scheduling plays a crucial role in maximizing accelerator efficiency, particularly as modern DNNs often present sparse computation.
Modern compiler techniques enable efficient representation and scheduling of DL workloads for hardware execution.
Interstellar~\cite{yang_asplos20} demonstrates that optimizing memory hierarchy can have a greater impact on energy efficiency than merely selecting an optimal dataflow, while SMAUG~\cite{xi_taco20} extends the analysis to full-system simulation, highlighting that data movement and software overhead often dominate execution time.
DeFiNES~\cite{mei_hpca23} and Sparseloop~\cite{wu_micro22} further address scheduling and sparse execution through analytical models that reduce the need for expensive cycle-accurate simulation.
%Interstellar~\cite{yang_asplos20} extends the Halide compiler to generate DNN accelerators, demonstrating that optimizing memory hierarchy can have a greater impact on energy efficiency than merely selecting an optimal dataflow. SMAUG~\cite{xi_taco20} is one of the first frameworks designed for full-system simulation of deep learning applications. By assessing total inference latency, it highlights that data movement and software overhead often dominate execution time. 
%DeFiNES~\cite{mei_hpca23} provides an analytical framework for evaluating depth-first scheduling strategies, commonly known as layer fusion or cascaded execution; by estimating energy and latency costs, it guides designers toward optimized scheduling decisions. Sparseloop~\cite{wu_micro22} facilitates the exploration of sparse tensor accelerators, leveraging analytical techniques to efficiently model performance and energy efficiency, reducing reliance on expensive cycle-level simulations. 
Nevertheless, handling dynamic sparsity patterns in real-time remains an open challenge, calling for more adaptive scheduling and mapping strategies.

As DL models scale in complexity, chiplet-based architectures are emerging as a promising solution to enhance scalability and performance. Gemini~\cite{cai_hpca24} introduces a novel approach to co-exploring accelerator mapping strategies and architecture design while considering monetary cost.
%Unlike traditional frameworks focused solely on performance and energy, Gemini provides a more holistic view of hardware deployment feasibility. 
Inter-chiplet communication overhead and design complexity remain significant hurdles~\cite{das_jetcs24}, necessitating further innovations in interconnect technologies and workload partitioning strategies.

Finally, accelerator generators bridge architectural exploration and implementation.
Gemmini~\cite{genc_dac21} is an open-source RISC-V-based generator designed for customizable DNN hardware, while QADAM and its evolution QAPPA~\cite{inci2022qadam,inci2022qappa} are parameterized RTL frameworks designed to model power, performance, and area of quantization-aware DNN accelerators.
Similarly, the design space exploration approach proposed in~\cite{juracy_tcs22} is integrated into CNN frameworks such as TensorFlow and provides estimates with an average error of less than 7\% for various factors, including area, performance, power, energy, and memory accesses.
These tools reduce the gap between high-level exploration and implementation, although extending them to emerging heterogeneous architectures remains an active research direction.
%Gemmini~\cite{genc_dac21} is an open-source accelerator generator designed for customizable DNN hardware. 
%With tight integration into the RISC-V ecosystem, it offers a multi-level software stack for efficient hardware/software co-design and accurate simulation.
%QADAM~\cite{inci2022qadam} and its evolution QAPPA~\cite{inci2022qappa} are parameterized RTL frameworks designed to model power, performance, and area of quantization-aware DNN accelerators.
%These frameworks support design space exploration and Pareto-efficiency analysis for a set of design choices, including bit precision, processing element (PE) type, scratchpad sizes of PEs, global buffer size, total number of PEs, and DNN configurations. 
%The design space exploration approach proposed in~\cite{juracy_tcs22} for CNNs employs an analytical model derived from the physical synthesis of hardware accelerators. This model is integrated into CNN frameworks such as TensorFlow, enabling precise outcomes. The analytical model provides estimates for various factors, including area, performance, power, energy, and memory accesses. The accuracy of the model was tested by comparing it to results obtained from physical synthesis, and the average error was less than 7\%.

The landscape of DL accelerator design is evolving rapidly, with an increasing emphasis on efficiency, scalability, and cost-effectiveness.
High-level modeling frameworks and accelerator generators play a pivotal role in shaping next-generation architectures, offering researchers powerful tools for informed decision-making. By leveraging these methodologies, designers can navigate the complexities of DL hardware and optimize performance across diverse application scenarios.

\subsection{Cycle-Accurate Simulation for Pre-RTL Accelerators}
\label{ssec:ca_simulators}

For accurate simulation of DL accelerators, it is crucial to model the hardware behavior cycle-by-cycle, accounting for all the interactions between different hardware components. 
Existing simulators differ mainly in the accelerator architectures they target and in the level of detail adopted for the memory subsystem.
SCALE-SIM (SystoliC AcceLErator SIMulator)~\cite{samajdar_ispass20} provides cycle-accurate energy/performance modeling for systolic-array-based DNN accelerators by considering various factors such as on-chip and off-chip memory accesses, and interface bandwidth. 
%It has two primary components: a compute unit that utilizes a systolic array that can be customized according to size and aspect ratio, and a memory system that features three double-buffered SRAM memories with user-specified sizes.
STONNE (Simulation Tool for Neural Network Engines)~\cite{munozmartinez2020stonne} is a highly modular and extensible simulation framework for the end-to-end evaluation of flexible accelerator DNN architectures with cycle accuracy. 
Like Timeloop~\cite{parashar_ispass19}, STONNE uses the Accelergy \cite{wu_iccad19} energy estimation tool to estimate energy and area.
SimuNN~\cite{cao_jestcs20} is a pre-RTL neural network simulator for early phase verification and fast prototyping before the design is converted into hardware, supporting multiple data precisions, quantization schemes, and hardware configurations while integrating with TensorFlow.
%It supports different data precisions and is compatible with TensorFlow. SimuNN provides multi-level trace results that can be used as a reference for the final hardware design. Additionally, it can evaluate the hardware performance under various quantization schemes, data flows, and configurations based on a generalized hardware model.
AccTLMSim~\cite{kim2020transactionlevel} is a pre-RTL simulation tool based on SystemC transaction-level modeling (TLM) to simulate CNN accelerators and DRAM transactions, enabling detailed bandwidth analysis.
%The tool includes a detailed model of the DRAM interface for precise tracking of each bus transaction between the accelerator and DRAM while considering the communication bandwidth. 

\subsection{FPGA Modeling and Profiling}
\label{ssec:prof_fpga}

Off-the-shelf FPGAs have been adopted by a large and increasing number of HPC systems~\cite{fpga-hpc,fpga-hpc-trends,fpga-hpc2} to implement highly parallel hardware accelerators and boost the performance and energy efficiency of software applications, including DL algorithms.
While GPUs are definitely the most common accelerators, some data centers have recently started to adopt FPGAs to speed up network interconnects~\cite{project-catapult} and specific workloads such as DNN inference~\cite{fpga-dl,project-brainwave}.
%The latest generations of FPGAs integrate thousands of programmable DSPs able to efficiently implement floating-point operations~\cite{fpga-dsp-fp,mapping-dsp,xilinx-dsp-flops}, leading to devices capable of reaching a performance in the same order of magnitude as commodity HPC processors (i.e., TFLOP/s), and in some cases, able to deliver a better energy-efficiency~\cite{fpga-opencl-hpc}.
%At the same time, recent improvements in synthesis tools and the development of new programming approaches such as HLS (Section ref{sec:HLS_Design_based_methodologies}) allow programmers to develop accelerators using high-level languages, further lowering the barrier to widespread FPGA use.
%Approaches like OpenCL~\cite{fpga-opencl-hpc} are very similar to those commonly used by HPC developers to target multi-core CPUs and other accelerators (e.g., OpenMP and OpenACC), guaranteeing a fair level of code portability~\cite{ompss-fpga2}.
In this context, application developers need to estimate the performance achievable on target FPGAs to decide whether an application kernel is worth being offloaded, or which FPGA better fits its computing requirements.
At the same time, system architects and engineers need to estimate the performance of FPGAs to feed performance models that tune, balance, and optimize the performance at the system level~\cite{exa-dataflow}.

Several research works have investigated FPGA performance modeling, using synthetic benchmarks to estimate the bandwidth of off-chip memories~\cite{shuhai,fpga-stream-opencl,intel-fpga-mem}, and OpenCL kernels to measure the FPGA computing performance~\cite{fpga-fp-eval,hpcc-fpga,hpcg-fpga}.
%The \textit{Shuhai} Verilog benchmark \cite{shuhai} was used to characterize the performance of High-Bandwidth Memory (HBM) and DDR off-chip memories embedded in the AMD/Xilinx Alveo U280 datacenter FPGA platform.
%An OpenCL implementation of the HPCChallenge Benchmark Suite~\cite{hpcc-fpga} allows the designer to collect results for different FPGAs on representative programs.
%Similarly, a C/HLS version of the HPCG benchmark was implemented to target FPGA performance modeling~\cite{hpcg-fpga}.
%The ERT benchmark has also been reported to run on 
%FPGAs using OpenCL~\cite{ert-opencl-fpga,ert-opencl-fpga2}.
The Roofline model \cite{roofline09} has been used in the past to evaluate the performance of specific applications ported to FPGAs~\cite{fpga-roofline-tsunami}; however, few works provide a generic application-independent extension of this model for FPGA architectures, mainly due to the difficulty in defining the maximum compute performance for a reconfigurable device.
The first comprehensive work extending the Roofline model to FPGAs focuses mainly on aiding developers to explore different options in the design space~\cite{roofline-fpga-hls}.
Building on the same principle, more recently, a semi-automated performance optimization methodology based on the Roofline model for FPGAs has been proposed~\cite{roofline-fpga-cad,roofline-fpga-cad2}.
Another work proposes a methodology for the performance analysis of FPGAs through Roofline plots and cross-architectural comparisons~\cite{roofline-multibench-fpga}.
%; the authors use OpenCL as a programming language to provide mini-apps, such as SHOCL0, LINPACK, and STREAM, to measure the computing performance and the memory bandwidth of the off-chip memory.
FER (FPGA Empirical Roofline)~\cite{parco19-fp,fer-fpl, fer} is the first open-source C/HLS benchmark tool capable of providing empirical Roofline plots for FPGAs, allowing for application-agnostic performance assessment of FPGA-based accelerators, and for cross-architectural comparisons of generic HPC kernels on a given device.
%and it supports the Xilinx Vitis workflow to allow for wider adoption.
%The tool, available in open-source\footnote{\url{https://baltig.infn.it/EuroEXA/FER}},  allows for application-agnostic performance assessment of FPGA-based accelerators aiming for comprehensive machine characterization; it also allows cross-architectural comparisons and performance estimation of generic HPC kernels on a given device.
%FER can estimate both the computing peak performance of FPGAs and the bandwidths of on-chip and off-chip memories.
%It is based on the Roofline Model, and it is implemented as a directive-annotated C/HLS kernel with tunable operational intensity and hardware resources usage.
%Moreover, it relies on a theoretical model aiming to strictly link the performance results to the available hardware resources.
%The choice of C/HLS allows to expose low-level fine-tuning knobs to the user, and at the same time uses a high-level programming paradigm that can easily be adopted by the HPC user community for development and porting.

\subsection{Modeling and Simulation Frameworks for In-Memory Computing}
\label{ssec:sim_emerging}

In-Memory Computing (IMC) is an emerging paradigm that uses either analog (non-volatile memories) or digital (SRAM and DRAM) components to offload and accelerate computation.
DNN+NeuroSim \cite{peng2019iedm} is an integrated framework to benchmark IMC accelerators for DNNs, with hierarchical design options spanning from the device level to the algorithm level.
%A Python wrapper is developed to interface NeuroSim with popular DL platforms such as PyTorch and TensorFlow. The framework supports the automatic mapping of algorithms to hardware and the evaluation of chip-level performance and inference accuracy with hardware constraints. 
SySCIM \cite{shadmehri2022date} considers the impact of the non-idealities of the IMC components, especially for analog approaches, including memristor devices, memristor crossbar (interconnects), analog-to-digital converters, and trans-impedance amplifiers, on the vector-matrix multiplication performed by the IMC unit.
%The IMC modules are described in SystemC and SystemC-AMS to reach a high simulation speed while maintaining simulation accuracy. 
A similar SystemC-AMS has been reprised in \cite{rizzi21}, with the simulation performance optimization goal in memory-consuming Monte Carlo simulations.
MemTorch \cite{Lammie2022} is an open-source framework for customized large-scale memristive DL simulations, with a refined focus on the co-simulation of device non-idealities. 
%MemTorch also facilitates the co-modeling of key crossbar peripheral circuitry. MemTorch adopts a modernized software engineering methodology and integrates directly with the well-known PyTorch DL library.
MNSIM \cite{xia2016date} proposes a simulation platform for the memristor-based neuromorphic system with a hierarchical structure and flexible interfaces for customization. 
%A behavioral computing accuracy model is incorporated to evaluate the computing error rate affected by interconnect lines and non-ideal device factors. Experimental results show that MNSIM is over 7000 times faster than SPICE simulation. MNSIM can optimize the design and estimate the tradeoff relationships among different performance metrics. 
A simulation framework is proposed in \cite{reiser2023newcas} together with suitable abstractions to propagate the effects of RRAM-based analog crossbar configuration parameters to their ultimate implications over inference performance stability, as the non-idealities of RRAM devices result in significant inference accuracy drops compared to software baseline accuracy.
\section{Deep Learning Compilers}
\label{sec:compilers}

\begin{figure}[tb]
	\centering
	{
    	\footnotesize
    	\begin{forest}
        	rounded/.style={ellipse,draw},
        	squared/.style={rectangle,draw},
        	qtree,
        	[{\textbf{Deep Learning compilers}}
            	[{\textit{DL compilers for HPC systems}}
                	[{
    Halide~\cite{ragan2013halide}, TVM~\cite{chen2018tvm}, Relay~\cite{roeschRelayNewIR2018}\\
    Relax~\cite{relaxASPLOS2025}, ONNC~\cite{linONNCCompilationFramework2019}, Glow~\cite{rotemGlowGraphLowering2019}\\
    XLA~\cite{XLA}, StableHLO~\cite{stablehlo}, MLIR~\cite{lattner2021mlir}\\
    ONNX-MLIR~\cite{jinCompilingONNXNeural2020}, CUDA Tile~\cite{cudatile}, IREE~\cite{IREE} \\
    ByteIR~\cite{ByteIR}, Triton~\cite{Triton}, TileLang~\cite{tilelang2025}\\
    Hexcute~\cite{Zhang2025HexcuteAC}, TorchDynamo/Inductor~\cite{torchdynamo}\\
    Kitsune~\cite{kitsune2025}, PolyBlocks~\cite{polyblocks2026}, Event Tensor~\cite{eventTensor2026}\\
    Nautilus~\cite{nautilus2026}, 
    Neptune~\cite{neptune2025}, Prism~\cite{prism2026}\\ 
    LLMCompiler~\cite{Cummingfirst,LLMCompile}, Compiler-R1~\cite{compilerR12025}\\
    Reasoning Compiler~\cite{reasoningCompiler2025}\\
                	}]
            	],
                [{\textit{DL compilers for edge systems}}
                	[{
    EXECUTORCH~\cite{ExecuTorch}\\
    TFLite Micro~\cite{david2020tensorflow}\\
                        %Larq Compute Engine~\cite{Larq}\\
                        X-CUBE-AI~\cite{CubeAI}\\
                        AutoTiler~\cite{flamand2018gap}\\
                        DORY~\cite{burrello2021dory}, Deeploy~\cite{Deeploy}\\
                        HTVM~\cite{vandelmHTVMEfficientNeural2023a}\\
                        MATCH~\cite{MATCH}, MATCHA~\cite{MATCHA}\\
                        TelaMalloc~\cite{TelaMalloc}\\
                        eIQ Neutron~\cite{eiqNeutron2025}\\
                        Hexagon-MLIR~\cite{hexagonmlir}
                	}],
                ],
            	  [{\textit{DL compilers for IMC architectures}}
                	[{
                    	MIG Compiler~\cite{Soeken16}\\
                    	AIM~\cite{Zhang25}\\
                    	QPIM~\cite{Long2020QPIM}\\
                        CINM~\cite{Khan25}\\
                        ComPRIME~\cite{comprime}\\
                        PIMCOMP~\cite{sun25}
                	}],
            	]
        	]
    	\end{forest}
	}
	\caption{Deep Learning compilers discussed in Section~\ref{sec:compilers}.}
	\label{fig:compiler_taxonomy}
\end{figure}

%Neural networks are complex workloads that have inspired a wide range of custom accelerators, but exploiting specialized hardware efficiently requires substantial developer effort.
Neural network models expressed in high-level frameworks such as TensorFlow~\cite{tensorflow2015}, PyTorch~\cite{NEURIPS2019_9015}, or ONNX~\cite{ONNX} must be mapped onto heterogeneous platforms while preserving numerical correctness, minimizing data movement, and exploiting target-specific compute units.
Deep learning compilers address this problem by automating the lowering of high-level neural network descriptions into optimized executable code for CPUs, GPUs, domain-specific accelerators, edge devices, and emerging in-memory/near-memory architectures.
%During this lowering process, compilers may apply graph-level transformations such as operator fusion, constant folding, quantization, layout propagation, algebraic simplification, and layer reordering.
%At lower abstraction levels, they also optimize tensor programs through loop tiling, vectorization, parallelization, memory promotion, tensorization, and instruction scheduling.
%In addition, modern compilers manage the placement of weights, activations, and intermediate buffers across complex memory hierarchies, including caches, scratchpads, local SRAMs, high-bandwidth memory, and accelerator-specific memories.
%This requires deciding not only which hardware resource should execute a given operation, but also how data should be laid out, which in turn impacts reuse and bandwidth utilization.
Figure~\ref{fig:compiler_taxonomy} distinguishes the three main classes of deep learning compilers discussed in this section: compilers for HPC and datacenter systems, compilers for edge and TinyML systems, and compilers for in-memory/processing-in-memory architectures.
We also include a discussion of AI-assisted optimization methods, as they can accelerate schedule search, guide code transformations, and improve the performance of kernels generated by existing compiler stacks.
Although all these frameworks aim to improve neural network deployment, they differ substantially in their optimization methodologies.
Graph-level compilers operate on whole-model representations and are well-suited for global transformations such as operator fusion, layout propagation, algebraic simplification, and backend dispatch.
Tensor-program compilers work at a lower abstraction level, exposing tensor operations as loop nests, tiled computations, or kernel schedules that can be mapped to hardware-specific memory hierarchies and instructions.
Edge compilers use many of the same lowering and code-generation mechanisms, but target a different deployment regime: rather than maximizing throughput on large accelerators, they prioritize predictable execution under tight memory, code-size, energy, and latency constraints, often through static memory planning, quantization, and explicit management of on-chip buffers.
IMC/PIM compilers introduce an additional placement problem because computation is tied to the physical location of weights inside memory arrays or banks.

\begin{table*}[t]
\centering
\scriptsize
\setlength{\tabcolsep}{3.5pt}
\renewcommand{\arraystretch}{1.15}

\caption{Representative DL compiler methodologies and frameworks for HPC systems.}
\label{tab:compiler_comparison_hpc}
\begin{tabular}{m{0.9cm}|m{1.9cm}|m{1.5cm}|m{3.95cm}|m{4.3cm}}

\hline

\textbf{Target Device} & \textbf{Framework} & \textbf{Abstraction level} & \textbf{Strengths} & \textbf{Limitations} \\
\hline

\multirow[c]{6}{1.1cm}{CPU, \newline GPU, accel.} & TVM~\cite{chen2018tvm}, \newline Relay~\cite{roeschRelayNewIR2018}, \newline Relax~\cite{relaxASPLOS2025}
& Graph, tensor
& Flexible scheduling, autotuning, tensorization, dynamic-shape support, extensible backends
& High autotuning cost; tensorization and hardware intrinsics must be manually defined for novel accelerators \\
\cline{2-5}

& XLA~\cite{XLA}, \newline OpenXLA~\cite{stablehlo} 
& Graph
& Strong whole-graph optimization, fusion, and layout propagation
& Less transparent to users; performance tied to HLO coverage and backend maturity \\
\cline{2-5}

& Glow~\cite{rotemGlowGraphLowering2019} 
& Graph
& Clear lowering pipeline; backend-friendly operator decomposition
& Limited dynamic-shape support; narrower frontend coverage than MLIR-based stacks \\
\cline{2-5}

& ONNC~\cite{linONNCCompilationFramework2019} 
& Graph
& Preserves coarse-grained ONNX operators, easing porting to accelerators
& Tied to the ONNX operator set; coarse operators are not reduced to a reusable basis, so each requires per-target code generation \\
\cline{2-5}

& MLIR~\cite{lattner2021mlir}, \newline ONNX-MLIR~\cite{jinCompilingONNXNeural2020}, \newline IREE~\cite{IREE}, \newline ByteIR~\cite{ByteIR} 
& Multiple levels
& Extensible dialect system and progressive lowering across abstraction levels
& Requires substantial compiler engineering for each new target \\
\cline{2-5}

& PolyBlocks~\cite{polyblocks2026} 
& Tensor
& Automatic analytical polyhedral code generation; reusable across device classes without autotuning
& Analytical cost models trade peak per-target performance for portable, search-free compilation \\
\cline{1-5}
\cline{2-5}

\multirow[c]{4}{1.1cm}{GPU} & Triton~\cite{Triton}, \newline TileLang~\cite{
tilelang2025}, \newline Hexcute~\cite{Zhang2025HexcuteAC}, \newline CUTLASS~\cite{cutlass}, \newline CUDA Tile~\cite{cudatile}
& Tensor
& High-performance custom kernels for GEMM, attention, and fused operators
& Requires hardware-aware, tile-level programming and target-specific tuning \\
\cline{2-5}

& Kitsune~\cite{kitsune2025} 
& Graph
& Reduces off-chip traffic and improves utilization through producer--consumer dataflow execution
& Requires hardware-architecture and runtime support; less portable than pure software optimization \\
\cline{2-5}

& Neptune~\cite{neptune2025}, \newline Prism~\cite{prism2026}
& Tensor
& Improves locality, reduces tuning cost, and expands optimization beyond conventional schedule search
& Benefits depend on workload structure, cost-model quality, and backend integration \\
\cline{2-5}

& Event Tensor~\cite{eventTensor2026}, Nautilus~\cite{nautilus2026} 
& Graph
& Persistent megakernels for dynamic shapes; autotuned tiled scheduling of optimizations
& Rely on hand-crafted heuristics tuned mainly on attention-like operators, leaving transfer to other operator families open \\
\hline

%\multirow[c]{2}{1.1cm}{Agnostic} & LLMCompiler / Reasoning Compiler~\cite{Cummingfirst,LLMCompile,reasoningCompiler2025}
%& LLM-/agentic-guided code and schedule optimization
%& Reduces manual optimization effort and proposes application-specific transformations
%& Correctness, reproducibility, and validation of the proposed transformations remain open \\

%& Compiler-R1~\cite{compilerR12025}
%& Agentic RL-based autotuning
%& Explores large optimization spaces with less exhaustive manual tuning
%& Generalization across programs and targets; costly reinforcement-learning training \\
%\hline
\end{tabular}
\end{table*}

\subsection{DL Compilers for HPC Systems}

%Deep neural network workloads on HPC and datacenter systems require compilers capable of targeting CPUs, GPUs, tensor accelerators, and heterogeneous accelerator clusters.
%In this setting, compilation goes beyond functional lowering: it must optimize throughput, latency, memory locality, communication, and accelerator utilization while supporting large and structurally complex workloads such as large language models and mixture-of-experts architectures.

Table~\ref{tab:compiler_comparison_hpc} summarizes representative compilers for HPC systems, highlighting their hardware targets, abstraction levels, strengths, and limitations.
%A first class of approaches operates at the graph level by lowering high-level neural network graphs and applying transformations such as operator fusion, constant folding, algebraic simplification, and layout propagation, along with backend dispatch.
Graph-level approaches maintain a global view of the model, allowing optimizations across operator boundaries and reducing redundant memory traffic.
Glow~\cite{rotemGlowGraphLowering2019}, for example, converts the input model into a strongly typed graph IR, applies graph-level optimizations, and then lowers it through a node-lowering phase that reduces high-level operators to a small set of linear-algebra primitives, so that new backends need only implement those primitives.
ONNC~\cite{linONNCCompilationFramework2019} adopts an IR with a one-to-one mapping to the ONNX operator set.
% this correspondence simplifies porting to hardware by using coarse-grained operators that are not part of the generic LLVM backend IRs.
XLA \cite{XLA} follows a similar whole-program philosophy, but represents a program as a dataflow graph of linear algebra and element-wise operations over which aggressive operator fusion and layout assignment are performed before generating code for CPUs, GPUs, and TPUs.
StableHLO \cite{stablehlo} promotes XLA's operation set into a portable dialect, decoupling frontends from backends and acting as a stable interchange layer between the two.
ONNX-MLIR~\cite{jinCompilingONNXNeural2020} builds on MLIR~\cite{lattner2021mlir} and progressively lowers ONNX operators through dialects toward executable code; similar MLIR-based approaches are adopted by IREE~\cite{IREE} and ByteIR~\cite{ByteIR}.
%IREE is an end-to-end compiler and runtime that ingests StableHLO and progressively lowers it through internal dialects (flow, stream, HAL) into target-specific code.
%ByteIR similarly couples frontends (TensorFlow, PyTorch, ONNX), an MLIR-based compiler, and a runtime, using StableHLO as the frontend--compiler interface; it utilizes generic graph-, loop-, and tensor-level optimizations in MHLO and Linalg, so that ASIC backend developers can reuse them and implement only the last-mile lowering for their target.
These systems are effective when the workload can be captured as a graph of tensor operations, but their final performance still depends on the backend's kernel quality and the IR's ability to represent target-specific memory layouts and execution constraints.

%A second class of methods focuses on tensor-program optimization.
%Halide~\cite{ragan2013halide} introduced the separation between algorithm and schedule, allowing transformations such as tiling, reordering, vectorization, unrolling, and parallelization to be expressed independently from the computation itself.
TVM~\cite{chen2018tvm} combines graph-level optimization with tensor-level scheduling; it extends the separation between algorithm and schedule proposed by Halide~\cite{ragan2013halide} to deep learning and allows autotuning through machine learning-based mechanism.
Relay~\cite{roeschRelayNewIR2018} and Relax~\cite{relaxASPLOS2025} further enrich this compilation stack by representing neural network programs at higher abstraction levels before lowering them to tensor programs and executable kernels.
A particularly important mechanism in tensor-level compilers is the mapping of loop-level tensor computations to hardware-specific tensor instructions.
In TVM, this is formalized through an extensible tensor-intrinsic mechanism: the compiler separates the target hardware intrinsic from the schedule, allowing the behavior of the intrinsic and its lowering rule to be declared in the tensor expression language.
Conceptually, this allows a recognized loop-level computation pattern to be replaced by a hardware-specific primitive.
%For example, an inner matrix-multiplication loop nest can be tiled into fragments compatible with tensor instructions such as NVIDIA Tensor Core WMMA or MMA operations, and the corresponding sub-computation can then be lowered to the target intrinsic.
This mechanism is essential for exploiting specialized tensor hardware, as maximum utilization can be achieved only if the computation and its data are precisely aligned with the hardware.
%Overall, tensor-program optimization mechanisms, while allowing decoupling between scheduling and computation, still require finding which schedule is good for a given architecture.

Domain-specific languages for AI kernel generation express computations over tiles rather than directly over tensors.
Triton~\cite{Triton} provides a tile-level programming abstraction in which programmers express GPU kernels over parts of tensors, leaving intra-tile thread mapping, memory layout, and access scheduling to the compiler.
Hexcute~\cite{Zhang2025HexcuteAC} retains the same tile-based structure but exposes finer-grained abstractions than Triton, making shared memory and registers visible so the programmer can control how tiles are laid out across the memory hierarchy.
%It targets the mixed-type (quantized) matmuls that heuristics handle poorly, and, to avoid the manual layout coding required by lower-level models such as CUTLASS~\cite{cutlass}, it treats layouts as functions and automates layout and task-mapping synthesis with a type-inference algorithm.
TileLang~\cite{tilelang2025} separates the dataflow expression of a kernel from scheduling decisions such as thread binding, pipelining, layout transformation, and tensorization.
%Overall, these DSLs trade performance against the architectural detail that the programmer must manage.
%None of them presents a definitive solution, but all represent different operating points of this tradeoff.
While DSLs expose the tile abstraction at the language level, a complementary direction pushes it into the compiler infrastructure itself.
NVIDIA's CUDA Tile IR is an MLIR-based tile-level IR for tile-based execution and tensor-core code generation~\cite{cudatile}; performance is still bounded by automatic memory hierarchy placement, and its scope is limited to NVIDIA tensor cores.
%The PyTorch ecosystem has also moved toward compiler-assisted execution.
TorchDynamo~\cite{torchdynamo} captures Python bytecode at runtime and extracts FX graphs while preserving PyTorch's define-by-run programming model.
TorchInductor~\cite{torchdynamo} lowers these graphs to optimized kernels, often through Triton on GPUs.
%This design allows compilation to be integrated into dynamic Python workflows, although graph breaks, dynamic shapes, unsupported operators, and side effects still limit the extent to which complete applications can be optimized as a single static graph.
Kitsune~\cite{kitsune2025} questions the execution model assumed by conventional GPU compilation, introducing primitives for dataflow execution on GPUs and an end-to-end compiler based on PyTorch Dynamo.
%Instead of executing one operator at a time under a bulk-synchronous execution model, Kitsune enables finer-grained producer-consumer execution and spatial pipelines across GPU resources.
%This is relevant because many performance bottlenecks are not caused only by the arithmetic cost of individual operators, but also by synchronization and memory-traffic overheads between dependent kernels.
%By enabling data to flow directly between producer and consumer computations, dataflow-oriented execution can reduce redundant off-chip memory accesses and improve utilization of otherwise idle GPU resources.
However, this kind of optimization requires both software and hardware support, making it difficult to generalize.

More recently, methods targeting multiple IR levels have also been developed, such as Event Tensor~\cite{eventTensor2026}, Neptune~\cite{neptune2025} and its extension Nautilus~\cite{nautilus2026}, Prism~\cite{prism2026}.
While they can achieve strong performance, they are bound by greedy heuristics and hand-crafted rewriting rules that limit general applicability: the search space they explore is almost entirely tuned on attention-like operators, leaving open how well the heuristics may transfer to operators outside that family.
%PolyBlocks~\cite{polyblocks2026} focuses on reusable compiler infrastructure for AI chips and programming frameworks, using MLIR-based abstractions, loop transformations, cost models, and hardware-aware mapping strategies to reduce the engineering effort required for new accelerators.
%Event Tensor~\cite{eventTensor2026} addresses dynamic megakernel compilation, the generation of a single kernel fusing multiple kernels, by representing dependencies between tiled tasks as first-class compiler objects, enabling the compiler to reason about shape-dependent and data-dependent execution inside persistent kernels.
%Neptune~\cite{neptune2025} focuses on advanced operator fusion, especially for sequences of reduction operators, by intentionally breaking dependencies and introducing algebraic correction expressions that restore correctness.
%Nautilus~\cite{nautilus2026} builds on top of Neptune IRs, adding a VR-tile IR (virtual-register tile IR) for expression rewriting and an autotuner to optimize scheduling of optimizations across the IRs.
%Prism~\cite{prism2026} uses symbolic superoptimization to represent families of tensor programs equivalent to the one given and prunes suboptimal regions of the search space before instantiating concrete implementations.
Relax~\cite{relaxASPLOS2025} is an outlier as it does contribute a new optimization method but rather a new abstraction, i.e., a cross-level representation with first-class symbolic shapes, whose value lies in unifying and enabling optimizations rather than in performance itself.
%Relax~\cite{relaxASPLOS2025} introduces composable abstractions for end-to-end dynamic machine learning by combining graph-level computation, tensor programs, external calls, and symbolic shape information in a single representation.
%This is useful when shapes or control-flow properties cannot be fully fixed at compile time.
PolyBlocks~\cite{polyblocks2026}, instead, is a complete analytical, polyhedral-based code-generating compiler that automatically emits target-specific code across device classes; by relying on analytical cost models rather than empirical autotuning, it trades peak per-target performance for portable, search-free compilation.

In recent years, compilers have also begun to incorporate LLMs to assist with optimization decisions.
Rather than relying solely on autotuning, LLMCompiler~\cite{Cummingfirst,LLMCompile} uses language models to propose or guide transformations, retaining up to 77\% of the gains of comprehensive autotuning.
Compiler-R1~\cite{compilerR12025} provides a two-stage end-to-end RL training framework along with a curated dataset for the task.
Reasoning Compiler~\cite{reasoningCompiler2025} couples an LLM with Monte Carlo tree search, where the LLM acts as a proposer of hardware-informed transformations conditioned on the current program state and accumulated performance feedback.
These methods are attractive for emerging hardware, where expert-written schedules and cost models may not yet exist, but the transformations they generate still require correctness validation and may not be reproducible or aligned with the true hardware cost model.
They are therefore best viewed as complementary to, not a replacement for, conventional compiler analysis, autotuning, and verification.

\subsection{DL Compilers for Edge and TinyML Systems}

Deep learning compilers for edge and TinyML systems target devices such as microcontrollers, low-power embedded processors, and heterogeneous edge SoCs equipped with small neural accelerators (NPUs), with strict constraints on memory capacity, energy consumption, code size, and real-time behavior.
Edge compilers use many of the same lowering and code generation mechanisms as HPC compilers, but their optimization objectives are different: rather than maximizing throughput, they prioritize predictable deployment under tight memory, energy, and latency constraints.
Table~\ref{tab:compiler_comparison_edge} presents representative compilers for edge systems with their strengths and limitations.

\begin{table*}[tb]
\centering
\scriptsize
\setlength{\tabcolsep}{3.5pt}
\renewcommand{\arraystretch}{1.15}
\caption{Representative DL compiler methodologies and frameworks for edge and TinyML systems.}
\label{tab:compiler_comparison_edge}
\begin{tabular}{m{0.9cm}|m{1.9cm}|m{1.5cm}|m{3.95cm}|m{4.3cm}}
\hline
\textbf{Target Device} & \textbf{Framework} & \textbf{Optimization level} & \textbf{Main strengths} & \textbf{Main limitations} \\
\hline

\multirow[c]{5}{1.2cm}{MCU} & TFLite Micro~\cite{david2020tensorflow}
& Runtime
& Small footprint, portability, and broad embedded ecosystem
& Interpreter overhead; limited dynamic-shape/control-flow support; performance depends on optimized backend kernels \\
\cline{2-5}

& ExecuTorch~\cite{ExecuTorch}
& Runtime
& Unified PyTorch-native deployment from MCUs to SoCs; AOT graph compilation, quantization, and memory planning; selective build for tight code size
& Delegates kernels to vendor backends; performance and debuggability depend on backend quality \\
\cline{2-5}

%& Larq Compute Engine~\cite{Larq} 
%& Runtime and optimized binary neural-network kernels
%& Efficient execution of binarized networks with reduced memory and arithmetic cost
%& Specialized to binary networks; limited operator coverage \\

& DORY~\cite{burrello2021dory} 
& Layer, memory
& Constraint-programming-based tiling for tight SRAM budgets
& Primarily static layer-wise execution; limited support for dynamic workloads \\
\cline{2-5}

& TelaMalloc~\cite{TelaMalloc} 
& Memory, scheduling
& Combines constraint solving and heuristics to reduce memory contention
& Focuses mainly on allocation; not a complete general compiler stack \\
\cline{2-5}

& X-CUBE-AI~\cite{CubeAI} 
& Code generation
& Integrated conversion, quantization, profiling, and deployment for STM32 devices
& Vendor-specific; limited compiler extensibility and portability \\
\hline

\multirow[c]{5}{1.2cm}{MCU + accel./ \newline NPU} & AutoTiler~\cite{flamand2018gap}
& Memory, code generation
& Explicit orchestration of L1/L2/L3 transfers and DMA-based tiling
& Vendor-specific; requires regular static workloads \\
\cline{2-5}

& Deeploy~\cite{Deeploy} 
& Memory, code generation
& Integrates target-specific kernels with generated memory orchestration; heterogeneous deployment
& Requires target-specific backend descriptions and kernel integration \\
\cline{2-5}

& HTVM~\cite{vandelmHTVMEfficientNeural2023a}, \newline MATCH~\cite{MATCH}, \newline MATCHA~\cite{MATCHA}
& Tensor
& Improves retargetability while exposing accelerator-specific mapping and memory information
& Depends on accurate hardware models, cost models, and custom backend descriptions \\
\cline{2-5}

& eIQ Neutron~\cite{eiqNeutron2025} 
& HW/SW co-design
& Co-designed compiler optimizes compute utilization and data movement on the target NPU
& Vendor-specific; closely tied to the accelerator architecture, limiting portability \\
\cline{2-5}

& Hexagon-MLIR~\cite{hexagonmlir} 
& Code generation
& No reliance on kernel libraries; operator fusion and data orchestration for mega-kernels
& Reaches up to $\sim$80\% of hand-written kernel performance; orchestration scoped to single mega-kernels, not the whole graph \\
\hline
\end{tabular}
\end{table*}

A first class of tools focuses on lightweight inference runtimes and model conversion.
TensorFlow Lite for Microcontrollers~\cite{david2020tensorflow} provides a small-footprint runtime for deploying quantized neural networks on microcontrollers;
%Larq Compute Engine~\cite{Larq} specializes deployment for binarized neural networks using optimized kernels that reduce arithmetic and memory cost.
X-CUBE-AI~\cite{CubeAI} provides a vendor-supported flow for converting models into optimized C code for STM32 microcontrollers and their accelerators.
These toolchains are practical for deployment because they integrate conversion, quantization, profiling, and runtime support, but they typically expose less compiler extensibility than research-oriented frameworks.

A second class of edge compilers explicitly optimizes the mapping of a network onto the target memory hierarchy.
%Within this space, two design philosophies coexist.
Portable, vendor-agnostic compilers can target multiple architectures; for example, for accelerator-free targets, DORY~\cite{burrello2021dory} casts tiling and memory allocation as a constraint-programming problem to deploy quantized layers under tight SRAM budgets, while TelaMalloc~\cite{TelaMalloc} combines constraint solving with heuristics to reduce on-chip memory contention.
Other compilers target heterogeneous platforms equipped with accelerators: Deeploy~\cite{Deeploy,Deeploy2} follows a bottom-up flow that wraps user-provided kernels with tiling, allocation, and target-specific orchestration; HTVM~\cite{vandelmHTVMEfficientNeural2023a} couples TVM with DORY-style code generation; MATCH~\cite{MATCH} generalizes the approach with a customizable hardware abstraction and cost model, and its extension MATCHA~\cite{MATCHA} adds support for multiple concurrent accelerators.
Finally, ExecuTorch \cite{ExecuTorch} captures a DNN model into a graph IR
% called the Edge Dialect, built from a small Core ATen\cite{NEURIPS2019_9015} operator set, 
over which it performs ahead-of-time quantization, memory planning, and graph partitioning, then routes each subgraph to a backend based on the export target.
Crucially, none of these frameworks generates the compute kernels itself: they orchestrate data movement around hand-written, target-optimized kernels, on which their end-to-end efficiency ultimately depends.
Other compilers trade portability for performance through vendor-specific co-design.
AutoTiler~\cite{flamand2018gap} generates tiled code for GAP processors
%by orchestrating transfers across L1, L2, and external memory, 
while eIQ Neutron~\cite{eiqNeutron2025} co-designs the NPU and its compiler;
%, using constraint programming to maximize compute utilization and minimize data movement on a single accelerator.
these proprietary toolchains typically extract higher performance than portable alternatives, but at the cost of being locked to one vendor's silicon, with no path to other targets.
One notable exception is Hexagon-MLIR~\cite{hexagonmlir}, which keeps an open compilation pipeline, using MLIR to lower Triton kernels directly into binary code for Qualcomm's Neural Processing Units rather than orchestrating data movement around a fixed kernel library.
Its scope, however, is narrower than that of the portable frameworks above: it optimizes orchestration only within a single mega-kernel, whereas compilers such as Deeploy generate orchestration code for the network as a whole.
Moreover, the generated kernels still fall short of hand-optimized ones, reaching only 80\% of their performance.

%In recent times, interest has also been shown toward the usage of deployment frameworks for the on-device training of neural networks.
%An example is given by Deeploy and MATCH, both originally inference compilers that have been extended to map full training graphs onto memory-constrained edge hardware.
%The MATCH-based flow~\cite{rusci2026odl} adds automatic differentiation and an external-memory-aware planner to deploy end-to-end training.
%TrainDeeploy~\cite{wang2026traindeeploy} instead extends Deeploy by jointly scheduling tiling and memory over the entire forward--backward graph and offloading GEMMs to an on-chip accelerator.
% enabling parameter-efficient fine-tuning of a small Transformer on-device via LoRA.
%Beyond compiler-based flows, standalone frameworks such as AIfES~\cite{wulfert2024aifes} bring on-device training to MCU-class targets through hand-written per-layer backward passes and a pluggable accelerator interface, without an operating system or automatic differentiation.

\subsection{DL Compilers for IMC Architectures}

%Traditional DL compilers focus heavily on caching, loop tiling, and latency-hiding to neutralize the "memory wall". However, IMC architectures utilizing emerging non-volatile memory crossbars such as ReRAM and PCM\cite{Mannocci2026}, or even digital volatile SRAM \cite{perri24} radically alter the execution paradigm.
Compiling for IMC architectures based on emerging non-volatile memory crossbars \cite{Mannocci2026,perri24} introduces paradigms completely distinct from traditional register-based or thread-based compilation \cite{Asifuzzman26}.
%Spatial mapping and data placement are leveraged in IMC architectures since data is the compute engine. 
Standard compilers optimize for temporal locality; IMC compilers must optimize for physical and spatial placement, as operands must be co-located within specific memory crossbars or wordlines.
Analog non-idealities are also a key point to address by compilers for IMC: many IMC architectures rely on analog current/voltage summation (e.g., crossbars for Matrix-Vector Multiplication) \cite{chen21}, and compilers must tolerate or compensate for hardware non-idealities like device variations, IR drop, and quantization errors \cite{Baroni22}.
Further, practical IMC chips are rarely homogeneous and largely based on resource heterogeneity; they combine IMC arrays with traditional digital logic (near-memory processing) and routing fabrics. The compiler must map parts of a data-flow graph to execute in-memory and parts on auxiliary digital cores.

The modern IMC compiler toolchain often relies on MLIR to capture hardware-specific semantics \cite{Khan25}. 
%The compilation pipeline starts from a high-level representation of DNNs or other neuromorphic tasks using PyTorch or C++. 
After a front-end parsing phase, high-level tensors are broken down into smaller sub-tensors that physically match the dimensions of the hardware memory arrays. The next step of the pipeline is crossbar partitioning and assignment; weights (for neural networks) or boolean logic variables are then statically mapped to specific memory cells. 
%Loop tiling is also performed. The final step of the pipeline is instruction scheduling. 
Finally, IMC scheduling controls the sequence of wordline activations, sensing cycles, and analog-to-digital converter (ADC) sampling windows to maximize parallelism while respecting strict power-delivery constraints.

To maximize the efficiency of IMC hardware, compilers employ several specialized optimization passes \cite{Mambu22}. At the algorithmic level, quantization-aware mapping \cite{Long2020QPIM} recognizes that some layers are highly sensitive to accuracy drops while others are incredibly resilient.
%When compiling for analog crossbars, this idea is particularly useful since high-bit tensors must be split into slices that match the narrow bit-width of physical devices. The technique also helps to relax ADC constraints, which turns out to be beneficial as they consume up to 80\% of an IMC chip’s total energy and physical area. 
%Regarding spatial-level optimization, we mention the work in \cite{Ankit19}, a foundational paper presenting a complete compiler and architecture toolchain for IMC.
Precise spatial optimization passes are needed to handle large neural network layers by creating matrix replication and pipelined crossbar cascading topologies, ensuring that inter-core routing bottlenecks are mitigated at compile-time \cite{Ankit19}.
%An important point for IMC compilers is the creation of hardware-aware strategies that mitigate the IR-drop. 
As signal wire lengths grow in IMC macros, parasitic resistance creates an IR voltage drop that corrupts analog computational accuracy. The work in \cite{Zhang25} introduces a cross-layer software-hardware co-design that models spatial voltage drops during the compilation phase to intelligently predict signal degradation. 
%On the software side, the compiler implements hardware-aware mapping and scheduling passes, restructuring matrix layouts and weight distributions to balance current loads across the memory array. On the hardware side, it pairs these compiler optimizations with dynamic peripheral circuit tuning, effectively mitigating accuracy drops while maximizing energy efficiency and crossbar utilization.
Bridging logic synthesis with digital IMC is also an important task: instead of compiling code into traditional AND/OR gates, the compiler can transform logic natively into 3-input majority gates and inverters, which can be executed in situ by emerging non-volatile memory crossbars using stateful logic primitives \cite{Soeken16}.
\section{HLS-based Design Methodologies}
\label{sec:HLS_Design_based_methodologies}

The possibility of quickly designing and evaluating FPGA/ASIC accelerators is crucial to obtaining efficient heterogeneous architectures for DL.
One methodology that significantly increases the productivity of hardware developers is High-Level Synthesis (HLS), i.e., the automated translation of algorithmic descriptions into Register-Transfer Level (RTL) accelerators.
HLS provides a variety of advantages to designers, who can not only work faster at a higher level of abstraction and perform easier functional verification but also create various solutions with different performance and area characteristics from the same source code by changing design constraints or tweaking optimization directives.
Such advantages make design space exploration much faster than with low-level hardware design in Verilog or VHDL.

Figure~\ref{fig:HLS_Design_based_methodologies_taxonomy} lists the HLS tools and HLS-based design methodologies reviewed in the following sub-sections.
A previous survey \cite{hlssurvey} analyzed the evolution of HLS tools and the challenges that lay ahead of them to support newer application domains and performance requirements; in this paper, we briefly describe the main academic and commercial HLS tools before diving into research that integrates HLS into design frameworks for DL acceleration.
We focus on \textit{methodologies that exploit HLS}, rather than accelerators and architectural templates for DL generated through HLS, which are addressed extensively in other surveys \cite{silvano2023survey}.
The emphasis is therefore placed on design flows, frameworks, and compilation methodologies that use HLS as a key enabling technology to translate pre-trained DL models into corresponding hardware implementations.

\begin{figure}[t]
\centering
{
    \footnotesize
    \begin{forest}
        qtree,
        [{\textbf{HLS-based design methodologies}},
            [{\textit{HLS tools}},
              [{Commercial},
                [{Vitis HLS~\cite{vitisug212},\\
                Altera HLS Compiler~\cite{IntelHLS2022},\\
                Catapult~\cite{CatapultHLS2022},\\ 
                Stratus HLS~\cite{StratusHLS2022},\\
                Libero (LegUp)~\cite{smartHLS} 
                }],
              ],
              [{Research/Academic},
                [{Bambu~\cite{ferrandi2021bambu},\\
                Dynamatic \cite{dynamatic},\\
                DASS \cite{dass},\\
                MLIR CIRCT~\cite{circt,circt-hls},\\
                HECTOR \cite{xu2022hector}}],
              ]
            ],
            [{\textit{HLS design flows for DL}}
              [{Library-based},
                [{hls4ml~\cite{duarte2018fast,fastml_hls4ml,schulte2026hls4ml,jiang2025low},\\
                FINN~\cite{blott2018finn,finnT},\\
                GNN libraries~\cite{gcnhls,gnnhls},\\
                LLM library~\cite{chen2024understanding}
                }],
              ],
[{Compiler-based},
  [{ScaleHLS~\cite{ye2022scalehls}, HIDA~\cite{hida}\\
    SODA~\cite{sodaMICRO,sodaSNN,TCdataflow},\\
    Stream-HLS~\cite{streamHLS}, BraggHLS~\cite{levental2024bragghls},\\
    Allo~\cite{chen2024allo}, ARIES~\cite{aries}
    }]
]
            ]
        ],
    \end{forest}}
    
\caption{Tools and methodologies discussed in Section~\ref{sec:HLS_Design_based_methodologies}.}
\label{fig:HLS_Design_based_methodologies_taxonomy}
\end{figure}

\subsection{HLS Tools}

The most popular commercial HLS tool is undoubtedly Vitis HLS \cite{vitisug212}, provided by AMD (formerly by Xilinx) as part of its FPGA design toolchain.
Vitis HLS allows users to describe their designs using high-level languages, such as C and C++, and targets embedded and data center FPGA platforms sold by AMD.
As is the case for most HLS tools, the input code must be written with a hardware implementation in mind: in fact, arbitrary software code written for a CPU target can achieve very low performance after synthesis since it typically does not expose enough parallelism to exploit the spatial concurrency available on FPGAs. 
%In the Vitis HLS flow, C/C++ is used to describe both the hardware functionality and the testbench.
%A report produced at the end of the synthesis estimates the operating frequency and resource usage of the accelerator and can be used as a reference for further refinement.
%C/RTL co-simulation can be performed to verify and validate the synthesized RTL design using the C/C++ testbench provided, and it also outputs the latency of the accelerator in terms of clock cycles.
The designer can specify optimization directives and constraints as annotations in the code (commonly referred to as \textit{pragmas}) or in a configuration script; these include loop-level transformations, indications to allow dataflow execution, constraints on hardware resources utilization, memory layout directives, and the choice of an interface protocol.

Other FPGA vendors provide HLS tools within a full design suite for their platforms.
The Altera HLS Compiler \cite{IntelHLS2022} is part of the Quartus design suite, although
%: it compiles C++ functions into an RTL implementation for Altera FPGAs and optimizes them through a simple command-line interface. 
Intel and Altera have discontinued the HLS Compiler and have suggested the oneAPI toolkit \cite{oneapi} instead to enable developers to seamlessly port OpenCL code across CPUs, GPUs, and FPGAs.
Catapult \cite{CatapultHLS2022} is a multi-target HLS and verification tool provided by Siemens, synthesizing C++ and SystemC code for FPGA and ASIC. 
Stratus HLS \cite{StratusHLS2022} from Cadence synthesizes SystemC code written with a lower-level perspective, i.e., requiring users to describe interface protocols between components explicitly. 
LegUp \cite{LegUpHLS2013} started as an open-source, LLVM-based HLS tool developed in academia; it was later acquired by Microchip and integrated in the Libero design suite \cite{smartHLS}.

Moving to research tools, Bambu \cite{ferrandi2021bambu} 
%is the most mature and up-to-date open-source HLS framework, developed and maintained by Politecnico di Milano.
is a command-line tool with extensive coverage of C/C++ constructs (including some that are not supported by Vitis HLS, such as pointer arithmetic and dynamic resolution of memory accesses), and it can also synthesize compiler Intermediate Representations (IR) generated through LLVM-based frameworks.
The open-source nature of Bambu allows developers to dive deep into the analysis and transformation passes at the core of the HLS process and to research novel HLS methodologies.
Bambu can generate optimized VHDL or Verilog code for various FPGA and ASIC targets through a library of pre-characterized functional units with area and delay information for each possible target, making it much more flexible than commercial tools tied to a single hardware vendor.
%Floating-point operations are supported in Bambu through the FloPoCo framework~\cite{DinechinPasca2011-DaT} or through an optimized soft-float library that allows generating non-standard floating-point encodings \cite{truefloat}.
%Recently, Bambu has been extended to support the synthesis of multi-threaded OpenMP applications \cite{svelto,gozzi24sparta}, and it has been fundamental in research projects dealing with the acceleration of big data analytics \cite{everest24date}, aerospace applications~\cite{hermes23date}, and AI \cite{pnrr26date}.

Most HLS tools (including Vitis HLS and Bambu) follow a static scheduling paradigm, i.e., they statically assign each operation to a clock cycle; the opposite choice is dynamic scheduling, where functional units decide when to execute operations at runtime based on the availability of data and on handshake signals.
Dynamatic \cite{dynamatic} is an academic HLS tool based on dynamic scheduling, shown to generate high-performance accelerators for irregular applications at the cost of additional area overhead.
DASS \cite{dass} splits the input representation and uses either static or dynamic scheduling depending on user-defined pragmas so that it is possible to combine the strengths of both methods in the same accelerator.
In general, dynamic scheduling and dataflow \cite{hida,streamHLS} are active research areas for HLS, as they can result in higher performance when synthesizing irregular applications.
While DNN and CNN operators are usually synthesized through static scheduling, other classes of DL algorithms, such as graph neural networks (GNNs), may benefit from accelerators that can respond at runtime to data-dependent computational patterns.

Another active research area engages with the MLIR infrastructure \cite{lattner2021mlir}.
%MLIR allows defining specialized IRs called \textit{dialects} to implement analysis and transformation passes at different levels of abstraction, and it can interface with multiple software programming frameworks, including the ones used to implement DL algorithms.
HECTOR \cite{xu2022hector} uses MLIR to implement dialects for scheduling and resource management, which can be used to achieve similar functionality as HLS tools.
The CIRCT project \cite{circt} aims at using MLIR to build a new generation of interoperable tools and compilers for hardware design, starting from the definition of circuit-level IRs and working upwards to higher levels of abstraction (e.g., dataflow models or finite state machines).
Part of the project is dedicated to HLS \cite{circt-hls}, particularly to implementing static and dynamic scheduling through MLIR and CIRCT dialects.
CIRCT could be an essential building block for future industrial and academic design flows; however, its degree of maturity is lower compared to HLS tools with optimized synthesis algorithms and resource libraries supported by decades of research.

\subsection{HLS-based Design Flows for Deep Learning}

\begin{figure}[tb]
    \centering
    \begin{subfigure}[b]{0.48\columnwidth}
        \centering
        \includegraphics[width=0.9\textwidth]{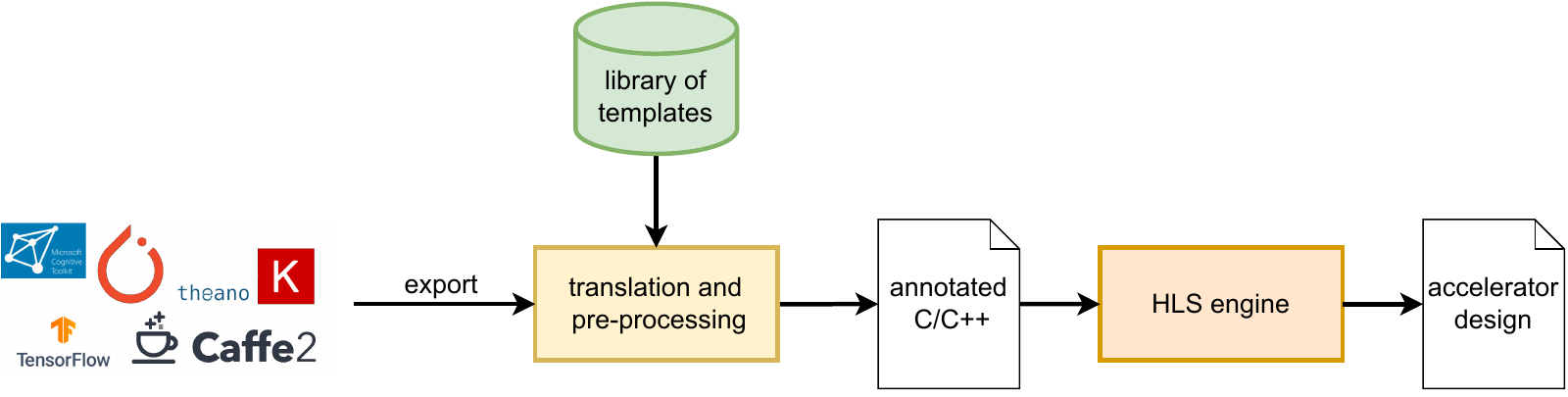} 
        \subcaption{Library-based HLS flow.}
        \label{fig:classicHLS}
    \end{subfigure}
    \begin{subfigure}[b]{0.48\columnwidth}
        \centering
        \includegraphics[width=0.9\textwidth]{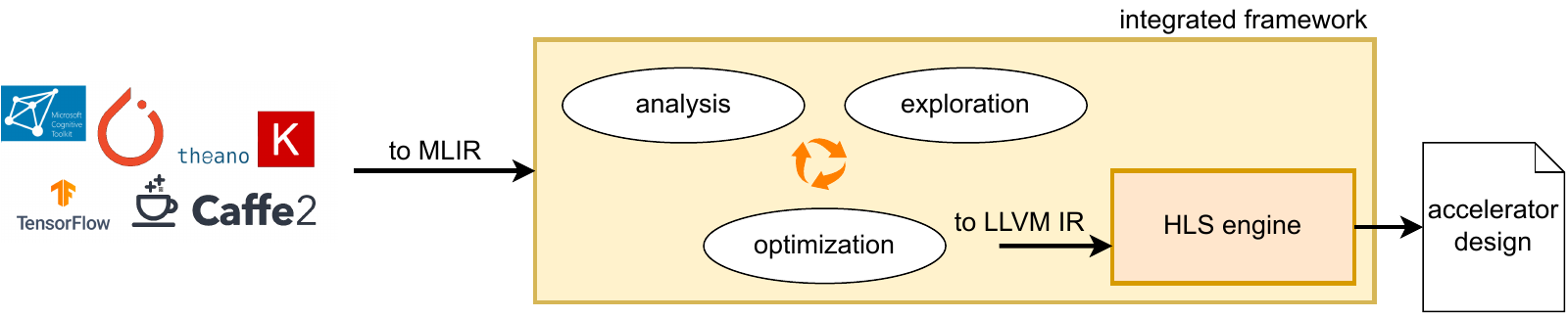}
        \subcaption{Compiler-based HLS flow.}
        \label{fig:modernHLS}    
    \end{subfigure}
    \caption{Two different ways of generating DL hardware accelerators through HLS \cite{posterCF22}.}     \label{fig:HLS}
\end{figure}

Tools and methodologies described in this section can be broadly divided into \textit{library-based} and \textit{compiler-based} approaches (Figure \ref{fig:HLS}).
Given a pre-trained DL model, library-based frameworks instantiate and connect HLS templates (usually corresponding to neural network layers), and send the resulting annotated C/C++ representation to an HLS backend.
Compiler-based approaches, instead, exploit common intermediate representations to translate the DL model through a series of progressive analysis and optimization passes until it is lowered to an abstraction that can be synthesized; tools that belong to this category may be presented as solutions to automate the generation of DL accelerators, but they are usually more flexible and extensible across different classes of applications.

\begin{table}[t]
\centering
\caption{Characteristics of library-based vs compiler-based HLS frameworks for DL.}
\label{tab:hls_methodologies}
\scriptsize

\begin{tabular}{p{0.25\textwidth}| p{0.25\textwidth}|p{0.4\textwidth}}

\multicolumn{3}{c}{\textbf{Library-based approaches}} \\
\midrule

\multicolumn{1}{@{}p{0.22\textwidth}}{
\textbf{Advantages}} & \multicolumn{2}{p{0.68\textwidth}@{}}{
$\bullet\,$ Mature and highly optimized operator implementations\newline
$\bullet\,$ Highest performance for supported workloads
} \\ 

\midrule

\multicolumn{1}{@{}p{0.22\textwidth}}{
\textbf{Disadvantages}} & \multicolumn{2}{p{0.68\textwidth}@{}}{
$\bullet\,$ Strongly dependent on HLS backend syntax \newline
$\bullet\,$ Optimizations are not portable to different applications
} \\ 

\hline

\textbf{Frameworks} & \textbf{Backends} & \textbf{DL models} \\
hls4ml \cite{duarte2018fast,fastml_hls4ml,jiang2025low,schulte2026hls4ml} & Vitis HLS, Catapult HLS (stable) \newline oneAPI, Bambu (experimental) & Small DNN/CNN, initial support for Transformer layers \\
FINN \cite{blott2018finn,finnT} & Vitis HLS & Quantized DNN/CNN, quantized Transformer (FINN-T) \\
GNN libraries \cite{gcnhls,gnnhls} & Vitis HLS & Graph DNN and CNN \\
Spatial acceleration for LLMs \cite{chen2024understanding} & Vitis HLS, Catapult HLS & Transformer \\

\hline

\multicolumn{3}{c}{} \\

\multicolumn{3}{c}{\textbf{Compiler-based approaches}} \\
\midrule

\multicolumn{1}{@{}p{0.22\textwidth}}{
\textbf{Advantages}} & \multicolumn{2}{p{0.68\textwidth}@{}}{
$\bullet\,$ Applicability to any input application even beyond DL \newline
$\bullet\,$ Easier design space exploration and extensibility
} \\ 

\midrule

\multicolumn{1}{@{}p{0.22\textwidth}}{
\textbf{Disadvantages}} & \multicolumn{2}{p{0.68\textwidth}@{}}{
$\bullet\,$ Possibly lower QoR than library-based approaches \newline
$\bullet\,$ Optimization requires compiler expertise
} \\ 

\hline

\textbf{Frameworks} & \textbf{Backends} & \textbf{DL models*} \\
ScaleHLS, HIDA \cite{ye2022scalehls,hida} & Vitis HLS & DNN/CNN \\
SODA \cite{sodaMICRO,sodaSNN,TCdataflow} & Bambu and Vitis HLS & DNN/CNN (stable), SNN (experimental) \\ 
Stream-HLS \cite{streamHLS} & Vitis HLS & DNN/CNN, Transformer \\ 
BraggHLS \cite{levental2024bragghls} & CIRCT & DNN \\
Allo~\cite{chen2024allo} & Vitis HLS & CNN/LLM \\
ARIES~\cite{aries} & Vitis HLS & CNN \\
\hline

\multicolumn{3}{p{0.9\textwidth}}{{\color{gray}* since these frameworks are, in theory, applicable to any workload, we report the models used as benchmarks in the corresponding papers}} \\

\end{tabular}
\end{table}

Two popular library-based frameworks that help automate the design of DL accelerators are hls4ml \cite{duarte2018fast,fastml_hls4ml,schulte2026hls4ml} and FINN \cite{blott2018finn}.
Both parse DNN models and replace operators with C/C++ functions taken from a library of templates that contain pragmas (Figure \ref{fig:classicHLS}); the HLS tool used as the backend (mainly Vivado or Vitis HLS) processes this intermediate C/C++ representation and produces a corresponding accelerator design without requiring further manual intervention.
The library of templates in hls4ml and FINN is necessarily tied to a specific HLS tool and a narrow set of models: this is required so that expert HLS developers can implement in advance the best version of all necessary operators for a pre-determined backend tool.
Given that each tool has its own coding patterns, annotations, and configuration directives that are not recognized by other tools, a new version of the library is needed to achieve efficient designs when switching to a new hardware target.
Library-based frameworks likely also have a narrow scope limited by application requirements, which can significantly affect resource utilization. 
For example, the original implementation of hls4ml was optimized for small, fully connected models under tight latency constraints, reflecting the needs of a high-energy physics experiment at CERN. 
To comply with those requirements, hls4ml proposed to store network weights inside on-chip logic and unroll all loops to increase parallelism, which quickly depletes FPGA resources when considering more complex models.
Other HLS libraries for DL implement kernels typically found in GNNs \cite{gcnhls,gnnhls}, with the additional challenge of the sparse and irregular nature of graph workloads.
LLM inference introduces new challenges, as transformer-based models follow different computational and memory access patterns that cannot be easily implemented by reusing library operators specialized for DNNs/CNNs; memory access latency often becomes a bottleneck as models scale dramatically in parameter count.
A first library of HLS kernels implementing LLM layers has been proposed \cite{chen2024understanding}, focused on maximizing on-chip data reuse through carefully tuned parallelism and dataflow optimizations.
hls4ml has added support for the multi-head attention (MHA) layer, SoftMax, and normalization for small low-latency transformers \cite{jiang2025low}.
FINN-T \cite{finnT} extends FINN to support quantized transformer kernels.

A different approach to the design of new HLS-based frameworks for DL relies on the possibilities offered by MLIR (Figure \ref{fig:modernHLS}).
ScaleHLS \cite{ye2022scalehls} exploits the multiple levels of abstraction provided by existing MLIR dialects to reason about graph-level, loop-level, and directive-level optimizations, and then it uses a custom dialect to translate MLIR into C++ with Vitis HLS pragmas.
A quality of results estimator and a design space exploration engine automatically identify the best combination of optimizations following user-defined constraints without requiring long simulation or synthesis runs to evaluate the effect of changes in the optimization directives.
HIDA \cite{hida} extends ScaleHLS to generate efficient dataflow accelerators.
% The MASE compiler \cite{MASE} also exploits MLIR passes to emit a C representation for HLS.
The SODA Synthesizer \cite{sodaMICRO} is an open-source, multi-level, modular, extensible, no-human-in-the-loop hardware compiler that translates high-level DL models into domain-specific accelerators through MLIR and HLS.
It comprises a compiler-based frontend that leverages MLIR (SODA-OPT \cite{sodaopt}) and a compiler-based backend that integrates state-of-the-art HLS tool Bambu; it generates highly specialized designs that can be synthesized with both commercial and open-source tools on FPGAs or ASICs, and it allows exploring design metrics through compilation passes and parameters, enabling the identification of architectural trade-offs depending on the target application requirements.
SODA has been extended to explore the synthesis of spiking neural network models \cite{sodaSNN} and to support a combination of static and dynamic scheduling in the HLS-generated accelerators \cite{TCdataflow}.
Both SODA-OPT and Stream-HLS~\cite{streamHLS} use MLIR also to generate a host application driving the execution of the accelerator; Stream-HLS introduces MLIR passes that optimize the performance of dataflow accelerators and presents results on linear algebra, DNN/CNN, and transformer kernels.
BraggHLS \cite{levental2024bragghls} implements a fully MLIR-based design flow for small DNNs by relying on the CIRCT backend to generate RTL code.
Allo~\cite{chen2024allo} is a framework that decouples computation and optimization decisions to enable faster design space exploration; it is built through MLIR and is one of the first works to tackle the HLS of a large transformer model.
ARIES \cite{aries} is an MLIR-based compilation flow for platforms that provide both AI engines and FPGA fabric, automatically generating C/C++ for Vitis HLS when a component of the accelerator needs to be implemented on the FPGA.

\section{Approximate Computing}
\label{sect:ApproxComp} 

In the last few years, the design of efficient hardware architectures for DL has received a great deal of attention, and several implementations have been proposed for both ASIC and FPGA-based platforms.
%Special efforts were dedicated to methodologies for the design of efficient computational units, later integrated as Intellectual Property (IP) blocks within accelerator architectures.
%To comply with the computational complexity of HPC applications, often including DL and DNNs, hardware designers first focused on solutions that could reach very high speeds.
%With this objective in mind, ever-increasing levels of parallelism have been introduced to realize increasingly complex computational architectures. 
%Unfortunately, this trend poses several concerns in power- and resource-constrained systems; several approaches were thus proposed to reduce the number of parameters and computations without renouncing acceptable accuracy. 
This section summarizes research into design methodologies for low-precision and approximate computing used in DL accelerators to improve their energy efficiency, as outlined in the taxonomy of Figure \ref{fig:approximate}.
Following the recent evolution from convolution-dominated DNNs to transformer-based workloads, the section also discusses how approximate computing can be applied to attention, normalization, and low-precision transformer inference, highlighting the corresponding accuracy/energy trade-offs.

\begin{figure}[t]
    \centering
    {
    \footnotesize
    \begin{forest}
        qtree,
        [{\textbf{Approximate computing approaches}}
            [{\textit{Algorithmic level}}
                [{SoftMax \cite{Zhu_2020,Cardarilli_2021,Spagnolo_2022}\\
                Sensors, memory,\\
                compute and communication \cite{Ghosh_2020}\\
                Pooling layers \cite{Sayal_2021, Spagnolo_2022_2}\\
                Super-Resolution \cite{Spagnolo_2023}, 
                Denoising \cite{Spagnolo_2023_2}\\
                Transformer SoftMax/LayerNorm \cite{Kim2025SoftmaxLayerNorm,LowPrecisionSoftmax2026,BAPS2026}}
                ]
            ],
        [{\textit{Architecture, gate- and transistor-level}}
            [{\textit{ASIC-based}}
                [{Multipliers \cite{Chen_2012,Cho_2004,Song_2007,Esposito_2017,Farshchi_2013,Strollo_2022}\\
                Dynamic \cite{Frustaci_2020}, 
                Compressors \cite{Esposito_2018,Strollo_2020}\\          
                Segmentation \cite{Strollo_2022}, 
                MACs \cite{Kim_2021}
                }],
            ],
            [{\textit{FPGA-based}}
                [{Adders {\cite{Perri_2020}}\\
                Multipliers \cite{Ullah_2021,Ullah_2022,Perri_2022,Waris_2021}\\
                }]
            ]
            ]
        ]
    \end{forest}
    }
    \caption{Approximate computing approaches discussed in Section \ref{sect:ApproxComp}.}
    \label{fig:approximate}
\end{figure}

\subsection{Compression Techniques}

To fit DNN models into hardware devices with constrained resources and a limited energy budget, designers typically exploit several compression techniques, such as quantization, pruning, and knowledge distillation \cite{Rokh_2023}. 
Quantization maps the floating-point weights and/or activations of a DNN to representations with lower bit-width and precision.
An alternative compression approach is pruning, which removes unnecessary or less critical connections within the network model and thus introduces a certain level of sparsity that reduces both memory usage and computational complexity with respect to the original DNN model. 
Knowledge distillation is another widely used technique to reduce the number of parameters and computations in DNNs.
In this case, the knowledge of a complex model is used to train a simpler model that will replace the complex one, thus ensuring reduced memory requirements and computational complexity at an accuracy level comparable to the complex model.
Although all the above methods lead to more resources- and energy-efficient hardware implementations, quantization offers several advantages over the other approaches \cite{Rokh_2023}.
In fact, quantization guarantees higher compression with lower accuracy loss, is independent of the network model, and can be exploited either during training (i.e., the network is trained with discrete quantized values) or as a post-training solution to accelerate the inference phase, even though this often causes a reduction in model accuracy.

\subsection{Approximate Computing for FPGA/ASIC Design}

Compression techniques are not the only design methodologies allowing the exploration of precision/performance/power consumption trade-offs; approximate computing has recently gained popularity as a powerful technique to reduce energy consumption and computational delay in error-resilient applications, such as multimedia processing, digital signal processing, wireless communications, and DL \cite{Alioto_2017}.
The basic principle of the approximate computing paradigm is straightforward, i.e., by relaxing the requirements of exact computation, it is possible to trade off the quality of the result for speed and energy dissipation.
Hardware accelerators for critical DL operators designed through approximate computing methodologies can be integrated as custom IPs in an overall system with high speed and energy efficiency.
Approximate computational IPs can be exploited in both ASIC and FPGA-based designs, with the common objectives of reducing the computational complexity of DL layers and optimizing speed and power consumption while introducing a reasonable accuracy loss.

Approximate computing can be exploited at different design levels, starting from the algorithm, passing through the architecture, down to gate- and transistor-level circuit topologies \cite{Jiang_2020}. 
Hardware-oriented approximation strategies at the algorithmic level re-formulate the mathematical functions of critical computational layers employed in DL models to significantly reduce the complexity and energy consumption.
For example, the approximate approach demonstrated in \cite{Spagnolo_2022} allows realizing the SoftMax layer in hardware by exploiting simple additions and logical bit-shifting operations to replace the computationally expensive exponentiations and divisions.
When adopted in the realization of DNN accelerators for popular models such as VGG-16 and ResNet-50, this approach guarantees computational times and power dissipation up to 7 and 12 times lower, respectively, than the accurate counterparts, introducing an accuracy penalty lower than 2\%. 
Another example of approximation at the algorithmic level is discussed in \cite{Spagnolo_2022_2}, where the computations performed within convolutional layers followed by down-sampling layers are approximated through a prediction method that identifies potential predominant features.
This approximation strategy has been customized for DNN inference hardware, and it leads to an overall energy saving of up to 70\% when applied to benchmark models with an accuracy loss lower than 3\%.
Recently, algorithmic approximation has also been exploited to reconstruct high-resolution images \cite{Spagnolo_2023} and to reduce the computational complexity of image denoising \cite{Spagnolo_2023_2}, as such applications have to meet tight frame rate and energy consumption constraints, making innovative and domain-specific design methodologies highly desirable.    
%Table~\ref{tab:approx_techniques_comparison} summarizes the techniques described so far.
It is worth noting that quantization and approximate arithmetic usually provide the most direct path toward hardware simplification, because they reduce datapath width and/or operator complexity. Conversely, pruning and knowledge distillation may provide substantial model-level compression, but their hardware impact depends on the regularity of the resulting model and on the availability of sparse or model-specific accelerator support. Approximate functions and runtime approximation occupy an intermediate position: they can offer large savings for expensive operators, but they require careful error analysis and calibration to preserve end-to-end accuracy.

\begin{comment}

\begin{table}[t]
\caption{Comparison across approximate computing techniques used in DL accelerators.}
\label{tab:approx_techniques_comparison}
\footnotesize
\resizebox{1\columnwidth}{!}{%
\begin{tabular}{lcccc}%{p{2.5cm} p{2.2cm} p{2.8cm} p{3.1cm} p{2.6cm}}
\toprule
\textbf{Technique} & \textbf{Design level} & \textbf{Main benefit} & \textbf{Main limitation} & \textbf{Hardware impact} \\
\midrule
Quantization & Algorithm & Reduces memory footprint & Accuracy loss & Narrower datapaths \\
 & & and arithmetic cost & & and smaller memories \\

Pruning & Algorithm & Removes redundancies & Irregular sparsity & Sparse-aware storage, \\
 & & & & and datapaths \\

Knowledge distillation& Algorithm & Transfers accuracy & Additional training & Hardware-friendly\\
 & & to a smaller model & & student model\\

Approximate functions& Algorithm & Simplifies costly & Operator-specific & Efficient computational\\
 & & nonlinear layers & error analysis & units \\

Runtime approximation& Architecture & Adaptable quality-energy  & Control, verification, and & Configurable accuracy \\
 & & trade-off & calibration overhead & modes \\

Approximate arithmetic& Gate/Transistor & Reduced area, delay,  & Accuracy loss & Adders, multipliers,\\
 & & and power & & and MAC units \\

\bottomrule
\end{tabular}
}
\end{table}
\end{comment}

When approximate computing is exploited at the architectural level, the mathematical functions of critical DL computational layers are kept unchanged, but they are implemented through approximate hardware operators.
Efficient solutions to exploit approximate computing at gate- and transistor-level have led to the design of approximate adders, multipliers, and multiply-accumulate units (MACs) for both ASIC \cite{Frustaci_2019, Frustaci_2020, Strollo_2020, Strollo_2022} and FPGA devices \cite{Prabakaran_2018, Ahmad_2021, Ullah_2022, Perri_2022}. 
These arithmetic operators receive a great deal of attention since they are the basic computational elements most extensively used in DL models.
ASIC- and FPGA-based designs are substantially different; in the former case, any circuit can be customized at both the gate and transistor level, while in the latter, it must be implemented using specific resources available within the FPGA chips, such as Look-Up Tables (LUTs) with a predefined number of input and output ports, Block RAMs with a pre-established memory capacity, Digital Signal Processing slices (DSPs) supporting a pre-established set of operations, and specific routing resources.
As hardware designers know, even optimized ASIC circuits may not be the best solutions to implement on an FPGA.
As an example, ASIC designs of adders can exploit efficient and fast topologies like the Carry-Look-Ahead (CLA), the carry-skip, and the Parallel-Prefix Architectures (PPA). 
On the other hand, the unique structure of FPGA devices and their reduced set of hardware primitives make it impossible to benefit from such topologies, while a basic Ripple-Carry logic can use the available computational and routing resources efficiently \cite{AMD}.

Typically, the operands to be processed by approximate arithmetic circuits are split into sub-words: some of the least significant bits are processed inaccurately, whereas the remaining most significant bits are passed to the accurate circuits.
Some strategies exploit static approximation that inflexibly sets the achieved accuracy at design time, while other solutions adopt dynamic approximation that allows tuning the quality target at runtime, thus leveraging the specificity of the data being processed with a graceful quality degradation. 
Similarly, several approximation techniques can be adopted in the design of multipliers \cite{Akbari_2017, Esposito_2018, Strollo_2020, Frustaci_2019, Frustaci_2020}. Some techniques use dynamic and static segmentation methods \cite{Hashemi_2015, Narayanamoorthy_2015, Strollo_2022, Di_Meo_2023}: the former downsizes the multiplier by selecting only a segment of the inputs starting from the leading one bit, whereas the latter processes only predefined portions of the multiplicands. 
Besides the approximation strategies oriented to ASIC designs, appropriate methodologies are available to achieve high-performance and low-power designs of approximate modular multipliers also on FPGAs \cite{Ullah_2022, Perri_2022}. 
Finally, efficient approaches suitable to approximate divisions are demonstrated in \cite{Saadat_2019, Imani_2019, Zendegani_2016} that either exploit approximate subtractors \cite{Chen_2015, Chen_2016} or implement the signal segmentation technique \cite{Hashemi_2016}.

%\noindent\textbf{Critical discussion and comparison.}
Table~\ref{tab:approx_levels} analyzes approximate computing paradigms from the perspective of the design abstraction level,
%While Table~\ref{tab:approx_techniques_comparison} focuses on individual techniques, Table~\ref{tab:approx_levels} 
clarifying where approximation is introduced in the design flow and how this choice affects reusability, accuracy control, and hardware efficiency.
%The main advantage of approximate computing is that it exposes quality as an additional design knob, enabling energy, delay, and area reductions beyond those achievable by exact low-power design alone.
%However, the achievable gain depends strongly on where approximation is introduced.
Algorithm-level approximation can produce large savings when it replaces expensive functions such as exponentiation, division, pooling, or normalization, but it is often domain-specific and must be validated at model level.
Architecture- and circuit-level approximation provides more direct control over hardware metrics, but it requires accurate error models because local arithmetic errors may be amplified by sensitive layers or accumulated across long inference pipelines.
Runtime-adaptive approximation is particularly attractive for heterogeneous and edge platforms, but the flexibility obtained through configurable accuracy modes comes at the cost of additional control logic, calibration, and verification effort.
%This view highlights that approximation is not a single design technique, but a family of strategies ranging from mathematical reformulation of DL operators to circuit-level simplification of arithmetic blocks.
High-level approximations are usually easier to relate to model accuracy and can simplify complete layers, but they are less reusable across applications. Low-level approximations, instead, are more directly connected to area, delay, and power reductions, but their effect on the final inference quality is harder to predict without workload-aware error propagation analysis. FPGA-oriented and runtime-adaptive methods further emphasize that the best approximation strategy depends not only on the workload, but also on the target technology and on whether the accelerator must support fixed or variable quality constraints.

\begin{table}[t]
\caption{Approximate-computing design levels and their relevance to DL accelerators.}
\label{tab:approx_levels}
\footnotesize
\resizebox{1\columnwidth}{!}{
\begin{tabular}{p{2.3cm} p{2.8cm} p{3.1cm} p{3.0cm} p{2.5cm}}
\toprule
\textbf{Approximation level} & \textbf{Examples} & \textbf{Advantages} & \textbf{Challenges} & \textbf{Most suitable target} \\
\midrule
Algorithmic & SoftMax, pooling, denoising, super-resolution & Large complexity reduction at operator level & Requires end-to-end accuracy validation & Layer-specific optimization \\
Architecture & Approximate MACs, segmented multipliers, configurable datapaths & Direct energy--delay trade-offs & Must account for layer sensitivity and data distribution & Custom DL accelerators \\
Gate/transistor & Approximate adders, compressors, voltage overscaling & Fine-grained area and power reduction & Technology-dependent error behavior & ASIC accelerators \\
FPGA-oriented & LUT-based approximate adders and multipliers & Exploits reconfigurable primitives & ASIC-oriented circuits may not map efficiently & FPGA accelerators \\
Runtime-adaptive & Dynamic approximation modes & Input- and layer-aware efficiency & Control and verification overhead & Energy-constrained edge AI \\
\bottomrule
\end{tabular}
}
\end{table}

%\noindent\textbf{Approximate computing for transformer-based workloads.}
Transformer-based workloads introduce new opportunities and challenges for approximate computing.
%Unlike CNNs, where convolutions dominate the computational cost, transformer models rely heavily on matrix multiplications, 
Multi-head self-attention, SoftMax, LayerNorm, and feed-forward networks exhibit different sensitivity to approximation: matrix multiplications can benefit from approximate multipliers and low-precision arithmetic, whereas SoftMax and LayerNorm require careful bounded-error approximation because numerical errors may affect attention scores and activation distributions.
%Recent works confirm this shift in the state of the art.
TransAxx \cite{TransAxx2024} investigates approximate arithmetic for Vision Transformers, including approximate multipliers, approximation-aware fine-tuning, and design-space exploration for approximate ViT accelerators. Runtime-adaptive FPGA accelerators for transformers further show that quantized and configurable datapaths are important when transformer models differ in size, tiling structure, and latency requirements \cite{Kabir2024ADAPTOR}. In addition, recent hardware-oriented SoftMax and LayerNorm designs exploit approximation to reduce the cost of nonlinear transformer operators, either through shared fixed-point-friendly architectures \cite{Kim2025SoftmaxLayerNorm}, low-precision SoftMax approximations for small FPGAs \cite{LowPrecisionSoftmax2026}, or block-aware precision rescaling in attention \cite{BAPS2026}.
These results suggest that approximate computing for transformers should be applied in a layer-aware and operator-aware manner. 
%Attention-score computation and feed-forward layers may tolerate approximate arithmetic under proper retraining or fine-tuning, whereas normalization and SoftMax often require tighter error bounds to avoid unstable behavior. 
Consequently, transformer-oriented approximate computing requires a systematic design-space exploration that jointly considers bit-width, arithmetic approximation, memory traffic, nonlinear-function approximation, and accuracy-recovery strategies.
\section{Emerging trends and research directions} \label{sec:open}

While this survey presented the application of design methodologies to deep learning workloads, one emerging trend in this field is the application of machine learning and deep learning techniques in the optimization of design tasks themselves~\cite{ML4EDA21,RenHu23}. However, the current understanding is that we still need to find the right combination of existing heuristic solutions and ML-based approaches.
The ability of DL-based techniques to extract meaningful knowledge from large amounts of data could be of help in driving the design automation process, in particular to improve predictions on the effects of applied optimizations, which are a fundamental component in many design automation tasks.
Moreover, machine learning can support the scalability requirements of modern design automation tools~\cite{Kahng23}. 
A challenge for the convergence of DL and electronic design automation (EDA) lies in the explainability of machine learning models, as the relationship between their outputs and the input design objectives and constraints is not always straightforward~\cite{SALEEM22}.
Preliminary results have also been achieved by applying Large Language Models to design automation tasks \cite{xu2024llmaidedefficienthardwaredesign,llmHLS25}, and we expect an increase in research activities on this subject.

Secondly, the carbon footprint of computing systems cannot be ignored, and it will have to be taken into consideration when designing the next generation of accelerators and systems \cite{sustainableHPC,wu2022sustainableai}.
Future EDA tools will therefore need to integrate methodologies to estimate the environmental impact of new designs \cite{act} and work under stricter energy efficiency constraints.
Existing \textbf{partitioning and mapping} strategies for distributed DL training and inference typically optimize performance objectives such as throughput, scalability, communication efficiency, and model accuracy.
However, there is a significant opportunity for energy savings, for example by improving utilization through GPU sharing~\cite{li2023green}.
Recent research has highlighted the importance of energy- and quantization-aware partitioning strategies for deploying DNNs on resource-constrained devices~\cite{nicosanti26}.
Future DL partitioning and mapping strategies should therefore be optimized for multiple non-functional requirements, including energy consumption, monetary cost, and sustainability constraints. 
Other promising research directions include automated, policy-driven model orchestration that can adapt deployment decisions to changing workload and hardware conditions. 
Existing orchestration systems such as INFaaS~\cite{romero21infaas} and GreenServ~\cite{ziller26greenserv} demonstrate the potential of automated decision-making for heterogeneous inference systems. 
Building on these efforts, future orchestration frameworks may incorporate more autonomous optimization mechanisms, enabling intelligent agents to jointly optimize model selection, partitioning, and placement decisions with minimal manual intervention under highly dynamic operating conditions.

%As DL applications increasingly span the edge-cloud continuum, mapping strategies must support runtime adaptation by re-partitioning workloads according to changes in network conditions, resource availability, and energy-aware constraints such as power availability and grid carbon intensity. 
%Furthermore, closer hardware-software co-design is required to enable partitioning frameworks to exploit interconnect topology and hardware telemetry, including thermal and power information, when making mapping decisions. 
%Finally, as model architectures evolve beyond transformers toward alternative sequence modeling paradigms such as State Space Models~\cite{gu24mamba}, existing parallelization and mapping techniques will need to be revisited to accommodate their distinct memory access patterns and communication characteristics.

% mau
Another emerging trend is the evolution of \textbf{design space exploration} methodologies, with recent efforts suggesting a transition from traditional analytical DSE frameworks toward AI-assisted exploration and architectural co-optimization approaches. At the same time, the growing adoption of heterogeneous, chiplet-based, and 3D-integrated systems is expanding the complexity of the design space that must be explored. Future DSE frameworks will therefore need to combine accurate modeling capabilities with intelligent exploration strategies capable of scaling to increasingly diverse accelerator architectures and workloads.

The evolution of deep learning has been driven by a tight co-design loop between hardware and software: new hardware architectures expose compute that demands new programming abstractions, and new models expose workloads that demand new hardware to be executed efficiently.
\textbf{DL compilers} sit in between:
%, and their central open problem is keeping pace with both fronts at once.
on the hardware front, they must continuously adapt to emerging devices to exploit the compute power that new platforms expose,
%This requires representing not only computation but also placement, dataflow, memory hierarchy, and arithmetic precision.
on the software front, they must support new models trying to avoid a full manual rewrite of schedules and intrinsic mappings.
In the data center, compiler frameworks have evolved to address these challenges
%Graph-level compilers optimize whole-model representations through fusion, layout propagation, and algebraic simplification; tensor-program compilers expose schedules separately from computation and search them with autotuning and cost models; tile-level DSLs raise the unit of programming from scalar threads to tiles of tensors, partitioning work to maximize data reuse and map it onto hardware accelerators leaving mapping to the compiler; dataflow and megakernel approaches restructure execution itself to handle the synchronization overheads and data-dependent shapes that transformers introduce; and, most recently, LLM- and search-guided agents have begun to automate the schedule search where expert-written schedules and cost models do not yet exist or are too costly to collect. Together, these form an increasingly automated stack, and 
by converging on shared MLIR-based substrates that enable techniques at multiple levels of abstraction to compose.
However, to fully exploit the lowering mechanism in MLIR would require maintaining both the current separation of concerns and a global view: a structure that decides when to switch between levels of abstraction and lets each level inform the others, rather than fixing each lowering in isolation. A possible path is through MLIR's transform dialect, which treats transformations as first-class IR. A cross-level optimizer is missing that would retain a global cost view while still respecting the modular boundaries that keep the infrastructure tractable. Reconciling those two demands, global optimality and modular separation, is the central unsolved problem that needs to be addressed in order to unlock new optimization opportunities.
While most edge compilation frameworks do not generate their own compute kernels but rather orchestrate data movement around hand-written, target-optimized kernels, 
%so end-to-end efficiency rests on code provided to the compiler.
%Autotuning is largely manual or absent, and, to the best of the authors' knowledge, no effort has been made to apply LLMs directly to edge-device compilation pipelines.
the core research challenge for edge compilers is the same as on the HPC front: to portably represent software and hardware, and to automatically and modularly optimize the mapping between them as each side evolves, extending these advances across the full range of targets while minimizing the burden on the programmer.
Transformer workloads introduce further challenges, as their primary kernel, softmax attention, 
%requires each output to depend on a reduction over all other elements in the input sequence, thus breaking locality; in its naive form, the full L $\times$ L attention matrix must be materialized, making the operator 
is memory- rather than compute-bound~\cite{flashattention}.
Furthermore, transformer models used in generative AI exhibit two distinct operating modes with different characteristics: a compute-bound prefill,
%which consumes the input sequence in parallel as a dense general matrix multiplication (GEMM), 
and a bandwidth-bound decode.
%which degenerates to general matrix vector (GEMV) as it generates one token at a time, reading every weight and streaming a growing KV cache on each step.
For larger models, the introduction of mixture-of-experts routing makes not only the tensor shapes but also the computational graph itself input-dependent.
%In HPC, the problem is serving both regimes with a single model;
% balancing compute-bound prefill with bandwidth-bound decode.
%at the edge, transformers also complicate memory planning.
%: variable-length state, the KV cache, activations, and weights must coexist in very small on-chip memory, shifting the task from static deployment to a dynamic execution whose cost scales with sequence length.
%For IMC/PIM, a conflict arises between static weight placement and data-dependent intermediates: projection matrices suit weight-stationary crossbars, but attention produces intermediates whose size grows with sequence length, and a KV cache that expands at every step, neither of which fits a fixed crossbar allocation, forcing capacity-aware remapping and hybrid in-memory/digital execution.
% IMC compilers

For \textbf{IMC accelerators}, several open challenges remain critical for the architecture and compiler community \cite{Mambu22}. There is a lack of a unified intermediate representation that can seamlessly target both digital IMC (SRAM-based boolean logic) and analog IMC (emerging non-volatile memory crossbars), calling therefore for a unified intermediate representation standard. Current compilers rely almost entirely on static, compile-time mapping. Developing lightweight runtime systems capable of handling dynamic memory allocation and fault-tolerant remapping when memory cells degrade is crucial. Finally, while deep learning workloads have robust compiler support, compiling general-purpose, non-deterministic programs (with complex control flow and pointer chasing) into IMC arrays remains an unsolved frontier.

\textbf{HLS-based} design tools have successfully bridged the abstraction gap between DL frameworks and hardware acceleration, making FPGAs and ASICs accessible to non-expert users.
Despite these advantages, several challenges remain to be addressed.
As DL models evolve and present more complex computational patterns, the static scheduling and fine-grained parallelism exploited by most HLS tools will cease to be effective; research into dataflow/streaming execution models and dynamic scheduling will then be fundamental to accommodate the synthesis of irregular and data-dependent applications such as GNNs and transformers.
Moreover, the increasing size of DL models in terms of number of layers and parameters often makes it impossible to synthesize a single accelerator kernel for the entire application, as it would deplete the available hardware resources and drastically reduce the operating frequency due to routing congestion.
Methodologies that generate smaller, reusable kernels will be required, as well as techniques to improve the management of external memory accesses, and tools that facilitate the integration of HLS-generated accelerators into a complete HPC system.
Finally, rather than using existing commercial tools as ``black boxes", future HLS research must embrace open-source development to foster broader adoption, reproducibility, and innovation.

Despite the energy and area savings offered by \textbf{approximate computing} approaches, their application remains challenging, as there is a lack of design tools and formal models for error analysis and energy/accuracy/delay trade-offs.
Future research directions in this field include combining approximate computing with the ability to switch between high-precision and low-precision operating modes, enabling approximate circuits to tune the energy consumption to the specific precision demands of different computational layers.
%To maximize the benefits of approximate computing in heterogeneous embedded platforms, a cross-layer co-design approach is desirable to introduce approximations at both hardware and software levels. 
%At a lower level of abstraction, 
One critical concern is ensuring data reliability and integrity: exploring the impact of process, voltage, and temperature variations, as well as transistor aging, on the errors of approximate computations is crucial. 
Finally, the potential of approximate computing extends beyond AI, HPC, and signal processing applications; it is reasonable to expect that it will gain popularity in any application context with stringent constraints on resource utilization, energy efficiency, and computational delays.

\section{Conclusion} \label{sec:end}

The design of deep learning accelerators for heterogeneous architectures requires a wide stack of methodologies and electronic design automation tools to assist the developer at different design, simulation, and verification stages.
Recent advancements in EDA tools enabled handling the increased complexity of heterogeneous architectures (which may include CPUs, GPUs, FPGAs, and many other specialized components), improving design productivity while achieving the desired speedup and energy efficiency.
In this paper, we aimed to provide a survey of design flows and frameworks developed in recent years to manage the complexity of designing DL accelerators at different levels of abstraction.
In particular, the paper discussed hardware-software partitioning tools that decompose a DL model and choose the best architecture for each component, methodologies for modeling, profiling, and exploring different architectures, domain-specific compilers, hardware generation through High-Level Synthesis, and techniques to design processing elements based on approximate computing.
We expect the themes we highlighted as emerging trends to remain prominent research directions, while recognizing that the rapid evolution of the field will continue to reshape both priorities and possible solutions.

\begin{acks}
This work has been partially supported by the Spoke 1 “FutureHPC \& BigData” of the Italian Research Center on High-Performance Computing, Big Data and Quantum Computing (ICSC), funded by MUR Missione 4 - Next Generation EU (NGEU).
\end{acks}

%% Elimina il commento alla riga successiva se vuoi produrre 
%% il documento senza DOI
\newcommand{\noDOI}{}

\ifdefined\noDOI
    \bibliographystyle{acm} % Without DOI
\else
    \bibliographystyle{ACM-Reference-Format} % With DOI
\fi

\bibliography{biblio}

@inproceedings{Alioto_2017,
	title        = {{Energy-quality scalable adaptive VLSI circuits and systems beyond approximate computing}},
	author       = {Alioto, Massimo},
	year         = 2017,
	booktitle    = {Design, Automation \& Test in Europe Conference \& Exhibition},
    series = {DATE '17},
    publisher    = {{IEEE}},
	pages        = {127-132}
}

@article{Rokh_2023,
	title        = {{A Comprehensive Survey on Model Quantization for Deep Neural Networks in Image Classification}},
	author       = {Babak, Rokh and Ali, Azarpeyvand and Alireza, Khanteymoori },
	year         = 2023,
	month        = {06},
	journal      = {ACM Transactions on Intelligent Systems and Technology},
	volume       = 14,
	pages        = {1--50}
}

@article{Jiang_2020,
	title        = {{Approximate Arithmetic Circuits: A Survey, Characterization, and Recent Applications}},
	author       = {Jiang, Honglan and Santiago, Francisco Javier Hernandez and Mo, Hai and Liu, Leibo and Han, Jie},
	year         = 2020,
	journal      = {Proceedings of the IEEE},
	volume       = 108,
	number       = 12,
	pages        = {2108--2135}
}

@article{Zhu_2020,
	title        = {{Efficient Precision-Adjustable Architecture for Softmax Function in Deep Learning}},
	author       = {Danyang Zhu and Siyuan Lu and Meiqi Wang and Jun Lin and Zhongfeng Wang},
	year         = 2020,
	journal      = {IEEE Transactions on Circuits and Systems II: Express Briefs},
	volume       = 67,
	pages        = {3382--3386}
}

@misc{AMD,
	title        = {{7 Series FPGAs Configurable Logic Block User Guide, UG474}},
	author       = {Xilinx/AMD},
	year         = 2020,
	note         = {Accessed: 2026-07-05},
	howpublished = {\url{https://www.xilinx.com/support/documentation/user_guides/ug474_7Series_CLB.pdf}}
}

@article{Spagnolo_2023,
	title        = {{Design of a Low-Power Super-Resolution Architecture for Virtual Reality Wearable Devices}},
	author       = {Spagnolo, Fanny and Corsonello, Pasquale and Frustaci, Fabio and Perri, Stefania},
	year         = 2023,
	journal      = {IEEE Sensors Journal},
	volume       = 23,
	number       = 8,
	pages        = {9009--9016}
}

@article{Cardarilli_2021,
	title        = {{A pseudo-softmax function for hardware-based high speed image classification}},
	author       = {Cardarilli, Gian Carlo and Di Nunzio, Luca and Fazzolari, Rocco and Giardino, Daniele and Nannarelli, Alberto and Re, Marco and Span{\`o}, Sergio},
	year         = {2021},
	day          = {28},
	journal      = {Scientific Reports},
	volume       = {11},
	number       = {1},
	pages        = {15307}
}

@article{Spagnolo_2022,
	title        = {{Aggressive Approximation of the SoftMax Function for Power-Efficient Hardware Implementations}},
	author       = {Spagnolo, Fanny and Perri, Stefania and Corsonello, Pasquale},
	year         = 2022,
	journal      = {IEEE Transactions on Circuits and Systems II: Express Briefs},
	volume       = 69,
	number       = 3,
	pages        = {1652--1656}
}

@article{Spagnolo_2023_2,
	title        = {{Design of Approximate Bilateral Filters for Image Denoising on FPGAs}},
	author       = {Spagnolo, Fanny and Corsonello, Pasquale and Frustaci, Fabio and Perri, Stefania},
	year         = 2023,
	journal      = {IEEE Access},
	volume       = 11,
	number       = {},
	pages        = {1990--2000}
}

@inproceedings{Ghosh_2020,
	title        = {{Approximate Inference Systems (AxIS): End-to-End Approximations for Energy-Efficient Inference at the Edge}},
    booktitle = {ACM/IEEE International Symposium on Low Power Electronics and Design},
    series = {ISLPED '20},    
	author       = {Ghosh, Soumendu and Raha, Arnab and Raghunathan, Vijay},
	year         = 2020,
	pages        = {7--12}
}

@article{Sayal_2021,
	title        = {{COMPAC: Compressed Time-Domain, Pooling-Aware Convolution CNN Engine With Reduced Data Movement for Energy-Efficient AI Computing}},
	author       = {Sayal, Aseem and Fathima, Shirin and Nibhanupudi, SS and Kulkarni, Jaydeep},
	year         = 2021,
	journal      = {IEEE Journal of Solid-State Circuits},
	volume       = {56},
    number = {7},
	pages        = {2205--2220}
}

@article{Spagnolo_2022_2,
	title        = {{Approximate Down-Sampling Strategy for Power-Constrained Intelligent Systems}},
	author       = {Spagnolo, Fanny and Perri, Stefania and Corsonello, Pasquale},
	year         = 2022,
	journal      = {IEEE Access},
	volume       = 10,
	number       = {},
	pages        = {7073--7081}
}

@inproceedings{Esposito_2017,
	title        = {{Low-power approximate MAC unit}},
	author       = {Esposito, Darjn and Strollo, Antonio G. M. and Alioto, Massimo},
	year         = 2017,
	booktitle    = {2017 13th Conference on Ph.D. Research in Microelectronics and Electronics (PRIME)},
	volume       = {},
	number       = {},
	pages        = {81--84}
}

@article{Esposito_2018,
	title        = {{Approximate Multipliers Based on New Approximate Compressors}},
	author       = {Esposito, Darjn and Strollo, Antonio Giuseppe Maria and Napoli, Ettore and De Caro, Davide and Petra, Nicola},
	year         = 2018,
	journal      = {IEEE Transactions on Circuits and Systems I: Regular Papers},
	volume       = 65,
	number       = 12,
	pages        = {4169--4182}
}

@article{Frustaci_2019,
	title        = {{Energy-Quality Scalable Adders Based on Nonzeroing Bit Truncation}},
	author       = {Frustaci, Fabio and Perri, Stefania and Corsonello, Pasquale and Alioto, Massimo},
    journal={IEEE Transactions on Very Large Scale Integration (VLSI) Systems}, 
    year={2019},
    volume={27},
    number={4},
    pages={964-968},
    doi={10.1109/TVLSI.2018.2881326}
}

@article{Frustaci_2020,
	title        = {{Approximate Multipliers With Dynamic Truncation for Energy Reduction via Graceful Quality Degradation}},
	author       = {Frustaci, Fabio and Perri, Stefania and Corsonello, Pasquale and Alioto, Massimo},
	journal      = {IEEE Transactions on Circuits and Systems II: Express Briefs},
    year={2020},
    volume={67},
    number={12},
    pages={3427--3431},
}

@article{Strollo_2020,
	title        = {{Comparison and Extension of Approximate 4-2 Compressors for Low-Power Approximate Multipliers}},
	author       = {Strollo, Antonio Giuseppe Maria and Napoli, Ettore and De Caro, Davide and Petra, Nicola and Meo, Gennaro Di},
	year         = 2020,
	journal      = {IEEE Transactions on Circuits and Systems I: Regular Papers},
	volume       = 67,
	number       = 9,
	pages        = {3021--3034}
}

@article{Kim_2021,
	title        = {{The Effects of Approximate Multiplication on Convolutional Neural Networks}},
	author       = {Kim, Min Soo and Del Barrio, Alberto A. and Kim, HyunJin and Bagherzadeh, Nader},
	year         = 2022,
	journal      = {IEEE Transactions on Emerging Topics in Computing},
	volume       = 10,
	number       = 2,
	pages        = {904--916}
}

@article{Strollo_2022,
	title        = {{Approximate Multipliers Using Static Segmentation: Error Analysis and Improvements}},
	author       = {Strollo, Antonio Giuseppe Maria and Napoli, Ettore and De Caro, Davide and Petra, Nicola and Saggese, Gerardo and Di Meo, Gennaro},
	year         = 2022,
	journal      = {IEEE Transactions on Circuits and Systems I: Regular Papers},
	volume       = 69,
	number       = 6,
	pages        = {2449--2462}
}

@article{Perri_2020,
	title        = {{Efficient Approximate Adders for FPGA-Based Data-Paths}},
	author       = {Perri, Stefania and Spagnolo, Fanny and Frustaci, Fabio and Corsonello, Pasquale},
	year         = 2020,
	journal      = {Electronics},
	volume       = 9,
	number       = 9
}

@inproceedings{Prabakaran_2018,
	title        = {{DeMAS: An efficient design methodology for building approximate adders for FPGA-based systems}},
	author       = {Prabakaran, Bharath Srinivas and Rehman, Semeen and Hanif, Muhammad Abdullah and Ullah, Salim and Mazaheri, Ghazal and Kumar, Akash and Shafique, Muhammad},
	year         = {2018},
	booktitle    = {2018 Design, Automation \& Test in Europe Conference \& Exhibition},
    series = {DATE '18},
	pages        = {917--920}
}

@article{Ahmad_2021,
	title        = {{Low Error Efficient Approximate Adders for FPGAs}},
	author       = {Ahmad, Waqar and Ayrancioglu, Berke and Hamzaoglu, Ilker},
	year         = 2021,
	journal      = {IEEE Access},
	volume       = 9,
	number       = {},
	pages        = {117232--117243}
}

@article{Waris_2021,
	title        = {{AxBMs: Approximate Radix-8 Booth Multipliers for High-Performance FPGA-Based Accelerators}},
	author       = {Waris, Haroon and Wang, Chenghua and Liu, Weiqiang and Lombardi, Fabrizio},
	year         = 2021,
	journal      = {IEEE Transactions on Circuits and Systems II: Express Briefs},
	volume       = 68,
	number       = 5,
	pages        = {1566--1570}
}

@article{Ullah_2021,
	title        = {{Area-Optimized Accurate and Approximate Softcore Signed Multiplier Architectures}},
	author       = {Ullah, Salim and Schmidl, Hendrik and Sahoo, Siva Satyendra and Rehman, Semeen and Kumar, Akash},
	year         = 2021,
	journal      = {IEEE Transactions on Computers},
	volume       = 70,
	number       = 3,
	pages        = {384--392}
}

@article{Ullah_2022,
	title        = {{High-Performance Accurate and Approximate Multipliers for FPGA-Based Hardware Accelerators}},
	author       = {Ullah, Salim and Rehman, Semeen and Shafique, Muhammad and Kumar, Akash},
	year         = 2022,
	journal      = {IEEE Transactions on Computer-Aided Design of Integrated Circuits and Systems},
	volume       = 41,
	number       = 2,
	pages        = {211--224}
}

@article{Perri_2022,
	title        = {{Designing Energy-Efficient Approximate Multipliers}},
	author       = {Perri, Stefania and Spagnolo, Fanny and Frustaci, Fabio and Corsonello, Pasquale},
	year         = 2022,
	journal      = {Journal of Low Power Electronics and Applications},
	volume       = 12,
	number       = 4
}

@article{Cho_2004,
	title        = {{Design of low-error fixed-width modified booth multiplier}},
	author       = {Kyung-Ju Cho and Kwang-Chul Lee and Jin-Gyun Chung and Parhi, K.K.},
	year         = 2004,
	journal      = {IEEE Transactions on Very Large Scale Integration (VLSI) Systems},
	volume       = 12,
	number       = 5,
	pages        = {522--531}
}

@article{Song_2007,
	title        = {{Adaptive Low-Error Fixed-Width Booth Multipliers}},
	author       = {Song, Min-An and Van, Lan-Da and Kuo, Sy-Yen},
	year         = 2007,
	journal      = {IEICE Transactions on Fundamentals of Electronics, Communications and Computer Sciences},
	volume       = {E90-A},
	number        = {6},
	pages        = {}
}

@inproceedings{Farshchi_2013,
	title        = {{New Approximate Multiplier for Low Power Digital Signal Processing}},
	author       = {Farshchi, Farzad and Abrishami, Muhammad Saeed and Fakhraie, Sied Mehdi},
	year         = 2013,
	booktitle    = {The 17th CSI International Symposium on Computer Architecture \& Digital Systems},
	series       = {CADS '13},
	pages        = {25--30}
}

@article{Chen_2012,
	title        = {{A High-Accuracy Adaptive Conditional-Probability Estimator for Fixed-Width Booth Multipliers}},
	author       = {Chen, Yuan-Ho and Chang, Tsin-Yuan},
	year         = 2012,
	journal      = {IEEE Transactions on Circuits and Systems I: Regular Papers},
	volume       = 59,
	number       = 3,
	pages        = {594--603}
}

@article{Akbari_2017,
	title        = {{Dual-Quality 4:2 Compressors for Utilizing in Dynamic Accuracy Configurable Multipliers}},
	author       = {Akbari, Omid and Kamal, Mehdi and Afzali-Kusha, Ali and Pedram, Massoud},
	year         = 2017,
	journal      = {IEEE Transactions on Very Large Scale Integration (VLSI) Systems},
	volume       = 25,
	number       = 4,
	pages        = {1352--1361}
}

@inproceedings{Hashemi_2015,
	title        = {{DRUM: A Dynamic Range Unbiased Multiplier for approximate applications}},
	author       = {Hashemi, Soheil and Bahar, R. Iris and Reda, Sherief},
	year         = 2015,
	booktitle    = {2015 IEEE/ACM International Conference on Computer-Aided Design},
	series       = {ICCAD '15},
	pages        = {418--425}
}

@article{Narayanamoorthy_2015,
	title        = {{Energy-Efficient Approximate Multiplication for Digital Signal Processing and Classification Applications}},
	author       = {Narayanamoorthy, Srinivasan and Moghaddam, Hadi Asghari and Liu, Zhenhong and Park, Taejoon and Kim, Nam Sung},
	year         = 2015,
	journal      = {IEEE Transactions on Very Large Scale Integration (VLSI) Systems},
	volume       = 23,
	number       = 6,
	pages        = {1180--1184}
}

@article{Di_Meo_2023,
	title        = {{Design of Generalized Enhanced Static Segment Multiplier with Minimum Mean Square Error for Uniform and Nonuniform Input Distributions}},
	author       = {Di Meo, Gennaro and Saggese, Gerardo and Strollo, Antonio G. M. and De Caro, Davide},
	year         = 2023,
	journal      = {Electronics},
	volume       = 12,
	number       = 2
}

@inproceedings{Chen_2015,
	title        = {{Design of Approximate Unsigned Integer Non-Restoring Divider for Inexact Computing}},
	author       = {Chen, Linbin and Han, Jie and Liu, Weiqiang and Lombardi, Fabrizio},
	year         = 2015,
	booktitle    = {25th Edition on Great Lakes Symposium on VLSI},
	series       = {GLSVLSI '15},
	pages        = {51--56},
	numpages     = 6,
}

@article{Chen_2016,
	title        = {{On the Design of Approximate Restoring Dividers for Error-Tolerant Applications}},
	author       = {Chen, Linbin and Han, Jie and Liu, Weiqiang and Lombardi, Fabrizio},
	year         = 2016,
	journal      = {IEEE Transactions on Computers},
	volume       = 65,
	number       = 8,
	pages        = {2522--2533}
}

@inproceedings{Hashemi_2016,
	title        = {{A Low-Power Dynamic Divider for Approximate Applications}},
	author       = {Hashemi, Soheil and Bahar, R. Iris and Reda, Sherief},
	year         = 2016,
	booktitle    = {2016 53nd ACM/EDAC/IEEE Design Automation Conference},
	series       = {DAC '16},
	number       = {},
	pages        = {1--6}
}

@inproceedings{Saadat_2019,
	title        = {{Approximate Integer and Floating-Point Dividers with Near-Zero Error Bias}},
    booktitle = {2019 56th ACM/IEEE Design Automation Conference},
    series = {DAC '19},
	author       = {Saadat, Hassaan and Javaid, Haris and Parameswaran, Sri},
	year         = 2019,
	pages        = {1--6}
}

@inproceedings{Imani_2019,
	title        = {{CADE: Configurable Approximate Divider for Energy Efficiency}},
	author       = {Imani, Mohsen and Garcia, Ricardo and Huang, Andrew and Rosing, Tajana},
	year         = 2019,
	booktitle    = {2019 Design, Automation \& Test in Europe Conference \& Exhibition},
	series       = {DATE '19},
	pages        = {586--589}
}

@inproceedings{Zendegani_2016,
	title        = {{SEERAD: A High Speed yet Energy-Efficient Rounding-Based Approximate Divider}},
	author       = {Zendegani, Reza and Kamal, Mehdi and Fayyazi, Arash and Afzali-Kusha, Ali and Safari, Saeed and Pedram, Massoud},
	year         = 2016,
	booktitle    = {2016 Design, Automation \& Test in Europe Conference \& Exhibition},
    series = {DATE '16},
	volume       = {},
	number       = {},
	pages        = {1481--1484}
}

@article{duarte2018fast,
	title        = {{Fast inference of deep neural networks in FPGAs for particle physics}},
	author       = {Duarte, Javier and Han, Song and Harris, Philip and Jindariani, Sergo and Kreinar, Edward and Kreis, Benjamin and others},
	year         = 2018,
	journal      = {Journal of Instrumentation},
	publisher    = {IOP Publishing},
	volume       = 13,
	number       = {07},
	pages        = {P07027}
}

@article{blott2018finn,
	title        = {{FINN-R: An end-to-end deep-learning framework for fast exploration of quantized neural networks}},
	author       = {Blott, Michaela and Preu{\ss}er, Thomas B and Fraser, Nicholas J and Gambardella, Giulio and O'brien, Kenneth and Umuroglu, Yaman and others},
	year         = 2018,
	journal      = {ACM Transactions on Reconfigurable Technology and Systems},
	publisher    = {ACM New York, NY, USA},
	volume       = 11,
	number       = 3,
	pages        = {1--23}
}

@inproceedings{lattner2021mlir,
	title        = {{MLIR: Scaling Compiler Infrastructure for Domain Specific Computation}},
	author       = {Lattner, Chris and Amini, Mehdi and Bondhugula, Uday and Cohen, Albert and Davis, Andy and Pienaar, Jacques and others},
	year         = 2021,
	booktitle    = {IEEE/ACM International Symposium on Code Generation and Optimization},
    series = {CGO '21},
	pages        = {2--14}
}

@INPROCEEDINGS{gnnhls,
  author={Zhao, Chenfeng and Dong, Zehao and Chen, Yixin and Zhang, Xuan and Chamberlain, Roger D.},
  booktitle={2023 IEEE 41st International Conference on Computer Design}, 
  series = {ICCD '23},
  title={{GNNHLS: Evaluating Graph Neural Network Inference via High-Level Synthesis}}, 
  year={2023},
  pages={574--577},
}

@INPROCEEDINGS{gcnhls,
  author={Lin, Yi Chien and Zhang, Bingyi and Prasanna, Viktor},
  booktitle={2021 IEEE High Performance Extreme Computing Conference},
  series = {HPEC '21},
  title={{GCN Inference Acceleration using High-Level Synthesis}}, 
  year={2021},
  pages={1--6}
}

@inproceedings{dynamatic,
  title={Dynamically scheduled high-level synthesis},
  author={Josipovi{\'c}, Lana and Ghosal, Radhika and Ienne, Paolo},
  booktitle={2018 ACM/SIGDA International Symposium on Field-Programmable Gate Arrays},
  pages={127--136},
  year={2018}
}

@inproceedings{dass,
  title={Combining dynamic \& static scheduling in high-level synthesis},
  author={Cheng, Jianyi and Josipovic, Lana and Constantinides, George A and Ienne, Paolo and Wickerson, John},
  booktitle={2020 ACM/SIGDA International Symposium on Field-Programmable Gate Arrays},
  pages={288--298},
  year={2020}
}

@inproceedings{xu2022hector,
  title={{HECTOR: A Multi-Level Intermediate Representation for Hardware Synthesis Methodologies}},
  author={Xu, Ruifan and Xiao, Youwei and Luo, Jin and Liang, Yun},
  booktitle={41st IEEE/ACM International Conference on Computer-Aided Design},
  series = {ICCAD '22},
  pages={1--9},
  year={2022}
}

@article{chen2024allo,
  title={{Allo: A Programming Model for Composable Accelerator Design}},
  author={Chen, Hongzheng and Zhang, Niansong and Xiang, Shaojie and Zeng, Zhichen and Dai, Mengjia and Zhang, Zhiru},
  journal={Proceedings of the ACM on Programming Languages},
  volume={8},
  number={PLDI},
  pages={593--620},
  year={2024},
  publisher={ACM New York, NY, USA}
}

@inproceedings{streamHLS,
author = {Basalama, Suhail and Cong, Jason},
title = {Stream-HLS: Towards Automatic Dataflow Acceleration},
year = {2025},
booktitle = {2025 ACM/SIGDA International Symposium on Field Programmable Gate Arrays},
pages = {103--114},
numpages = {12},
location = {Monterey, CA, USA},
series = {FPGA '25}
}

@article{chen2024understanding,
  title={Understanding the potential of fpga-based spatial acceleration for large language model inference},
  author={Chen, Hongzheng and Zhang, Jiahao and Du, Yixiao and Xiang, Shaojie and Yue, Zichao and Zhang, Niansong and Cai, Yaohui and Zhang, Zhiru},
  journal={ACM Transactions on Reconfigurable Technology and Systems},
  volume={18},
  number={1},
  pages={1--29},
  year={2024},
  publisher={ACM New York, NY}
}

@article{jiang2025low,
  title={{Low Latency Transformer Inference on FPGAs for Physics Applications with hls4ml}},
  author={Jiang, Zhixing and Yin, Dennis and Chen, Yihui and Khoda, Elham E and Hauck, Scott and Hsu, Shih-Chieh and others},
  journal={Journal of Instrumentation},
  volume={20},
  number={4},
  pages={P04014},
  year={2025},
  publisher={IOP Publishing}
}

@inproceedings{aries,
author = {Zhuang, Jinming and Xiang, Shaojie and Chen, Hongzheng and Zhang, Niansong and Yang, Zhuoping and Mao, Tony and Zhang, Zhiru and Zhou, Peipei},
title = {{ARIES: An Agile MLIR-Based Compilation Flow for Reconfigurable Devices with AI Engines}},
year = {2025},
doi = {10.1145/3706628.3708870},
booktitle = {2025 ACM/SIGDA International Symposium on Field Programmable Gate Arrays},
pages = {92--102},
numpages = {11},
series = {FPGA '25}
}

@INPROCEEDINGS{finnT,
  author={Berganski, Christoph and Jentzsch, Felix and Platzner, Marco and Kuhmichel, Max and Giefers, Heiner},
  booktitle={2024 International Conference on Field Programmable Technology}, 
  series = {ICFPT '24},
  title={{FINN-T: Compiling Custom Dataflow Accelerators for Quantized Transformers}}, 
  year={2024},
  volume={},
  number={},
  pages={01-10},
  doi={10.1109/ICFPT64416.2024.11113391}
  }

@inproceedings{levental2024bragghls,
  title={{Bragghls: High-Level Synthesis for Low-Latency Deep Neural Networks for Experimental Science}},
  author={Levental, Maksim and Khan, Arham and Chard, Ryan and Yoshii, Kazutomo and Chard, Kyle and Foster, Ian},
  booktitle={14th International Symposium on Highly Efficient Accelerators and Reconfigurable Technologies},
  series = {HEART '24},
  pages={10--17},
  year={2024}
}

@inproceedings{hida,
author = {Ye, Hanchen and Jun, Hyegang and Chen, Deming},
title = {{HIDA: A Hierarchical Dataflow Compiler for High-Level Synthesis}},
year = {2024},
booktitle = {29th ACM International Conference on Architectural Support for Programming Languages and Operating Systems},
volume ={1},
pages = {215--230},
numpages = {16},
location = {La Jolla, CA, USA},
series = {ASPLOS '24}
}

@inproceedings{circt,
  title={{CIRCT: Lifting Hardware Development out of the 20th Century}},
  author={Lenharth, Andrew and Lattner, Chris},
  booktitle={LLVM Developer Meeting},
  year={2021}
}

@misc{circt-hls,
	title        = {{HLS from PyTorch to System Verilog with MLIR and CIRCT}},
	author       = {Urbach, Mike and Petersen, Morten B},
	year         = 2022,
	note         = {2nd Workshop on Languages, Tools, and Techniques for Accelerator Design (LATTE)}
}

@inproceedings{ye2022scalehls,
	title        = {{ScaleHLS: A New Scalable High-Level Synthesis Framework on Multi-Level Intermediate Representation}},
	author       = {Ye, Hanchen and Hao, Cong and Cheng, Jianyi and Jeong, Hyunmin and Huang, Jack and Neuendorffer, Stephen and Chen, Deming},
	year         = 2022,
	booktitle    = {2022 IEEE International Symposium on High-Performance Computer Architecture},
    series = {HPCA '22},
	pages        = {741--755},
	organization = {IEEE}
}

@article{sodaMICRO,
	title        = {{Bridging Python to Silicon: The SODA Toolchain}},
	author       = {Bohm Agostini, Nicolas and Curzel, Serena and Zhang, Jeff Jun and Limaye, Ankur and Tan, Cheng and Amatya, Vinay and others},
	year         = 2022,
	journal      = {IEEE Micro},
	volume       = 42,
	number       = 5,
	pages        = {78--88}
}

@inproceedings{sodaopt,
	title        = {{An MLIR-based Compiler Flow for System-Level Design and Hardware Acceleration}},
	author       = {Bohm Agostini, Nicolas and Curzel, Serena and Amatya, Vinay and Tan, Cheng and Minutoli, Marco and Castellana, Vito Giovanni and others},
	year         = {2022},
	booktitle    = {IEEE/ACM International Conference On Computer Aided Design}, 
    series = {ICCAD '22},
	pages        = {1--9}
}

@inproceedings{david2020tensorflow,
 author = {David, Robert and Duke, Jared and Jain, Advait and Janapa Reddi, Vijay and Jeffries, Nat and others},
 booktitle = {Proceedings of Machine Learning and Systems},
 editor = {A. Smola and A. Dimakis and I. Stoica},
 pages = {800--811},
 title = {TensorFlow Lite Micro: Embedded Machine Learning for TinyML Systems}, volume = {3},
 year = {2021}
}

@incollection{NEURIPS2019_9015,
	title        = {{PyTorch: An Imperative Style, High-Performance Deep Learning Library}},
	author       = {Paszke, Adam and Gross, Sam and Massa, Francisco and Lerer, Adam and Bradbury, James and Chanan, Gregory and others},
	year         = 2019,
	booktitle    = {33rd Conference on Neural Information Processing Systems},
    series = {NeurIPS '19},
	pages        = {8024--8035},
    publisher = {Curran Associates Inc.},
}

@inproceedings{tensorflow2015,
	title        = {{ TensorFlow: Large-Scale Machine Learning on Heterogeneous Systems}},
	author       = {Mart\'{\i}n~Abadi and Ashish~Agarwal and Paul~Barham and Eugene~Brevdo and Zhifeng~Chen and Craig~Citro and others},
    booktitle    = {12th USENIX Symposium on Operating Systems Design and Implementation},
    series = {OSDI '16},
  pages        = {265--283},
  publisher    = {{USENIX} Association},
  year         = {2016},
  url          = {https://www.usenix.org/conference/osdi16/technical-sessions/presentation/abadi},
}

@inproceedings{flamand2018gap,
	title        = {{GAP-8: A RISC-V SoC for AI at the Edge of the IoT}},
	author       = {Flamand, Eric and Rossi, Davide and Conti, Francesco and Loi, Igor and Pullini, Antonio and Rotenberg, Florent and Benini, Luca},
	year         = 2018,
	booktitle    = {2018 IEEE 29th International Conference on Application-specific Systems, Architectures and Processors},
    series = {ASAP '18},
	pages        = {1--4},
	organization = {IEEE}
}

@inproceedings{flashattention,
  author={Dao, Tri and Fu, Daniel Y. and Ermon, Stefano and Rudra, Atri and R{\'e}, Christopher},
  title={{Flash{A}ttention: Fast and Memory-Efficient Exact Attention with {IO}-Awareness}},
  booktitle = {36th International Conference on Neural Information Processing Systems},
  series = {NIPS '22},
  year={2022}
}

@misc{CubeAI,
	title        = {{X-CUBE-AI}},
	author       = {{{ST Microelectronics}}},
	year         = 2017,
	url          = {https://www.st.com/en/embedded-software/x-cube-ai.html},
	urldate      = {2025-03-07}
}

@misc{cutlass,
  author  = {Vijay Thakkar and Pradeep Ramani and Cris Cecka and Aniket Shivam
             and Honghao Lu and Ethan Yan and others},
  title   = {{CUTLASS}},
  howpublished = {\url{https://github.com/NVIDIA/cutlass}},
  version = {3.0.0},
  year    = {2023}
}

@inproceedings{relaxASPLOS2025,
author = {Lai, Ruihang and Shao, Junru and Feng, Siyuan and Lyubomirsky, Steven and Hou, Bohan and Lin, Wuwei and others},
title = {{Relax: Composable Abstractions for End-to-End Dynamic Machine Learning}},
year = {2025},
isbn = {9798400710797},
publisher = {ACM},
doi = {10.1145/3676641.3716249},
booktitle = {30th ACM International Conference on Architectural Support for Programming Languages and Operating Systems},
volume = {2},
pages = {998–1013},
numpages = {16},
location = {Rotterdam, Netherlands},
series = {ASPLOS '25}
}

@misc{stablehlo,
  author       = {{OpenXLA Project}},
  title        = {StableHLO},
  howpublished = {\url{https://openxla.org/stablehlo}},
  note         = {Accessed 2026-06-07}
}

@misc{cudatile,
  author       = {{NVIDIA}},
  title        = {{CUDA} {Tile} {IR}},
  howpublished = {GitHub repository},
  year         = {2025},
  note         = {\url{https://github.com/NVIDIA/cuda-tile}, accessed 2026-06-20}
}

@misc{IREE,
  author       = {{IREE Project}},
  title        = {IREE: Intermediate Representation Execution Environment},
  howpublished = {\url{https://iree.dev/}},
  note         = {Accessed 2026-06-07}
}

@misc{ByteIR,
  author       = {{ByteDance}},
  title        = {ByteIR},
  howpublished = {\url{https://github.com/bytedance/byteir}},
  note         = {Accessed 2026-06-07}
}

@article{transformerSurvey2023,
  title={{A Comprehensive Survey on Applications of Transformers for Deep Learning Tasks}},
  author={Islam, Saidul and Elmekki, Hanae and Elsebai, Ahmed and Bentahar, Jamal and Drawel, Nagat and Rjoub, Gaith and Pedrycz, Witold},
  journal={Expert Systems with Applications},
  volume={241},
  pages={122666},
  year={2024},
  publisher={Elsevier}
}

@ARTICLE{9222299,
  author={Li, Mingzhen and Liu, Yi and Liu, Xiaoyan and Sun, Qingxiao and You, Xin and Yang, Hailong and Luan, Zhongzhi and Gan, Lin and Yang, Guangwen and Qian, Depei},
  journal={IEEE Transactions on Parallel and Distributed Systems}, 
  title={{The Deep Learning Compiler: A Comprehensive Survey}}, 
  year={2021},
  volume={32},
  number={3},
  pages={708--727},
  doi={10.1109/TPDS.2020.3030548}}

@article{zhang2023compiler,
  title={Compiler Technologies in Deep Learning Co-Design: A Survey},
  author={Zhang, Hongbin and Xing, Mingjie and Wu, Yanjun and Zhao, Chen},
  journal={Intelligent Computing},
  volume={2},
  year={2023},
  publisher={AAAS}
}

@inproceedings{
tilelang2025,
title={{TileLang: Bridge Programmability and Performance in Modern Neural Kernels}},
author={Lei Wang and Yu Cheng and Yining Shi and Zhiwen Mo and Zhengju Tang and Wenhao Xie and others},
booktitle={14th International Conference on Learning Representations},
series = {ICLR '26}, 
year={2026},
url={https://openreview.net/forum?id=Jb1WkNSfUB}
}

@article{kitsune2025,
author = {Davies, Michael and Crago, Neal and Sankaralingam, Karthikeyan and Keckler, Stephen},
title = {Kitsune: Enabling Dataflow Execution on GPUs with Spatial Pipelines},
year = {2025},
volume = {22},
number = {4},
url = {https://doi.org/10.1145/3777466},
doi = {10.1145/3777466},
journal = {ACM Trans. Archit. Code Optim.},
articleno = {146},
numpages = {22}
}

@article{polyblocks2026,
  author={Uday Bondhugula and Akshay Baviskar and Navdeep Katel and Vimal Patel and Anoop JS and Arnab Dutta},
  title   = {{PolyBlocks: A Compiler Infrastructure for AI Chips and Programming Frameworks}},
  journal = {arXiv preprint arXiv:2603.06731},
  year    = {2026}
}

@inproceedings{eventTensor2026,
title={{Event Tensor: A Unified Abstraction for Compiling Dynamic Megakernel}},
author={Hongyi Jin and Bohan Hou and Guanjie Wang and Ruihang Lai and Jinqi Chen and Zihao Ye and others},
booktitle={9th Conference on Machine Learning and Systems},
year={2026},
url={https://openreview.net/forum?id=PJqFhAbUHa}
}

@article{nautilus2026,
  author  = {Y. Zhao and Y. Yang and M. Budiu and S. Misailovic},
  title   = {{Nautilus: An Auto-Scheduling Tensor Compiler for Efficient Tiled GPU Kernels}},
  journal = {arXiv preprint arXiv:2604.14825},
  year    = {2026}
}

@article{neptune2025,
author = {Zhao, Yifan and Johnson, Egan and Chatarasi, Prasanth and Adve, Vikram and Misailovic, Sasa},
title = {{Neptune: Advanced ML Operator Fusion for Locality and Parallelism on GPUs}},
year = {2026},
publisher = {Association for Computing Machinery},
address = {New York, NY, USA},
volume = {10},
url = {https://doi.org/10.1145/3808298},
doi = {10.1145/3808298},
journal = {Proceedings of the ACM on Programming Languages},
articleno = {220},
numpages = {25}
}

@inproceedings{compilerR12025,
title={{Compiler-R1: Towards Agentic Compiler Auto-tuning with Reinforcement Learning}},
author={Haolin Pan and Hongyu Lin and Haoran Luo and Yang Liu and Kaichun Yao and Libo Zhang and Mingjie Xing and Yanjun Wu},
booktitle={39th Annual Conference on Neural Information Processing Systems},
year={2026},
url={https://openreview.net/forum?id=tY8ctrD4W2}
}

@inproceedings{reasoningCompiler2025,
title={{REASONING COMPILER: LLM-Guided Optimizations for Efficient Model Serving}},
author={Annabelle Sujun Tang and Christopher Priebe and Rohan Mahapatra and Lianhui Qin and Hadi Esmaeilzadeh},
booktitle={39th Annual Conference on Neural Information Processing Systems},
series = {NeurIPS '25},
year={2025},
url={https://openreview.net/forum?id=2D4TuZyNnr}
}

@article{prism2026,
      title={{Prism: Symbolic Superoptimization of Tensor Programs}}, 
      author={Mengdi Wu and Xiaoyu Jiang and Oded Padon and Zhihao Jia},
      year={2026},
      journal = {arXiv preprint arXiv:2604.15272},
}

@inproceedings{Zhang2025HexcuteAC,
  title={{Hexcute: A Compiler Framework for Automating Layout Synthesis in GPU Programs}},
  author={Xiao Zhang and Yaoyao Ding and Bolin Sun and Yang Hu and Tatiana Shpeisman and Gennady Pekhimenko},
  booktitle={2026 IEEE/ACM International Symposium on Code Generation and Optimization},
  series = {CGO '25},
  year={2025},
  pages={630--643},
  url={https://api.semanticscholar.org/CorpusID:278000417}
}

@inproceedings{chen2018tvm,
	title        = {{TVM: An automated End-to-End optimizing compiler for deep learning}},
	author       = {Chen, Tianqi and Moreau, Thierry and Jiang, Ziheng and Zheng, Lianmin and Yan, Eddie and Shen, Haichen and others},
	booktitle={13th USENIX Symposium on Operating Systems Design and Implementation (OSDI 18)},
  pages={578--594},
  year={2018}
}

@article{hexagonmlir,
      title="{Hexagon-MLIR: An AI Compilation Stack For Qualcomm's Neural Processing Units (NPUs)}", 
      author={Mohammed Javed Absar and Muthu Baskaran and Abhikrant Sharma and Abhilash Bhandari and Ankit Aggarwal and Arun Rangasamy and others},
      year={2026},
      journal={arXiv preprint arXiv:2602.19762},
      url={https://arxiv.org/abs/2602.19762}, 
}

@article{burrello2021dory,
	title        = {{DORY:} Automatic End-to-End Deployment of Real-World DNNs on Low-Cost {IoT MCUs}},
	author       = {Burrello, Alessio and Garofalo, Angelo and Bruschi, Nazareno and Tagliavini, Giuseppe and Rossi, Davide and Conti, Francesco},
	year         = 2021,
	journal      = {IEEE Trans Comput.}, 
    volume       = {70},
    number       = {8},
    pages        = {1253--1268},
}

@article{ragan2013halide,
	title        = {{Halide: A Language and Compiler for Optimizing Parallelism,. Locality, and Recomputation in Image Processing Pipelines}},
	author       = {Ragan-Kelley, Jonathan and Barnes, Connelly and Adams, Andrew and Paris, Sylvain and Durand, Fr{\'e}do and Amarasinghe, Saman},
	year         = {2013},
	journal      = {ACM SIGPLAN Notices},
	volume       = {48},
	number       = {6},
	pages        = {519--530}
}

@inproceedings{LLMCompile,
author = {Cummins, Chris and Seeker, Volker and Grubisic, Dejan and Roziere, Baptiste and Gehring, Jonas and Synnaeve, Gabriel and Leather, Hugh},
title = {{LLM Compiler: Foundation Language Models for Compiler Optimization}},
year = {2025},
booktitle = {34th ACM SIGPLAN International Conference on Compiler Construction},
pages = {141–153},
numpages = {13},
series = {CC '25}
}

@article{Cummingfirst,
      title={Large Language Models for Compiler Optimization}, 
      author={Chris Cummins and Volker Seeker and Dejan Grubisic and Mostafa Elhoushi and Youwei Liang and Baptiste Roziere and others},
      year={2023},
      journal = {arXiv preprint arXiv:2309.07062},
      url={https://arxiv.org/abs/2309.07062}, 
}

@inproceedings{TelaMalloc,
author = {Maas, Martin and Beaugnon, Ulysse and Chauhan, Arun and Ilbeyi, Berkin},
title = {{TelaMalloc: Efficient On-Chip Memory Allocation for Production Machine Learning Accelerators}},
year = {2022},
booktitle = {28th ACM International Conference on Architectural Support for Programming Languages and Operating Systems, Volume 1},
pages = {123--137},
numpages = {15},
series = {ASPLOS '23}
}

@article{MATCH,
      author={Hamdi, Mohamed Amine and Daghero, Francesco and Sarda, Giuseppe Maria and Delm, Josse Van and Symons, Arne and Benini, Luca and others},
  journal={IEEE Transactions on Computer-Aided Design of Integrated Circuits and Systems}, 
  title={{MATCH: Model-Aware TVM-Based Compilation for Heterogeneous Edge Devices}}, 
  year={2025},
  volume={44},
  number={10},
  pages={3844-3857},

}

@article{jinCompilingONNXNeural2020,
	title        = {{Compiling ONNX Neural Network Models Using MLIR}},
	author       = {Jin, Tian and Bercea, Gheorghe-Teodor and Le, Tung D. and Chen, Tong and Su, Gong and Imai, Haruki and others},
	year         = 2020,
	  journal = {arXiv preprint arXiv:2008.08272},
}

@inproceedings{linONNCCompilationFramework2019,
	title        = {{ONNC: A Compilation Framework Connecting ONNX to Proprietary Deep Learning Accelerators}},
	shorttitle   = {{{ONNC}}},
	author       = {Lin, Wei-Fen and Tsai, Der-Yu and Tang, Luba and Hsieh, Cheng-Tao and Chou, Cheng-Yi and Chang, Ping-Hao and Hsu, Luis},
	year         = 2019,
	booktitle    = {2019 IEEE International Conference on Artificial Intelligence Circuits and Systems},
    series = {AICAS '19},
	pages        = {214--218}
}

@inproceedings{roeschRelayNewIR2018,
	title        = {{Relay: A New IR for Machine Learning Frameworks}},
	shorttitle   = {Relay},
	author       = {Roesch, Jared and Lyubomirsky, Steven and Weber, Logan and Pollock, Josh and Kirisame, Marisa and Chen, Tianqi and Tatlock, Zachary},
	year         = 2018,
	booktitle    = {2nd {{ACM SIGPLAN International Workshop}} on {{Machine Learning}} and {{Programming Languages}}},
	publisher    = {{ACM}},
	series       = {{{MAPL}} '18},
	pages        = {58--68}
}

@article{rotemGlowGraphLowering2019,
	title        = {{Glow: Graph Lowering Compiler Techniques for Neural Networks}},
	author       = {Rotem, Nadav and Fix, Jordan and Abdulrasool, Saleem and Catron, Garret and Deng, Summer and Dzhabarov, Roman and others},
	year         = 2019,
	journal      = {arXiv preprint arXiv:1805.00907},
}

@INPROCEEDINGS{vandelmHTVMEfficientNeural2023a,
  author={Van Delm, Josse and Vandersteegen, Maarten and Burrello, Alessio and Sarda, Giuseppe Maria and Conti, Francesco and Pagliari, Daniele Jahier and others},
  booktitle={2023 60th ACM/IEEE Design Automation Conference (DAC)}, 
  title={{HTVM: Efficient Neural Network Deployment On Heterogeneous TinyML Platforms}}, 
  year={2023},
  pages={1-6}
}

@inproceedings{torchdynamo,
author = {Ansel, Jason and Yang, Edward and He, Horace and Gimelshein, Natalia and Jain, Animesh and Voznesensky, Michael and others},
title = {PyTorch 2: Faster Machine Learning Through Dynamic Python Bytecode Transformation and Graph Compilation},
year = {2024},
booktitle = {29th ACM International Conference on Architectural Support for Programming Languages and Operating Systems, Volume 2},
pages = {929–947},
series = {ASPLOS '24}
}

@misc{ONNX,
    author = {Bai, Junjie and Lu, Fang and Zhang, Ke and others},
    title = {ONNX: Open Neural Network Exchange},
    year = {2019},
    howpublished = {\url{https://github.com/onnx/onnx}}
}

@misc{oneapi,
	title        = {{oneAPI Programming Model}},
	author       = {{Unified Acceleration Foundation}},
	year         = 2026,
	howpublished = {\url{https://www.oneapi.io}}
}

@misc{smartHLS,
	title        = {{Libero SoC Design Suite}},
	author       = {Microchip},
	year         = 2026,
	howpublished = {\url{https://www.microchip.com/en-us/products/fpgas-and-plds/fpga-and-soc-design-tools/fpga/libero-software-later-versions}}
}

@misc{fastml_hls4ml,
  author       = {{FastML Team}},
  title        = {fastmachinelearning/hls4ml},
  year         = 2025,
  publisher    = {Zenodo},
  version      = {v1.3.0},
  doi          = {10.5281/zenodo.1201549},
  howpublished = {\url{https://github.com/fastmachinelearning/hls4ml}}
}

@article{schulte2026hls4ml,
  title={{hls4ml: A Flexible, Open Source Platform for Deep Learning Acceleration on Reconfigurable Hardware}},
  author={Schulte, Jan-Frederik and Ramhorst, Benjamin and Sun, Chang and Mitrevski, Jovan and Ghielmetti, Nicol{\`o} and Lupi, Enrico and others},
  journal={ACM Transactions on Reconfigurable Technology and Systems},
  volume={19},
  number={2},
  pages={1--35},
  year={2026},
}

@inproceedings{ferrandi2021bambu,
	title        = {{Invited: Bambu: An Open-Source Research Framework for the High-Level Synthesis of Complex Applications}},
	author       = {Ferrandi, Fabrizio and Castellana, Vito Giovanni and Curzel, Serena and Fezzardi, Pietro and Fiorito, Michele and Lattuada, Marco and others},
	year         = 2021,
	booktitle    = {2021 58th ACM/IEEE Design Automation Conference},
    series = {DAC '21},
	publisher    = {{IEEE}},
	pages        = {1327--1330}
}

@manual{vitisug212,
	title        = {{Vitis High-Level Synthesis User Guide}},
	author       = {{AMD}},
	year         = 2026,
	url          = {https://docs.amd.com/r/en-US/ug1399-vitis-hls}
}

@manual{IntelHLS2022,
	title        = {{Altera High Level Synthesis Compiler Pro Edition Reference Manual}},
	author       = {{Altera}},
	year         = 2025,
	url          = {https://docs.altera.com/r/docs/683349/current}
}

@manual{CatapultHLS2022,
	title        = {{Catapult C++/Systemc Synthesis}},
	author       = {{Siemens}},
	year         = 2026,
	url          = {https://eda.sw.siemens.com/en-US/ic/catapult-high-level-synthesis/hls/c-cplus/}
}

@manual{StratusHLS2022,
	title        = {{Stratus High-Level Synthesis}},
	author       = {{Cadence}},
	year         = 2026,
	url          = {https://www.cadence.com/en_US/home/tools/digital-design-and-signoff/synthesis/stratus-high-level-synthesis.html}
}

@article{LegUpHLS2013,
	title        = {{LegUp: An open-source high-level synthesis tool for FPGA-based processor/accelerator systems}},
	author       = {Andrew Canis and Jongsok Choi and Mark Aldham and Victor Zhang and Ahmed Kammoona and Tomasz S. Czajkowski and others},
	year         = 2013,
	journal      = {{ACM} Trans. Embed. Comput. Syst.},
	volume       = 13,
	number       = 2,
	pages        = {24:1--24:27},
}

@inproceedings{ABVG+19,
	title        = {{Placeto: Learning Generalizable Device Placement Algorithms for Distributed Machine Learning}},
	author       = {Addanki, Ravichandra and Venkatakrishnan, Shaileshh Bojja and Gupta, Shreyan and Mao, Hongzi and Alizadeh, Mohammad},
	year         = 2019,
	booktitle    = {33rd International Conference on Neural Information Processing Systems},
	publisher    = {Curran Associates Inc.},
	pages        = {3981–3991},
	articleno    = 358,
	numpages     = 11
}

@inproceedings{NHP+19,
  title={{PipeDream: Generalized Pipeline Parallelism for {DNN} Training}},
  author={Narayanan, Deepak and Harlap, Aaron and Phanishayee, Amar and Seshadri, Vivek and Devanur, Nikhil R and Ganger, Gregory R and others},
  booktitle={27th ACM Symposium on Operating Systems Principles},
  series = {SOSP '19},
  pages={1--15},
  year={2019}
}

@article{KL70,
	title        = {{An Efficient Heuristic Procedure for Partitioning Graphs}},
	author       = {Kernighan, B. W. and Lin, S.},
	year         = 1970,
	journal      = {The Bell System Technical Journal},
	volume       = 49,
	number       = 2,
	pages        = {291--307}
}

@incollection{LLKS93,
	title        = {{Chapter 9 Sequencing and scheduling: Algorithms and complexity}},
	author       = {Eugene L. Lawler and Jan Karel Lenstra and Alexander H.G. {Rinnooy Kan} and David B. Shmoys},
	year         = 1993,
	booktitle    = {Logistics of Production and Inventory},
	publisher    = {Elsevier},
	series       = {Handbooks in Operations Research and Management Science},
	volume       = 4,
	pages        = {445--522}
}

@inproceedings{MM17,
	title        = {{The Tensorflow Partitioning and Scheduling Problem: It's the Critical Path!}},
	author       = {Mayer, Ruben and Mayer, Christian and Laich, Larissa},
	year         = 2017,
	booktitle    = {1st Workshop on Distributed Infrastructures for Deep Learning},
	publisher    = {ACM},
	series       = {DIDL '17},
	numpages     = 6,
    pages={1--6},
}

@inproceedings{MPL+17,
	title        = {{Device Placement Optimization with Reinforcement Learning}},
	author       = {Mirhoseini, Azalia and Pham, Hieu and Le, Quoc V. and Steiner, Benoit and Larsen, Rasmus and Zhou, Yuefeng and others},
	year         = 2017,
	booktitle    = {34th International Conference on Machine Learning},
    Volume = {70},
    publisher    = {{PMLR}},
	series       = {ICML '17},
	pages        = {2430--2439},
	numpages     = 10
}

@inproceedings{MGP+18,
	title        = {{A Hierarchical Model for Device Placement}},
	author       = {Azalia Mirhoseini and Anna Goldie and Hieu Pham and Benoit Steiner and Quoc V. Le and Jeff Dean},
	year         = {2018},
	booktitle    = {6th International Conference on Learning Representations},
	series       = {ICLR '18},
	pages = {1--3}
}

@article{PY90,
	title        = {{Towards an Architecture-Independent Analysis of Parallel Algorithms}},
	author       = {Christos H. Papadimitriou and Mihalis Yannakakis},
	year         = 1990,
	journal      = {SIAM Journal on Computing},
	volume       = 19,
	number       = 2,
	pages        = {322--328}
}

@inproceedings{PGN+20,
	title        = {{Reinforced Genetic Algorithm Learning for Optimizing Computation Graphs}},
	author       = {Aditya Paliwal and Felix Gimeno and Vinod Nair and Yujia Li and Miles Lubin and Pushmeet Kohli and Oriol Vinyals},
	year         = 2020,
	booktitle    = {8th International Conference on Learning Representations},
	series       = {ICLR '20},
	numpages     = 24,
    pages = {1--24}
}

@article{ST93,
	title        = {{An Approximation Algorithm for the Generalized Assignment Problem}},
	author       = {Shmoys, David B. and Tardos, {\'E}va},
	year         = 1993,
	journal      = {Mathematical Programming},
	volume       = 62,
	number       = 1,
	pages        = {461--474}
}

@inproceedings{SW99,
	title        = {{A PTAS for Minimizing the Weighted Sum of Job Completion Times on Parallel Machines}},
	author       = {Skutella, Martin and Woeginger, Gerhard J.},
	year         = 1999,
	booktitle    = {31 Annual ACM Symposium on Theory of Computing},
	publisher    = {ACM},
	series       = {STOC '99},
	pages        = {400--407}
}

@inproceedings{YY19,
	title        = {{GPipe: Efficient Training of Giant Neural Networks Using Pipeline Parallelism}},
	author       = {Huang, Yanping and Cheng, Youlong and Bapna, Ankur and Firat, Orhan and Chen, Mia Xu and Chen, Dehao and others},
	year         = 2019,
	booktitle    = {33rd International Conference on Neural Information Processing Systems},
	publisher    = {Curran Associates Inc.},
	series       = {NeurIPS '19},
	numpages     = 10,
    pages = {1--10}
}

@inproceedings{JQLA18,
	title        = {{Exploring Hidden Dimensions in Parallelizing Convolutional Neural Networks}},
	author       = {Zhihao Jia and Sina Lin and Charles R. Qi and Alex Aiken},
	year         = 2018,
	booktitle    = {35th International Conference on Machine Learning},
	publisher    = {{PMLR}},
	series       = {ICML '18},
	volume       = 80,
	pages        = {2279--2288},
}

@inproceedings{JZA19,
	title        = {{Beyond Data and Model Parallelism for Deep Neural Networks}},
	author       = {Jia, Zhihao and Zaharia, Matei and Aiken, Alex},
	year         = 2019,
	booktitle    = {Proceedings of Machine Learning and Systems 1},
	series       = {MLSys '19},
	volume       = 1,
	pages        = {1--13}
}

@article{ZRA+19,
  author={Zhou, Yanqi and Roy, Sudip and Abdolrashidi, Amirali and Wong, Daniel Lin-Kit and Ma, Peter and Xu, Qiumin and others},
  journal={IEEE Micro}, 
  title={{A Single-Shot Generalized Device Placement for Large Dataflow Graphs}}, 
  year={2020},
  volume={40},
  number={5},
  pages={26-36}
}

@inproceedings{han2016deep,
	title        = {{Deep Compression: Compressing Deep Neural Networks with Pruning, Trained Quantization and Huffman Coding}},
	author       = {Han, Song and Mao, Huizi and Dally, William J},
	year         = 2016,
	series       = {ICLR '16},
	booktitle  = {4th International Conference on Learning Representations},
    pages = {1--14}
}

@inproceedings{jacob2018quantization,
	title        = {{Quantization and Training of Neural Networks for Efficient Integer-arithmetic-only Inference}},
	author       = {Jacob, Benoit and Kligys, Skirmantas and Chen, Bo and Zhu, Menglong and Tang, Matthew and Howard, Andrew and others},
	year         = 2018,
	booktitle    = {2018 IEEE Conference on Computer Vision and Pattern Recognition},
    series = {CVPR '18},
	pages        = {2704--2713}
}

@inproceedings{krizhevsky2012,
	title        = {{ImageNet Classification with Deep Convolutional Neural Networks}},
	author       = {Krizhevsky, Alex and Sutskever, Ilya and Hinton, Geoffrey E.},
	year         = 2012,
	booktitle    = {25th International Conference on Neural Information Processing Systems},
    volume = {1},
	publisher    = {Curran Associates Inc.},
	series       = {NIPS'12},
	pages        = {1097–1105},
	numpages     = 9
}

@inproceedings{fan2021dapple,
	title        = {{DAPPLE: A Pipelined Data Parallel Approach for Training Large Models}},
	author       = {Fan, Shiqing and Rong, Yi and Meng, Chen and Cao, Zongyan and Wang, Siyu and Zheng, Zhen and others},
	year         = 2021,
	booktitle    = {26th ACM SIGPLAN Symposium on Principles and Practice of Parallel Programming},
	publisher    = {ACM},
	series       = {PPoPP '21},
	pages        = {431--445}
}

@inproceedings{wu2022sustainableai,
	title        = {{Sustainable AI: Environmental Implications, Challenges and Opportunities}},
	author       = {Carole{-}Jean Wu and Ramya Raghavendra and Udit Gupta and Bilge Acun and Newsha Ardalani and Kiwan Maeng and others},
	year         = 2022,
	booktitle    = {Proceedings of Machine Learning and Systems},
    series = {MLSys '22},
    volume = {4},
    pages={795--813},
}

@inproceedings{li2023green,
author = {Li, Baolin and Samsi, Siddharth and Gadepally, Vijay and Tiwari, Devesh},
title = {{Clover: Toward Sustainable AI with Carbon-Aware Machine Learning Inference Service}},
year = {2023},
doi = {10.1145/3581784.3607034},
booktitle = {International Conference for High Performance Computing, Networking, Storage and Analysis},
articleno = {20},
numpages = {15},
series = {SC '23}
}

@inproceedings{tang_dossa18,
	title        = {{MLPAT: A Power Area Timing Modeling Framework for Machine Learning Accelerators}},
	author       = {Tang, T. and Y. Xie.},
	year         = 2018,
	booktitle    = {IEEE International Workshop on Domain Specific System Architecture},
    series = {DOSSA '18},
  pages={1--3},
}

@article{kwon_micro20,
	title        = {{MAESTRO: A Data-Centric Approach to Understand Reuse, Performance, and Hardware Cost of DNN Mappings}},
	author       = {Kwon, Hyoukjun and Chatarasi, Prasanth and Sarkar, Vivek and Krishna, Tushar and Pellauer, Michael and Parashar, Angshuman},
	year         = 2020,
	journal      = {IEEE Micro},
	volume       = 40,
	number       = 3,
	pages        = {20--29}
}

@inproceedings{parashar_ispass19,
	title        = {{Timeloop: A Systematic Approach to DNN Accelerator Evaluation}},
	author       = {Parashar, Angshuman and Raina, Priyanka and Shao, Yakun Sophia and Chen, Yu-Hsin and Ying, Victor A. and Mukkara, Anurag and others},
	year         = {2019},
	booktitle    = {2019 IEEE International Symposium on Performance Analysis of Systems and Software},
    series = {ISPASS '19},
	pages        = {304--315},
	organization = {IEEE}
}

@inproceedings{wu_iccad19,
	title        = {{Accelergy: An Architecture-Level Energy Estimation Methodology for Accelerator Designs}},
	author       = {Wu, Yannan Nellie and Emer, Joel S. and Sze, Vivienne},
	year         = 2019,
	booktitle    = {2019 IEEE/ACM International Conference on Computer-Aided Design},
    series = {ICCAD '19},
	pages        = {1--8}
}

@inproceedings{zhao2021dnnchip,
  author       = {Yang Zhao and
                  Chaojian Li and
                  Yue Wang and
                  Pengfei Xu and
                  Yongan Zhang and
                  Yingyan Lin},
  title        = {{DNN-Chip Predictor: An Analytical Performance Predictor for {DNN} Accelerators with Various Dataflows and Hardware Architectures}},
  booktitle    = {2020 {IEEE} International Conference on Acoustics, Speech and Signal Processing}, 
  series = {ICASSP '20},
  pages        = {1593--1597},
  publisher    = {{IEEE}},
  year         = {2020},
  doi          = {10.1109/ICASSP40776.2020.9053977},
}

@inproceedings{genc_dac21,
	title        = {{Gemmini: Enabling Systematic Deep-Learning Architecture Evaluation via Full-Stack Integration}},
	author       = {Genc, Hasan and Kim, Seah and Amid, Alon and Haj-Ali, Ameer and Iyer, Vighnesh and Prakash, Pranav and others},
	year         = 2021,
	booktitle    = {58th ACM/IEEE Design Automation Conference},
    series = {DAC '21},
	pages        = {769--774}
}

@inproceedings{yang_asplos20,
	title        = {{Interstellar: Using Halide's Scheduling Language to Analyze DNN Accelerators}},
	author       = {Yang, Xuan and Gao, Mingyu and Liu, Qiaoyi and Setter, Jeff and Pu, Jing and Nayak, Ankita and others},
	year         = 2020,
	booktitle    = {25th International Conference on Architectural Support for Programming Languages and Operating Systems},
	location     = {Lausanne, Switzerland},
	publisher    = {ACM},
	series       = {ASPLOS '20},
	pages        = {369--383}
}

@inproceedings{samajdar_ispass20,
	title        = {{A Systematic Methodology for Characterizing Scalability of DNN Accelerators using SCALE-sim}},
	author       = {Samajdar, Ananda and Joseph, Jan Moritz and Zhu, Yuhao and Whatmough, Paul and Mattina, Matthew and Krishna, Tushar},
	year         = 2020,
	booktitle    = {2020 IEEE International Symposium on Performance Analysis of Systems and Software},
    series = {ISPASS '20},
	pages        = {58--68},
	organization = {IEEE}
}

@article{munozmartinez2020stonne,
	title        = {{STONNE: A Detailed Architectural Simulator for Flexible Neural Network Accelerators}},
	author       = {Francisco Muñoz-Martínez and José L. Abellán and Manuel E. Acacio and Tushar Krishna},
	year         = 2020,
	journal = {arXiv preprint arXiv:2006.07137},
}

@article{cao_jestcs20,
	title        = {{SimuNN: A Pre-RTL Inference, Simulation and Evaluation Framework for Neural Networks}},
	author       = {Cao, Shan and Deng, Wei and Bao, Zhenyi and Xue, Chengbo and Xu, Shugong and Zhang, Shunqing},
	year         = 2020,
	journal      = {IEEE Journal on Emerging and Selected Topics in Circuits and Systems},
	volume       = 10,
	number       = 2,
	pages        = {217--230}
}

@article{kim2020transactionlevel,
	title        = {{Transaction-level Model Simulator for Communication-Limited Accelerators}},
	author       = {Sunwoo Kim and Jooho Wang and Youngho Seo and Sanghun Lee and Yeji Park and Sungkyung Park and Chester Sungchung Park},
	year         = 2020,
	journal       = {arXiv preprint arXiv:2007.14897},
}

@article{inci2022qadam,
	title        = {{QADAM: Quantization-Aware DNN Accelerator Modeling for Pareto-Optimality}},
	author       = {Ahmet Inci and Siri Garudanagiri Virupaksha and Aman Jain and Venkata Vivek Thallam and Ruizhou Ding and Diana Marculescu},
	year         = 2022,
    journal      = {arXiv preprint arXiv:2205.13045},
}

@article{inci2022qappa,
	title        = {{QAPPA: Quantization-Aware Power, Performance, and Area Modeling of DNN Accelerators}},
	author       = {Ahmet Inci and Siri Garudanagiri Virupaksha and Aman Jain and Venkata Vivek Thallam and Ruizhou Ding and Diana Marculescu},
	year         = 2022,
    journal      = {arXiv preprint arXiv:2205.08648},
}

@article{juracy_tcs22,
	title        = {{A Fast, Accurate, and Comprehensive PPA Estimation of Convolutional Hardware Accelerators}},
	author       = {Juracy, Leonardo Rezende and de Morais Amory, Alexandre and Moraes, Fernando Gehm},
	year         = 2022,
	journal      = {IEEE Transactions on Circuits and Systems I: Regular Papers},
	volume       = 69,
	number       = 12,
	pages        = {5171--5184}
}

@inproceedings{russo_percom21,
	title        = {{LAMBDA: An Open Framework for Deep Neural Network Accelerators Simulation}},
	author       = {Russo, Enrico and Palesi, Maurizio and Monteleone, Salvatore and Patti, Davide and Ascia, Giuseppe and Catania, Vincenzo},
	year         = 2021,
	booktitle    = {2021 IEEE International Conference on Pervasive Computing and Communications Workshops and other Affiliated Events},
    series = {PerCom Workshops '21},
	pages        = {161--166}
}

@article{Lammie2022,
	title        = {{MemTorch: An Open-source Simulation Framework for Memristive Deep Learning Systems}},
	author       = {Corey Lammie and Wei Xiang and Bernabé Linares-Barranco and Mostafa Rahimi Azghadi},
	year         = 2022,
	journal      = {Neurocomputing},
    volume = {485}, 
    pages = {124--133},
}

@inproceedings{peng2019iedm,
	title        = {{DNN+NeuroSim: An End-to-End Benchmarking Framework for Compute-in-Memory Accelerators with Versatile Device Technologies}},
	author       = {Peng, Xiaochen and Huang, Shanshi and Luo, Yandong and Sun, Xiaoyu and Yu, Shimeng},
	year         = 2019,
	booktitle    = {{2019 IEEE International Electron Devices Meeting (IEDM)}},
	pages        = {32.5.1--32.5.4}
}

@inproceedings{shadmehri2022date,
	title        = {{SySCIM: SystemC-AMS Simulation of Memristive Computation In-Memory}},
	author       = {Shadmehri, Seyed Hossein Hashemi and BanaGozar, Ali and Kamal, Mehdi and Stuijk, Sander and Afzali-Kusha, Ali and Pedram, Massoud and Corporaal, Henk},
	year         = 2022,
	booktitle    = {2022 Design, Automation and Test in Europe Conference and Exhibition},
    series = {DATE '22},
	pages        = {1467--1472}
}

@inproceedings{xia2016date,
	title        = {{MNSIM: Simulation Platform for Memristor-based Neuromorphic Computing System}},
	author       = {Xia, Lixue and Li, Boxun and Tang, Tianqi and Gu, Peng and Yin, Xiling and Huangfu, Wenqin and others},
	year         = 2016,
	booktitle    = {2016 Design, Automation and Test in Europe Conference and Exhibition},
    series = {DATE '16},
	pages        = {469--474}
}

@inproceedings{reiser2023newcas,
  author={Reiser, Daniel and Reichenbach, Marc and Rizzi, Tommaso and Baroni, Andrea and Fritscher, Markus and Wenger, Christian and others},
  booktitle={2023 21st IEEE Interregional NEWCAS Conference},
  series = {NEWCAS '23},
  title={{Technology-Aware Drift Resilience Analysis of RRAM Crossbar Array Configurations}}, 
  year={2023},
  pages={1-5}
}

@INPROCEEDINGS{rizzi21,
  author={Rizzi, T. and Quesada, E. Pérez-Bosch and Wenger, Ch. and Zambelli, C. and Bertozzi, D.},
  booktitle={2021 XXXVI Conference on Design of Circuits and Integrated Systems},
  series = {DCIS '21},
  title={{Comparative Analysis and Optimization of the SystemC-AMS Analog Simulation Efficiency of Resistive Crossbar Arrays}}, 
  year={2021},
  pages={1-6}
}

@inproceedings{hpcc-fpga,
	title        = {{Evaluating FPGA Accelerator Performance with a Parameterized OpenCL Adaptation of Selected Benchmarks of the HPCChallenge Benchmark Suite}},
	author       = {M. {Meyer} and T. {Kenter} and C. {Plessl}},
	year         = 2020,
	booktitle    = {2020 IEEE/ACM International Workshop on Heterogeneous High-performance Reconfigurable Computing},
    series = {H2RC '20},
	pages        = {10--18}
}

@inproceedings{exa-dataflow,
	title        = {{Performance Estimation for Exascale Reconfigurable Dataflow Platforms}},
	author       = {R. {Yasudo} and J. {Coutinho} and A. {Varbanescu} and W. {Luk} and H. {Amano} and T. {Becker}},
	year         = 2018,
	booktitle    = {2018 International Conference on Field-Programmable Technology (FPT)},
	volume       = {},
	number       = {},
	pages        = {314--317}
}

@article{fpga-roofline-tsunami,
	title        = {{FPGA-based Tsunami Simulation: Performance Comparison with GPUs, and Roofline Model for Scalability Analysis}},
	author       = {Kohei Nagasu and Kentaro Sano and Fumiya Kono and Naohito Nakasato},
	year         = 2017,
	journal      = {Journal of Parallel and Distributed Computing},
	volume       = 106,
	pages        = {153--169}
}

@inproceedings{fpga-hpc-trends,
	title        = {{Trends of CPU, GPU and FPGA for High-Performance Computing}},
	author       = {M. {V\'estias} and H. {Neto}},
	year         = 2014,
	booktitle    = {2014 24th International Conference on Field Programmable Logic and Applications},
    series = {FPL '14},
	pages        = {1--6}
}

@inproceedings{fpga-fp-eval,
	title        = {{Evaluation of a Floating-Point Intensive Kernel on FPGA}},
	author       = {Jin, Zheming and Finkel, Hal and Yoshii, Kazutomo and Cappello, Franck},
	year         = 2018,
	booktitle    = {Euro-Par 2017: Parallel Processing Workshops},
	pages        = {664--675}
}

@article{fpga-dl,
	title        = {{FPGA-Based Accelerators of Deep Learning Networks for Learning and Classification: A Review}},
	author       = {A. {Shawahna} and S. M. {Sait} and A. {El-Maleh}},
	year         = 2019,
	journal      = {IEEE Access},
	volume       = 7,
	number       = {},
	pages        = {7823--7859}
}

@article{project-brainwave,
	title        = {{Inside Project Brainwave's Cloud-Scale, Real-Time AI Processor}},
	author       = {J. {Fowers} and K. {Ovtcharov} and M. K. {Papamichael} and T. {Massengill} and M. {Liu} and D. {Lo} and others},
	year         = 2019,
	journal      = {IEEE Micro},
	volume       = 39,
	number       = 3,
	pages        = {20--28}
}

@article{project-catapult,
	title        = {{A Reconfigurable Fabric for Accelerating Large-Scale Datacenter Services}},
	author       = {Putnam, Andrew and Caulfield, Adrian M. and Chung, Eric S. and Chiou, Derek and Constantinides, Kypros and Demme, John and others},
	year         = 2015,
	journal      = {IEEE Micro},
	volume       = 35,
	number       = 3,
	pages        = {10--22}
}

@book{fpga-hpc,
	title        = {{High-Performance Computing using FPGAs}},
	author       = {Vanderbauwhede, Wim and Benkrid, Khaled},
	year         = 2013,
	publisher    = {Springer},
	volume       = 3
}

@article{fpga-hpc2,
	title        = {{Suitability Analysis of FPGAs for Heterogeneous Platforms in HPC}},
	author       = {F. A. {Escobar} and X. {Chang} and C. {Valderrama}},
	year         = 2016,
	journal      = {IEEE Transactions on Parallel and Distributed Systems},
	volume       = 27,
	number       = 2,
	pages        = {600--612}
}

@inproceedings{shuhai,
	title        = {{Shuhai: Benchmarking High Bandwidth Memory On FPGAs}},
	author       = {Z. {Wang} and H. {Huang} and J. {Zhang} and G. {Alonso}},
	year         = 2020,
	booktitle    = {2020 IEEE 28th Annual International Symposium on Field-Programmable Custom Computing Machines (FCCM)},
	volume       = {},
	number       = {},
	pages        = {111--119}
}

@inproceedings{ert-opencl-fpga,
	title        = {{The Performance and Energy Efficiency Potential of FPGAs in Scientific Computing}},
	author       = {T. {Nguyen} and S. {Williams} and M. {Siracusa} and C. {MacLean} and D. {Doerfler} and N. J. {Wright}},
	year         = 2020,
	booktitle    = {2020 IEEE/ACM Performance Modeling, Benchmarking and Simulation of High Performance Computer Systems (PMBS)},
	volume       = {},
	number       = {},
	pages        = {8--19}
}

@article{ert-opencl-fpga2,
	title        = {{FPGA-based HPC accelerators: An evaluation on performance and energy efficiency}},
	author       = {Nguyen, Tan and MacLean, Colin and Siracusa, Marco and Doerfler, Douglas and Wright, Nicholas J. and Williams, Samuel},
	year         = 2021,
	journal      = {Concurrency and Computation: Practice and Experience},
	volume       = {n/a},
	number       = {n/a},
	pages        = {e6570}
}

@inproceedings{roofline-fpga-cad,
	title        = {{A CAD-based Methodology to Optimize HLS Code via the Roofline Model}},
	author       = {M. {Siracusa} and M. {Rabozzi} and E. {Del Sozzo} and L. {Di Tucci} and S. {Williams} and M. D. {Santambrogio}},
	year         = 2020,
	booktitle    = {2020 IEEE/ACM International Conference On Computer Aided Design},
	series       = {ICCAD '20},
	pages        = {1--9},
}

@article{roofline-fpga-cad2,
	title        = {{A Comprehensive Methodology to Optimize FPGA Designs via the Roofline Model}},
	author       = {Siracusa, Marco and Delsozzo, Emanuele and Rabozzi, Marco and Di Tucci, Lorenzo and Williams, Samuel and Sciuto, Donatella and Santambrogio, Marco Domenico},
    year={2022},
    volume={71},
    number={8},
    pages={1903--1915},
	journal      = {IEEE Transactions on Computers},
}

@article{roofline-fpga-hls,
	title        = {{Performance modeling for FPGAs: extending the roofline model with high-level synthesis tools}},
	author       = {Da Silva, Bruno and Braeken, An and D’Hollander, Erik H and Touhafi, Abdellah},
	year         = 2013,
	journal      = {International Journal of Reconfigurable Computing},
	publisher    = {Hindawi},
	volume       = 2013
}

@inproceedings{roofline-multibench-fpga,
	title        = {{A Semi-Automated Tool Flow for Roofline Anaylsis of OpenCL Kernels on Accelerators}},
	author       = {Muralidharan, Servesh and O'Brien, Kenneth and Lalanne, Christian},
	year         = 2015,
	booktitle    = {1st International Workshop on Heterogeneous High-performance Reconfigurable Computing},
    pages = {1--8}
}

@inproceedings{hpcg-fpga,
	title        = {{Optimized Implementation of the HPCG Benchmark on Reconfigurable Hardware}},
	author       = {Zeni, Alberto and O'Brien, Kenneth and Blott, Michaela and Santambrogio, Marco D.},
	year         = 2021,
	booktitle    = {Euro-Par 2021: Parallel Processing},
	pages        = {616--630}
}

@inproceedings{intel-fpga-mem,
	title        = {{The Memory Controller Wall: Benchmarking the Intel FPGA SDK for OpenCL Memory Interface}},
	author       = {Zohouri, Hamid Reza and Matsuoka, Satoshi},
	year         = 2019,
	booktitle    = {2019 IEEE/ACM International Workshop on Heterogeneous High-performance Reconfigurable Computing},
    series = {H2RC '19},
	pages        = {11--18}
}

@inproceedings{fpga-stream-opencl,
	title        = {{MP-STREAM: A Memory Performance Benchmark for Design Space Exploration on Heterogeneous HPC Devices}},
	author       = {S. W. {Nabi} and W. {Vanderbauwhede}},
	year         = 2018,
	booktitle    = {2018 IEEE International Parallel and Distributed Processing Symposium Workshops},
	series       = {IPDPSW '18},
	pages        = {194--197}
}

@article{fer,
	title        = {{FER: A Benchmark for the Roofline Analysis of FPGA Based HPC Accelerators}},
	author       = {Calore, Enrico and Schifano, Sebastiano Fabio},
	year         = 2022,
	journal      = {IEEE Access},
	volume       = 10,
	number       = {},
	pages        = {94220--94234}
}

@inproceedings{fer-fpl,
	title        = {{Performance assessment of FPGAs as HPC accelerators using the FPGA Empirical Roofline}},
	author       = {Calore, Enrico and Schifano, Sebastiano Fabio},
	year         = 2021,
	booktitle    = {2021 31st International Conference on Field-Programmable Logic and Applications (FPL)},
	volume       = {},
	number       = {},
	pages        = {83--90}
}

@inproceedings{parco19-fp,
	title        = {{Energy-efficiency evaluation of FPGAs for floating-point intensive workloads}},
	author       = {Calore, E. and Schifano, S.F.},
	year         = 2020,
	booktitle    = {Parallel Computing is Everywhere},
	series       = {Advances in Parallel Computing},
	volume       = 36,
	pages        = {555--564}
}

@article{hlssurvey,
	title        = {{FPGA HLS Today: Successes, Challenges, and Opportunities}},
	author       = {Cong, Jason and Lau, Jason and Liu, Gai and Neuendorffer, Stephen and Pan, Peichen and Vissers, Kees and Zhang, Zhiru},
	year         = 2022,
	journal      = {ACM Trans. Reconfigurable Technol. Syst.},
	publisher    = {Association for Computing Machinery},
	address      = {New York, NY, USA},
	volume       = 15,
	number       = 4,
	issue_date   = {December 2022},
	articleno    = 51
}

@article{chen2020engineering,
	title        = {{A Survey of Accelerator Architectures for Deep Neural Networks}},
	author       = {Yiran Chen and Yuan Xie and Linghao Song and Fan Chen and Tianqi Tang},
	year         = 2020,
	journal      = {Engineering},
	volume       = 6,
	number       = 3,
	pages        = {264--274},
}

@article{akkad2023embedded,
  title={{Embedded Deep Learning Accelerators: A Survey on Recent Advances}},
  author={Akkad, Ghattas and Mansour, Ali and Inaty, Elie},
  journal={IEEE Transactions on Artificial Intelligence},
  volume={5},
  number={5},
  pages={1954--1972},
  year={2023},
  publisher={IEEE},
}

@article{silvano2023survey,
    title={{A Survey on Deep Learning Hardware Accelerators for Heterogeneous HPC Platforms}}, 
    author={Cristina Silvano and Daniele Ielmini and Fabrizio Ferrandi and Leandro Fiorin and Serena Curzel and Luca Benini and others},
    year = {2025},
    publisher = {Association for Computing Machinery},
    journal = {ACM Computing Surveys},
    volume = {57},
    number = {11},
    pages = {1--39},
}

@article{ML4EDA21,
author = {Huang, Guyue and Hu, Jingbo and He, Yifan and Liu, Jialong and Ma, Mingyuan and Shen, Zhaoyang and others},
title = {{Machine Learning for Electronic Design Automation: A Survey}},
year = {2021},
issue_date = {September 2021},
publisher = {Association for Computing Machinery},
address = {New York, NY, USA},
volume = {26},
number = {5},
journal = {ACM Transactions on Design Automation of Electronic Systems},
articleno = {40},
pages = {1--46},
numpages = {46}
}

@book{RenHu23,
	title        = {{Machine Learning Applications in Electronic Design Automation}},
	author       = {},
	year         = 2023,
	publisher    = {Springer Cham},
	series       = {Mathematics and Statistics, Mathematics and Statistics (R0)},
	volume       = {},
	number       = {},
	pages        = {583},
	note         = {},
	editor       = {Haoxing Ren, Jiang Hu},
	subtitle     = {},
	edition      = {1},
	chapter      = {},
	type         = {}
}

@ARTICLE{Kahng23,
  author={Kahng, Andrew B.},
  journal={IEEE Design \& Test}, 
  title={{Machine Learning for {CAD}/{EDA}: The Road Ahead}}, 
  year={2023},
  volume={40},
  number={1},
  pages={8-16}
}

@article{SALEEM22,
author = {Rabia Saleem and Bo Yuan and Fatih Kurugollu and Ashiq Anjum and Lu Liu},
title = {{Explaining Deep Neural Networks: A Survey on the Global Interpretation Methods}},
journal = {Neurocomputing},
volume = {513},
pages = {165-180},
year = {2022},
}

@INPROCEEDINGS{sodaSNN,
    author={Curzel, Serena and Bohm Agostini, Nicolas and Song, Shihao and Dagli, Ismet and Limaye, Ankur and Tan, Cheng and others},
    booktitle={2021 IEEE/ACM International Conference On Computer Aided Design (ICCAD)},
    title={{Automated Generation of Integrated Digital and Spiking Neuromorphic Machine Learning Accelerators}},
    year={2021},
    pages={1-7}
}

@ARTICLE{TCdataflow,
  author={Curzel, Serena and Bohm Agostini, Nicolas and Castellana, Vito Giovanni and Minutoli, Marco and Limaye, Ankur and Manzano, Joseph and others},
  journal={IEEE Transactions on Computers}, 
  title={End-to-End Synthesis of Dynamically Controlled Machine Learning Accelerators}, 
  year={2022},
  volume={71},
  number={12},
  pages={3074-3087}
}

@ARTICLE{mei_tc21,
  author={Mei, Linyan and Houshmand, Pouya and Jain, Vikram and Giraldo, Sebastian and Verhelst, Marian},
  journal={IEEE Transactions on Computers}, 
  title={ZigZag: Enlarging Joint Architecture-Mapping Design Space Exploration for DNN Accelerators}, 
  year={2021},
  volume={70},
  number={8},
  pages={1160-1174}
}

@article{xi_taco20,
  author = {Xi, Sam (Likun) and Yao, Yuan and Bhardwaj, Kshitij and Whatmough, Paul and Wei, Gu-Yeon and Brooks, David},
  title = {{SMAUG: End-to-End Full-Stack Simulation Infrastructure for Deep Learning Workloads}},
  year = {2020},
  volume = {17},
  number = {4},
  articleno = {39},
  numpages = {26},
  journal = {ACM Transactions on Architecture and Code Optimization},
}

@INPROCEEDINGS{mei_hpca23,
  author={Mei, Linyan and Goetschalckx, Koen and Symons, Arne and Verhelst, Marian},
  booktitle={2023 IEEE International Symposium on High-Performance Computer Architecture}, 
  series = {HPCA '23},
  title={{DeFiNES: Enabling Fast Exploration of the Depth-first Scheduling Space for DNN Accelerators through Analytical Modeling}}, 
  year={2023},
  pages={570--583},
}

@inproceedings{wu_micro22,
  author = {Wu, Yannan Nellie and Tsai, Po-An and Parashar, Angshuman and Sze, Vivienne and Emer, Joel S.},
  title = {{Sparseloop: An Analytical Approach to Sparse Tensor Accelerator Modeling}},
  year = {2023},
  booktitle = {55th Annual IEEE/ACM International Symposium on Microarchitecture},
  pages = {1377--1395}
}

@INPROCEEDINGS{cai_hpca24,
  author={Cai, Jingwei and Wu, Zuotong and Peng, Sen and Wei, Yuchen and Tan, Zhanhong and Shi, Guiming and others},
  booktitle={2024 IEEE International Symposium on High-Performance Computer Architecture}, 
  series = {HPCA '24}, 
  title={{Gemini: Mapping and Architecture Co-exploration for Large-scale DNN Chiplet Accelerators}}, 
  year={2024},
  pages={156--171}
}

@ARTICLE{das_jetcs24,
  author={Das, Abhijit and Palesi, Maurizio and Kim, John and Pratim Pande, Partha},
  journal={IEEE Journal on Emerging and Selected Topics in Circuits and Systems}, 
  title={Chip and Package-Scale Interconnects for General-Purpose, Domain-Specific, and Quantum Computing Systems—Overview, Challenges, and Opportunities}, 
  year={2024},
  volume={14},
  number={3},
  pages={354--370},
}

@ARTICLE{Deeploy,
  author={Scherer, Moritz and Macan, Luka and Jung, Victor J. B. and Wiese, Philip and Bompani, Luca and Burrello, Alessio and others},
  journal={IEEE Transactions on Computer-Aided Design of Integrated Circuits and Systems}, 
  title={{Deeploy: Enabling Energy-Efficient Deployment of Small Language Models on Heterogeneous Microcontrollers}}, 
  year={2024},
  volume={43},
  number={11},
  pages={4009--4020},
}

@misc{XLA,
title	= {{XLA : Compiling Machine Learning for Peak Performance}},
author	= {Amit Sabne},
year	= {2020}
}

@inproceedings{ExecuTorch,
title={{ExecuTorch - A Unified PyTorch Solution to Run {ML} Models On-Device}},
author={Mergen Nachin and Digant Desai and Stephen Jia and Chen Lai and Mengwei Liu and Jacob Szwejbka and others},
booktitle={9th Conference on Machine Learning and Systems},
series = {MLSys '26},
year={2026},
url={https://openreview.net/forum?id=jmE5nwC9kb}
}

@article{MATCHA,
author = {Russo, Enrico and Hamdi, Mohamed and Ottaviano, Alessandro and Conti, Francesco and Garofalo, Angelo and Jahier Pagliari, Daniele and others},
year = {2026},
title = {{MATCHA: Efficient Deployment of Deep Neural Networks on Multi-Accelerator Heterogeneous Edge SoCs}},
journal = {arXiv preprint arXiv:2604.09124},
doi = {10.48550/arXiv.2604.09124}
}

@article{eiqNeutron2025,
      title={{eIQ Neutron: Redefining Edge-AI Inference with Integrated NPU and Compiler Innovations}}, 
      author={Lennart Bamberg and Filippo Minnella and Roberto Bosio and Fabrizio Ottati and Yuebin Wang and Jongmin Lee and others},
      year = {2025},
      journal = {arXiv preprint arXiv:2509.14388},
      archivePrefix={arXiv},
      primaryClass={cs.AR},
      url={https://arxiv.org/abs/2509.14388}, 
}

@ARTICLE{Deeploy2,
  author={Wiese, Philip and İslamoğlu, Gamze and Scherer, Moritz and Macan, Luka and Jung, Victor Jean-Baptiste and Burrello, Alessio and Conti, Francesco and Benini, Luca},
  journal={IEEE Design \& Test}, 
  title={{Toward Attention-Based TinyML: A Heterogeneous Accelerated Architecture and Automated Deployment Flow}}, 
  year={2025},
  volume={42},
  number={5},
  pages={63--72},
  doi={10.1109/MDAT.2025.3527371}}

@inproceedings{Triton,
author = {Tillet, Philippe and Kung, H. T. and Cox, David},
title = {{Triton: An Intermediate Language and Compiler for Tiled Neural Network Computations}},
year = {2019},
booktitle = {3rd ACM SIGPLAN International Workshop on Machine Learning and Programming Languages},
pages = {10–19},
numpages = {10},
series = {MAPL '19}
}

@article{roofline09,
author = {Williams, Samuel and Waterman, Andrew and Patterson, David},
title = {{Roofline: An Insightful Visual Performance Model for Multicore Architectures}},
year = {2009},
volume = {52},
number = {4},
journal = {Communications of the ACM},
pages = {65--76},
numpages = {12}
}

@inproceedings{posterCF22,
author = {Curzel, Serena and Agostini, Nicolas Bohm and Tumeo, Antonino and Ferrandi, Fabrizio},
title = {Hardware acceleration of complex machine learning models through modern high-level synthesis},
year = {2022},
booktitle = {19th ACM International Conference on Computing Frontiers},
pages = {209–210},
numpages = {2},
series = {CF '22}
}

@article{xu2024llmaidedefficienthardwaredesign,
      title={{LLM-Aided Efficient Hardware Design Automation}}, 
      author={Kangwei Xu and Ruidi Qiu and Zhuorui Zhao and Grace Li Zhang and Ulf Schlichtmann and Bing Li},
      year={2024},
      journal = {arXiv preprint arXiv:2410.18582},
      url={https://arxiv.org/abs/2410.18582}, 
}

@inproceedings{llmHLS25,
author = {Gai, Jiahao and Chen, Hao and Wang, Zhican and Zhou, Hongyu and Zhao, Wanru and Lane, Nicholas and Fan, Hongxiang},
title = {Exploring Code Language Models for Automated HLS-based Hardware Generation: Benchmark, Infrastructure and Analysis},
year = {2025},
booktitle = {30th Asia and South Pacific Design Automation Conference},
pages = {988–994},
numpages = {7}
}

@inproceedings{act,
author = {Gupta, Udit and Elgamal, Mariam and Hills, Gage and Wei, Gu-Yeon and Lee, Hsien-Hsin S. and Brooks, David and Wu, Carole-Jean},
title = {ACT: designing sustainable computer systems with an architectural carbon modeling tool},
year = {2022},
booktitle = {49th Annual International Symposium on Computer Architecture},
pages = {784–799},
numpages = {16},
location = {New York, New York},
series = {ISCA '22}
}

@inproceedings{sustainableHPC,
author = {Li, Baolin and Basu Roy, Rohan and Wang, Daniel and Samsi, Siddharth and Gadepally, Vijay and Tiwari, Devesh},
title = {{Toward Sustainable HPC: Carbon Footprint Estimation and Environmental Implications of HPC Systems}},
year = {2023},
booktitle = {International Conference for High Performance Computing, Networking, Storage and Analysis},
articleno = {19},
numpages = {15},
pages = {1-15},
series = {SC '23}
}

@inproceedings{factorflow,
author = {Ronzani, Marco and Silvano, Cristina},
title = {FactorFlow: Mapping GEMMs on Spatial Architectures through Adaptive Programming and Greedy Optimization},
year = {2025},
booktitle = {30th Asia and South Pacific Design Automation Conference},
pages = {706–712},
numpages = {7},
series = {ASPDAC '25}
}

@INPROCEEDINGS{perri24,
  author={Perri, Stefania and Zambelli, Cristian and Ielmini, Daniele and Silvano, Cristina},
  booktitle={2024 IEEE International Parallel and Distributed Processing Symposium Workshops}, 
  series = {IPDPSW '24},
  title={{Digital In-Memory Computing to Accelerate Deep Learning Inference on the Edge}}, 
  year={2024},
  pages={130-133},
  doi={10.1109/IPDPSW63119.2024.00037}
}

@article{TransAxx2024,
  author = {Danopoulos, Dimitrios and Zervakis, Georgios and Soudris, Dimitrios and Henkel, J{\"o}rg},
  title = {{TransAxx}: Efficient Transformers with Approximate Computing},
  year = {2024},
  journal = {arXiv preprint arXiv:2402.07545},
  url = {https://arxiv.org/abs/2402.07545}
}

@article{Kabir2024ADAPTOR,
  author = {Kabir, Ehsan and Bakos, Jason D. and Andrews, David and Huang, Miaoqing},
  title = {{A Runtime-Adaptive Transformer Neural Network Accelerator on FPGAs}},
  year = {2026},
  journal = {Microprocessors and Microsystems},
  volume = {120},
  pages = {105223},
  doi = {10.1016/j.micpro.2025.105223},
}

@article{Kim2025SoftmaxLayerNorm,
  author = {Kim, Raehyeong and Lee, Dayoung and Kim, Jinyeol and Park, Joungmin and Lee, Seung Eun},
  title = {{Hardware Accelerator for Approximation-Based Softmax and Layer Normalization in Transformer Models}},
  journal = {Electronics},
  year = {2025},
  volume = {14},
  number = {12},
  pages = {2337},
  doi = {10.3390/electronics14122337}
}

@article{LowPrecisionSoftmax2026,
  author = {Samuel, Aboagye and  Lujun, Zhai and Suxia, Cui},
  title = {{A Generalizable Low-Precision Softmax Approximation for Small-FPGA Devices}},
  journal = {Electronics},
  year = {2026},
  volume = {15},
  number = {9},
  pages = {1774},
  doi = {10.3390/electronics15091774}
}

@article{BAPS2026,
  author = {Ye, Zisheng and He, Xiaoyu and Song, Maoyuan and Qiu, Guoliang and Liao, Chao and Wu, Chen and others},
  title = {{BAPS: A Fine-Grained Low-Precision Scheme for Softmax in Attention via Block-Aware Precision reScaling}},
  year = {2026},
  journal = {arXiv preprint arXiv:2602.02071},
  url = {https://arxiv.org/abs/2602.02071}
}

@INPROCEEDINGS{luo_iccad25,
  author={Luo, Donger and Li, Tianyi and Li, Xinheng and Sun, Qi and Zhuo, Cheng and Yu, Bei and others},
  booktitle={2025 IEEE/ACM International Conference On Computer Aided Design}, 
  series = {ICCAD '25},
  title={{LLM-Augmented Multi-Modal Fusion for SoC Design Space Exploration}}, 
  year={2025},
  pages={1--8},
  doi={10.1109/ICCAD66269.2025.11240720}
}

@INPROCEEDINGS{youssef_iccad25,
  author={Youssef, Ismael and Yang, Hang and Hao, Cong},
  booktitle={2025 IEEE/ACM International Conference On Computer Aided Design}, 
  series = {ICCAD '25}, 
  title={{LaZagna: An Open-Source Framework for Flexible 3D FPGA Architectural Exploration}}, 
  year={2025},
  doi={10.1109/ICCAD66269.2025.11240893}
}

@inproceedings{Long2020QPIM,
  author    = {Long, Yun and Lee, Edward A. and Mukhopadhyay, Saibal},
  title     = {{Q-PIM: A Genetic Algorithm based Flexible DNN Quantization Method and Application to Processing-In-Memory Platform}},
  booktitle = {57th ACM/IEEE Design Automation Conference},
  series = {DAC '20},
  pages     = {1--6},
  year      = {2020},
  publisher = {IEEE},
  doi       = {10.1109/DAC18072.2020.9218693}
}

@inproceedings{Ankit19,
author = {Ankit, Aayush and Hajj, Izzat El and Chalamalasetti, Sai Rahul and Ndu, Geoffrey and Foltin, Martin and others},
title = {PUMA: A Programmable Ultra-efficient Memristor-based Accelerator for Machine Learning Inference},
year = {2019},
doi = {10.1145/3297858.3304049},
booktitle = {24th International Conference on Architectural Support for Programming Languages and Operating Systems},
series = {ASPLOS '19},
pages = {715--731},
}

@inproceedings{Zhang25,
author = {Zhang, Yuanpeng and Hu, Xing and Chen, Xi and Yuan, Zhihang and Li, Cong and Zhu, Jingchen and others},
title = {{AIM: Software and Hardware Co-design for Architecture-level IR-drop Mitigation in High-performance PIM}},
year = {2025},
isbn = {9798400712616},
doi = {10.1145/3695053.3730987},
booktitle = {52nd Annual International Symposium on Computer Architecture},
series = {ISCA '25},
pages = {849–866},
}

@INPROCEEDINGS{Soeken16,
  author={Soeken, Mathias and Shirinzadeh, Saeideh and Gaillardon, Pierre-Emmanuel and Amarú, Luca Gaetano and Drechsler, Rolf and De Micheli, Giovanni},
  booktitle={2016 53nd ACM/EDAC/IEEE Design Automation Conference}, 
  series = {DAC '16},
  title={{An MIG-based compiler for programmable logic-in-memory architectures}}, 
  year={2016},
  pages={1-6},
  doi={10.1145/2897937.2897985}
}

@Article{Mambu22,
AUTHOR = {Mambu, Kevin and Charles, Henri-Pierre and Kooli, Maha and Dumas, Julie},
TITLE = {{Towards Integration of a Dedicated Memory Controller and Its Instruction Set to Improve Performance of Systems Containing Computational SRAM}},
JOURNAL = {Journal of Low Power Electronics and Applications},
VOLUME = {12},
YEAR = {2022},
NUMBER = {1},
DOI = {10.3390/jlpea12010018}
}

@Article{Mannocci2026,
author={Mannocci, Piergiulio
and Larelli, Giacomo
and Bonomi, Marco
and Ielmini, Daniele},
title={{Achieving High Precision in Analog In-Memory Computing Systems}},
journal={npj Unconventional Computing},
year={2026},
volume={3},
number={1},
pages={1},
doi={10.1038/s44335-025-00044-2},
}

@ARTICLE{Asifuzzman26,
  author={Asifuzzaman, Kazi and He, Yuan and Zhang, Tianyun and Tang, Eric and Rao Miniskar, Narasinga and Teranishi, Keita and Vetter, Jeffrey S.},
  journal={IEEE Access}, 
  title={A Survey on the Expanding Scope and Interdisciplinary Opportunities for Processing-in-Memory Techniques}, 
  year={2026},
  volume={14},
  pages={18408-18430},
  doi={10.1109/ACCESS.2026.3659051}
}

@article{chen21,
title = {{Multiply Accumulate Operations in Memristor Crossbar Arrays for Analog Computing}},
journal = {Journal of Semiconductors},
volume = {42},
number = {1},
pages = {013104},
year = {2021},
doi = {10.1088/1674-4926/42/1/013104},
author = {Chen, Jia and Li, Jiancong and Li, Yi and Miao, Xiangshui},
}

@ARTICLE{Baroni22,
AUTHOR={Baroni, Andrea  and Glukhov, Artem  and Pérez, Eduardo  and Wenger, Christian  and Calore, Enrico  and Schifano, Sebastiano Fabio  and others},
TITLE={An energy-efficient in-memory computing architecture for survival data analysis based on resistive switching memories},      
JOURNAL={Frontiers in Neuroscience},
VOLUME={16},
YEAR={2022},
DOI={10.3389/fnins.2022.932270},
}

@inproceedings{Khan25,
author = {Khan, Asif Ali and Farzaneh, Hamid and Friebel, Karl Friedrich Alexander and Fournier, Cl\'{e}ment and Chelini, Lorenzo and Castrillon, Jeronimo},
title = {{CINM (Cinnamon): A Compilation Infrastructure for Heterogeneous Compute In-Memory and Compute Near-Memory Paradigms}},
year = {2025},
doi = {10.1145/3622781.3674189},
booktitle = {29th ACM International Conference on Architectural Support for Programming Languages and Operating Systems},
series = {ASPLOS '24},
volume = {4},
pages = {31–46},
}

@INPROCEEDINGS{comprime,
  author={Frerix, Steffen and Shirinzadeh, Saeideh and Fröhlich, Saman and Drechsler, Rolf},
  booktitle={2019 IEEE/ACM International Symposium on Nanoscale Architectures}, 
  series = {NANOARCH '19},
  title={{ComPRIMe: A Compiler for Parallel and Scalable ReRAM-based In-Memory Computing}}, 
  year={2019},
  pages={1-6},
  doi={10.1109/NANOARCH47378.2019.181285}
}

@article{sun25,
author = {Sun, Xiaotian and Wang, Xinyu and Li, Wanqian and Han, Yinhe and Chen, Xiaoming},
title = {{PIMCOMP: An End-to-End DNN Compiler for Processing-In-Memory Accelerators}},
year = {2025},
volume = {44},
number = {5},
doi = {10.1109/TCAD.2024.3496847},
journal = {IEEE Transactions on Computer-Aided Design of Integrated Circuits and Systems},
pages = {1745--1759},
}

@inproceedings{narayanan21megatron,
  author       = {Deepak Narayanan and
                  Mohammad Shoeybi and
                  Jared Casper and
                  Patrick LeGresley and
                  Mostofa Patwary and
                  Vijay Korthikanti and
                  others},
  title        = {{Efficient Large-scale Language Model Training on GPU Clusters using Megatron-LM}},
  booktitle    = {International Conference for High Performance Computing, Networking, Storage and Analysis}, 
  series = {SC '21},
  pages        = {58},
  publisher    = {{ACM}},
  year         = {2021},
  doi          = {10.1145/3458817.3476209},
}

@inproceedings{kwon23paged,
author = {Kwon, Woosuk and Li, Zhuohan and Zhuang, Siyuan and Sheng, Ying and Zheng, Lianmin and Yu, Cody Hao and others},
title = {{Efficient Memory Management for Large Language Model Serving with PagedAttention}},
booktitle = {29th Symposium on Operating Systems Principles},
series = {SOSP '23},
pages = {611--626},
year = {2023},
publisher = {ACM},
doi = {10.1145/3600006.3613165},
}

@article{zeng25survey,
  author       = {Fanlong Zeng and
                  Wensheng Gan and
                  Yongheng Wang and
                  Philip S. Yu},
  title        = {{Distributed Training of Large Language Models: {A} Survey}},
  journal      = {Natural Language Processing Journal},
  volume       = {12},
  pages        = {100174},
  year         = {2025},
  doi          = {10.1016/J.NLP.2025.100174},
}

@ARTICLE{shuvo23, 
author={Shuvo, Md. Maruf Hossain and Islam, Syed Kamrul and Cheng, Jianlin and Morshed, Bashir I.}, 
journal={Proceedings of the IEEE}, 
title={{Efficient Acceleration of Deep Learning Inference on Resource-Constrained Edge Devices: A Review}}, 
year={2023}, 
volume={111}, 
number={1}, 
pages={42--91}, 
doi={10.1109/JPROC.2022.3226481}}

@inproceedings{nicosanti26,
author = {Nicosanti, Simone and Russo~Russo, Gabriele and Cardellini, Valeria},
title = {{Energy- and Quantization-aware DNN Partitioning in the Edge-Cloud Continuum (Work In Progress Paper)}},
year = {2026},
publisher = {ACM},
doi = {10.1145/3777911.3801106},
booktitle = {Companion of the 17th ACM/SPEC International Conference on Performance Engineering},
series = {ICPE Companion '26},
pages = {47–54},
}

@inproceedings{patel23splitwise,
author = {Patel, Pratyush and Choukse, Esha and Zhang, Chaojie and Shah, Aashaka and Goiri, \'{I}\~{n}igo and Maleki, Saeed and Bianchini, Ricardo},
title = {{Splitwise: Efficient Generative LLM Inference Using Phase Splitting}},
year = {2025},
doi = {10.1109/ISCA59077.2024.00019},
booktitle = {51st Annual International Symposium on Computer Architecture},
series = {ISCA '24},
pages = {118–132},
}

@inproceedings{romero21infaas,
  author       = {Francisco Romero and
                  Qian Li and
                  Neeraja J. Yadwadkar and
                  Christos Kozyrakis},
  title        = {{INFaaS: Automated Model-less Inference Serving}},
  booktitle    = {2021 {USENIX} Annual Technical Conference},
  series = {USENIX ATC '21},  
  pages        = {397-411},
  year         = {2021},
  url          = {https://www.usenix.org/conference/atc21/presentation/romero},
}

@article{ziller26greenserv,
  author       = {Thomas Ziller and
                  Shashikant Ilager and
                  Alessandro Tundo and
                  Ezio Bartocci and
                  Leonardo Mariani and
                  Ivona Brandic},
  title        = {{GreenServ: Energy-Efficient Context-Aware Dynamic Routing for Multi-Model {LLM} Inference}},
  journal      = {arXiv preprint arXiv:2601.17551},
  year         = {2026},
  doi          = {10.48550/ARXIV.2601.17551},
}

@inproceedings{zhong24distserve,
  author       = {Yinmin Zhong and
                  Shengyu Liu and
                  Junda Chen and
                  Jianbo Hu and
                  Yibo Zhu and
                  Xuanzhe Liu and
                  others},
  title        = {{DistServe: Disaggregating Prefill and Decoding for Goodput-optimized
                  Large Language Model Serving}},
  booktitle    = {18th {USENIX} Symposium on Operating Systems Design and Implementation},
  series       = {OSDI '24},
  pages        = {193--210},
  year         = {2024},
  url          = {https://www.usenix.org/conference/osdi24/presentation/zhong-yinmin},
 }

@inproceedings{kang17neurosurgeon,
  author       = {Yiping Kang and
                  Johann Hauswald and
                  Cao Gao and
                  Austin Rovinski and
                  Trevor N. Mudge and
                  Jason Mars and
                  Lingjia Tang},
  title        = {{Neurosurgeon: Collaborative Intelligence Between the Cloud and Mobile Edge}},
  booktitle    = {22nd International Conference on Architectural Support for Programming Languages and Operating Systems},
  series = {ASPLOS '17},
  pages        = {615--629},
  publisher    = {{ACM}},
  year         = {2017},
  doi          = {10.1145/3037697.3037698},
}

@article{wang26scalpipe,
  author={Wang, Nianfu and Wang, Wanyou and Zhong, Xiaoxiong and Liu, Jingyu and Shi, Gaotao and Li, Zhijun},
  journal={IEEE Transactions on Mobile Computing}, 
  title={{ScalPipe: Scalable Collaborative Pipeline Inference for Distributed Heterogeneous Devices}}, 
  year={2026},
  volume={},
  number={},
  pages={1-15},
  doi={10.1109/TMC.2026.3693694}}

@inproceedings{hu22pipeedge,
  author       = {Yang Hu and
                  Connor Imes and
                  Xuanang Zhao and
                  Souvik Kundu and
                  Peter A. Beerel and
                  Stephen P. Crago and
                  John Paul Walters},
  title        = {{PipeEdge: Pipeline Parallelism for Large-Scale Model Inference on
                  Heterogeneous Edge Devices}},
  booktitle    = {25th Euromicro Conference on Digital System Design},
  series = {DSD '22}, 
  pages        = {298--307},
  publisher    = {{IEEE}},
  year         = {2022},
  doi          = {10.1109/DSD57027.2022.00048},
}

@article{li26apdrl,
      title={{AP-DRL: A Synergistic Algorithm-Hardware Framework for Automatic Task Partitioning of Deep Reinforcement Learning on Versal ACAP}}, 
      author={Enlai Li and Zhe Lin and Sharad Sinha and Wei Zhang},
      year={2026},
      journal = {arXiv preprint arXiv:2603.29369},
      url={https://arxiv.org/abs/2603.29369}, 
}

@inproceedings{danopoulos26aie4ml,
  author={Danopoulos, Dimitrios and Lupi, Enrico and Sun, Chang and Dittmeier, Sebastian and Kagan, Michael and Loncar, Vladimir and Pierini, Maurizio},
  booktitle={2026 IEEE 34th Annual International Symposium on Field-Programmable Custom Computing Machines}, 
  series = {FCCM '26}, 
  title={{AIE4ML: An End-to-End Framework for Compiling Neural Networks for the Next Generation of AMD AI Engines}}, 
  year={2026},
  pages={176--184},
  doi={10.1109/FCCM68464.2026.00035}
}

@misc{nvidia_fastertransformer,
  author       = {{NVIDIA}},
  title        = {{FasterTransformer}},
  year         = {2021},
  howpublished = {GitHub repository}, 
  note         = {\url{https://github.com/NVIDIA/FasterTransformer}, accessed: 2026-07-07},
}

@misc{nvidia_tensorrt_llm,
  author       = {{NVIDIA}},
  title        = {{TensorRT-LLM}},
  year         = {2023},
  howpublished = {GitHub repository}, 
  note         = {\url{https://github.com/NVIDIA/TensorRT-LLM}, accessed: 2026-07-07},
}

@inproceedings{sheng23flexgen,
  author       = {Ying Sheng and
                  Lianmin Zheng and
                  Binhang Yuan and
                  Zhuohan Li and
                  Max Ryabinin and
                  Beidi Chen and
                  others},
  title        = {{FlexGen: High-Throughput Generative Inference of Large Language Models with a Single GPU}},
  booktitle    = {2023 International Conference on Machine Learning}, 
  series = {ICML '23},
  volume       = {202},
  pages        = {31094--31116},
  publisher    = {{PMLR}},
  year         = {2023},
  url          = {https://proceedings.mlr.press/v202/sheng23a.html},
}

@inproceedings{mei25helix,
author = {Mei, Yixuan and Zhuang, Yonghao and Miao, Xupeng and Yang, Juncheng and Jia, Zhihao and Vinayak, Rashmi},
title = {{Helix: Serving Large Language Models over Heterogeneous GPUs and Network via Max-Flow}}, 
year = {2025},
doi = {10.1145/3669940.3707215},
booktitle = {Proceedings of the 30th ACM International Conference on Architectural Support for Programming Languages and Operating Systems}, 
volume = {1},
pages = {586--602},
series = {ASPLOS '25}
}

\end{document}